\documentclass[iop]{emulateapj}

\usepackage{savesym}
\usepackage{natbib}
\usepackage{amsmath,amssymb,amsthm,mathrsfs,dsfont}
\usepackage{graphicx}
\usepackage{subfigure}
\usepackage{captcont}
\usepackage{float}
\usepackage{booktabs}
\usepackage{epstopdf}
\usepackage{color}
\usepackage[colorlinks,linkcolor=blue,citecolor=blue,urlcolor=blue]{hyperref}

\newcommand{\yr}{{\rm \,yr}}

\newcommand{\gyr}{{\rm \,Gyr}}
\newcommand{\pyr}{{\rm \,yr^{-1}}}

\newcommand{\pc}{{\rm \,pc}}
\newcommand{\kpc}{{\rm \, kpc}}
\newcommand{\mpc}{{\rm \, Mpc}}

\newcommand{\kms}{{\rm \,km\,s^{-1}}}
\newcommand{\hz}{{\rm \, Hz}}
\newcommand{\mhz}{{\rm \, mHz}}
\newcommand{\nhz}{{\rm \, nHz}}
\newcommand{\msun}{{\rm M_{\odot}}}
\newcommand{\lsun}{{\rm L_{\odot}}}
\newcommand{\rmb}{_{\rm b}}
\newcommand{\rmc}{_{\rm c}}
\newcommand{\rms}{_{\rm s}}

\newcommand{\rmr}{_{\rm r}}
\newcommand{\gw}{_{\rm gw}}
\newcommand{\gr}{_{\rm gr}}
\newcommand{\infl}{_{\rm infl}}
\newcommand{\calE}{\mathcal{E}}
\newcommand{\calF}{{\mathcal{F}}}
\newcommand{\calI}{\mathcal{I}}
\newcommand{\calM}{\mathcal{M}}
\newcommand{\peak}{_{\rm peak}}
\newcommand{\evol}{_{\rm evol}}

\newcommand{\atlas}{{\rm ATLAS^{3D}}}
\newcommand{\bulge}{_{\rm *}}
\newcommand{\na}{New Astronomy}
\newcommand{\hst}{HST}

\newcommand{\obs}{_{\rm obs}}
\newcommand{\ayr}{{A_{\rm yr}}}
\newcommand{\pair}{{_{\rm pair}}}
\newcommand{\mrg}{{_{\rm mrg}}}
\newcommand{\tobs}{{\langle T\mrg\rangle}}
\newcommand{\proj}{_{\rm p}}

\newcommand{\mpri}{{M_{\rm gal,1}}}
\newcommand{\msec}{{M_{\rm gal,2}}}
\newcommand{\gal}{{_{\rm gal}}}
\newcommand{\mgal}{{M_{\rm gal}}}
\newcommand{\qgal}{{q_{\rm gal}}}
\newcommand{\mugal}{{\mu_{\rm gal}}}
\newcommand{\bgal}{{_{*}}}

\newcommand{\intact}{{_{\rm intact}}}

\newcommand{\bbh}{{_{\rm BBH}}}
\newcommand{\bh}{{_{\rm BH}}}
\newcommand{\coal}{{_{\rm coal}}}
\newcommand{\prog}{{_{\rm prog}}}
\newcommand{\rmh}{{_{\rm h}}}
\newcommand{\be}{\begin{equation}}
\newcommand{\ee}{\end{equation}}
\newcommand{\calR}{{\cal R}}

\begin{document}

\title{Dynamical evolution of cosmic supermassive binary black holes and
their gravitational wave radiation}
\shorttitle{Supermassive binary black holes and gravitational wave radiation}
\shortauthors{Chen, Yu, \& Lu}

\author{Yunfeng Chen$^1$, Qingjuan Yu$^{1,\dagger}$, and Youjun Lu$^{2,3}$}
\affil{$^1$~Kavli Institute for Astronomy and Astrophysics, and School of Physics, Peking University, Beijing, 100871, China; $^\dagger$yuqj@pku.edu.cn \\
       $^2$~National Astronomical Observatories, Chinese Academy of
       Sciences, Beijing, 100012, China\\
       $^3$~School of Astronomy and Space Science, University of Chinese
Academy of Sciences, Beijing 100049, China}

\begin{abstract}
We investigate the evolution of supermassive binary black holes (BBHs) in
galaxies with realistic property distributions and the gravitational-wave (GW)
radiation from the cosmic population of these BBHs. We incorporate a
comprehensive treatment of the dynamical interactions of the BBHs with their
environments by including the effects of galaxy triaxial shapes and inner
stellar distributions, and generate a large number of BBH evolution tracks. By
combining these BBH evolution tracks, galaxy mass functions, galaxy merger
rates, and supermassive black hole--host galaxy relations into our model, we
obtain the statistical distributions of surviving BBHs, BBH coalescence rates,
the strength of their GW radiation, and the stochastic GW background (GWB)
contributed by the cosmic BBH population. About $\sim$1\%--3\% (or $\sim 10\%$)
of supermassive BHs at nearby galactic centers are expected to be binaries with
mass ratio $>1/3$ (or $>1/100$).  The characteristic strain amplitude of the
GWB at frequency $1\pyr$ is estimated to be $\sim 2.0^{+1.4}_{-0.8}\times
10^{-16}$, and the upper bound of its results obtained with the different
BH--host galaxy relations can be up to $5.4\times 10^{-16}$, which await
testing by future experiments (e.g., the Square Kilometer Array, FAST,
Next-Generation Very Large Array).  The turnover frequency of the GWB spectrum
is at $\sim 0.25\nhz$. The uncertainties on the above estimates and prospects
for detecting individual sources are also discussed. The application of the
cosmic BBH population to the Laser Interferometer Space Antenna (LISA) band
provides a lower limit to the detection rate of BBHs by LISA, ${\sim}0.9\pyr$.
\end{abstract}


\section{Introduction} 
\label{sec:introduction}

In the modern paradigm of hierarchical galaxy formation and evolution, a natural
consequence of mergers of galaxies with central supermassive black holes (SMBHs,
hereafter BHs) is the formation of supermassive binary black holes (SMBBHs,
hereafter BBHs) at galactic centers (e.g., \citealt{Begelman80,Yu02,Benson10}).
The formed binaries interact with surrounding stars and gas, which may
distinguish the systems with some unique astronomical phenomena different from
those with only one BH. If the separation of the two BHs can decay to be
sufficiently close through the interactions, significant gravitational waves
(GWs) will be radiated, taking away energy and angular momentum, leading to
mergers of the BBHs, and contributing to the GW background (GWB) at the
relatively low-frequency band [e.g., the detection frequency bands of the
Pulsar Timing Array (PTA) experiments \citep{Hobbs13, McLaughlin13, Desvignes16,
Verbiest16}, Laser Interferometer Space Antenna (LISA;
\url{https://www.lisamission.org}), and Taiji/Tianqin
\citep{Ruanetal19,Wangetal19}].  Studies of the strength of the GW radiation
caused by BBH mergers, the characteristics of BBH systems and their statistical
properties, would provide an important probe to the hierarchical structure
formation and evolution model, BH physics, and the gravitation theory.

In this work, we investigate the expected statistical distributions of BBHs
surviving in the nearby universe and the strength of the GWs radiated by merging
BBHs. To determine how many BBHs can survive or coalesce, the dynamical
evolution timescales of BBHs play a vital role. For example, \citet{Yu02} shows
that the evolution of BBHs in a gas-poor system depends on BH mass ratio and
host galaxy type.  BBHs with very low mass ratios (say, $\la 0.001$) are hardly
formed by mergers of galaxies, because the dynamical friction timescale is too
long for the smaller BH to sink into the galactic center within a Hubble time.
For BBHs with moderate mass ratios, one essential factor to determine their
evolution timescales is how many stars can pass by the vicinity of the BBH to
interact strongly with the BBH and make it lose energy, and it has been shown
that BBHs in low-dispersion galaxies or in highly flattened or triaxial galaxies
have relatively low evolution timescales \citep{Yu02}. Over-simplified
treatments on the interaction of a BBH with its environments may lead to some
uncertainty in the current estimation of the strength of the GW radiation
emitted by the BBHs, as the interactions affect how many BBHs are surviving at a
given separation, and when and where gravitational radiation becomes dominant in
making a BBH lose energy; and a realistic treatment of the BBH interaction with
their environments is necessary for the BBH dynamical evolution models and the
GW radiation estimation \citep{Shannon15,Arzoumanian16,Lentati15}.  To obtain
the coalescence rate of merging BBHs in the realistic universe and the realistic
statistical distribution of surviving BBHs, in this work we model the dependence
of the BBH dynamical evolution on stellar distributions and galaxy shapes, and
we incorporate the BBH dynamical evolution model with the realistic
distributions of galaxy properties (e.g., stellar distributions, galaxy shapes,
galaxy merger rates).

This paper is organized as follows. In Section 2, we review the dynamical
processes and the model for the evolution of a BBH in gas-poor systems. The
detailed model is mainly adopted from \citet{Yu02}. In Section 3, we describe
the modeling of the distribution of surviving BBHs, BBH coalescence rates, and
the strength of stochastic GWB.  In Section 4, we describe the galaxy samples,
the realistic galaxy distributions and the galaxy merger rates used in the
model.  In Section 5, we present our results for the distribution of surviving
BBHs and the GW strength radiated by merging BBHs. Discussion is given in
Section~\ref{sec:discussion} and conclusions are given in
Section~\ref{sec:conclusion}.

\section{Orbital evolution of BBHs}
\label{sec:dynamics}

In a gas-poor galaxy merger, the formation and evolution of a BBH can be divided
into the following stages, according to the dominant physical mechanisms that
act to drain the energy of the orbit of two BHs \citep{Begelman80, Yu02}. The
relative importance of each mechanism is determined by the spatial scale of the
separation between the two BHs. (1) Dynamical friction stage: after two galaxies
merge, each BH in its progenitor galaxy will sink into the center of the common
potential well of the merging galaxy, due to dynamical friction, from a
separation of several tens of $\kpc$ to ${\sim}10\pc$. (2) Non-hard binary
stage: the two BHs with masses $M_1$ and $M_2$ ($M_2\le M_1$) form a bound
system with semi-major axis
\begin{eqnarray}
a \lesssim  R\infl & \equiv & \frac{GM\bh}{\sigma\rmc^2} \nonumber \\
& \simeq & 10\left(\frac{M\bh}{10^8\msun}\right)
\left(\frac{\sigma\rmc}{200\kms}\right)^{-2}\pc,
\label{eq:rinfl}
\end{eqnarray}
where $M\bh= M_1 + M_2 $, $\sigma\rmc$ is the one-dimensional velocity
dispersion of the merged galaxy core. The dynamical friction continues to make
the BBH lose energy but becomes less and less effective as the binary orbital
velocity increases and its orbital period decreases. Meanwhile, from time to
time a star walks to the vicinity of the BBH and the three-body interactions
between the BBH and the stars gradually become the dominant mechanism to make
the BBH lose energy. This stage is an intermediate stage between the previous
dynamical friction stage and the next hard binary stage. (3) Hard binary stage:
the BBH becomes hard when its semi-major axis
\begin{eqnarray}
a \la  a\rmh & \equiv & \frac{G M_2 }{4\sigma\rmc^2} \nonumber \\
& \simeq & 2.8\left(\frac{ M_2 }{10^8\msun}\right)
\left(\frac{\sigma\rmc}{200\kms}\right)^{-2}\pc.  \label{eq:ah}
\end{eqnarray}
A hard binary loses energy mainly by interacting with stars passing in their
vicinity, most of which will be expelled from the BBH with an energy gain after
one or more encounters with it. In such expulsions, the relative change in the
BBH orbital energy is independent of the energy \citep{Heggie75}, which can
simplify the analysis of the BBH evolution timescale at this stage. (4)
Gravitational radiation stage: when the semimajor axis of the BBH is small
enough (e.g., $<10^{-2}\pc$), gravitational radiation becomes the dominant
effect to drive the BBH orbital decay. This stage is likely to occur before the
BBH becomes hard if $M_2\ll M_1$.

In this paper, we obtain the orbital evolution of a BBH by following the model
presented in \citet{Yu02}, supplemented with a numerical method to calculate
the stellar orbital motion and obtain the stars that can move to the vicinity
the BBH in triaxial gravitational potentials. We summarize the model of the
evolution timescales of
the BBH at the four stages in Section 2.1-2.4, respectively.  Note that the
evolution of the BBHs at the non-hard and the hard stages is critical to
determine how the BBHs are distributed at pc and sub-pc scales, and when and
where the gravitational stages start. One essential factor to determine the BBH
evolution timescales in those stages is how many stars on low-angular momentum
orbits, which can pass through the vicinity of the BBH, are available to
interact strongly with the BBH. Even if the low-angular momentum stars initially
in the stellar system are expelled away by interactions with the BBH, the stars
on high-angular momentum orbits can still move onto low-angular momentum orbits
by the two-body relaxation or by orbital precession in non-spherical (e.g.,
flattened or triaxial) gravitational potentials \citep{Magorrian99, Yu02,
Berczik06, Khan11, Preto11, Khan13, Vasiliev15}. In this work, we include the
effect of the non-spherical potentials, as well as that of the two-body
relaxation. We define a ``loss region'' by the (specific energy, specific
angular momentum) phase space of the stars that can precess onto low-angular
momentum orbits with pericenter distances $\sim a$. The stars in the loss region
can pass by the BBH vicinity and then escape from the BBH after strong
interactions with it. We obtain the loss region by numerical simulations, and
the method is described in Section 2.5.

In the model, we assume that the BHs are on nearly circular orbits and ignore their
eccentricity evolution at the least for the following reasons. (1) At the
dynamical friction stage, the orbital decay of a star cluster or a secondary BH
generally does not result in a highly eccentric orbit when the two BHs become
bound \citep{PR94}. (2) If the BBH eccentricity when the two BHs become bound is
small (e.g., $\la 0.3$), \citet{Quinlan96} shows that the BBH eccentricity
hardly grows as the BBH hardens. (3) The BBH eccentricity decays quickly at the
gravitational radiation stage \citep{Peters63, Baker06}.  (4) In kinematically
anisotropic spherical systems with stars mostly counter-rotating with the BBH
orbit, some simulations show that the BBH eccentricity could increase
significantly \citep{Amaro-Seoane10, Sesana11, Holley-Bockelmann15}, however it
is expected that the significant increase in BBH eccentricity can be decreased
or erased if the systems are triaxial, by applying the same reasoning as for the
alignment-erasing effect obtained in \citet{Cui14}. (5) Evolution of triple or
multiple BHs can result in BBHs with extreme eccentricity; however, the
contribution of those cases to the stochastic GWB at the PTA bands is shown to
be insignificant (see the left panel of Fig.~4 in \citealt{Bonetti18}). (6) It
was suggested that a significant fraction of BBHs should have extreme
eccentricities close to 1 so that the strength of gravitational waves radiated
by mergers of those BBHs can be relatively low (e.g., see Fig.~1 in
\citealt{Chen16}), in order to relieve the tension between the non-detection of
the stochastic GWBs and the model expectation. According to the results of this
work below, the assumption of BBHs with extreme eccentricities are not
necessarily required. In addition, exploration on detection prospects for
individual BBH sources specifically with extreme eccentricities formed in special
conditions are beyond the scope of this work.

In the model, the evolution of the BBH orbital semimajor axis is described by
$a(t)$ or $a(\tau)$, where $a$ is the orbital semimajor axis of the BBH or the
separation of the two BHs (if they are not bound, yet), $t$ is the cosmic time,
and $\tau\equiv t-t'$ is the period since an event at cosmic time $t'$
(e.g, galaxy merger).  The BBH evolution timescale at a given $a$ is defined by 
\be
t_{\rm evol}\equiv |a/\dot{a}|,
\label{eq:tevol}
\ee
where $\dot{a}\equiv da/dt=da/d\tau$.

At the dynamical friction stage, $\dot{a}$ comes mainly from the contribution of
dynamical friction, and we denote the corresponding evolution timescale by the
dynamical friction timescale $t_{\rm df}$. After the BBH becomes bound, if we
only include the contribution of gravitation radiation to $\dot{a}$, the
corresponding evolution timescale is denoted by $t\gr$
(see Eq.~\ref{eq:gw_timescale} below); and if we only
include the contribution of the BBH interaction with stars to $\dot{a}$, the
corresponding evolution timescale is denoted by the BBH hardening timescale
$t_{\rm h}$. We obtain the BBH evolution timescale after the BBH becomes bound
by
\begin{equation}
t_{\rm evol}=(t_{\rm h}^{-1}+t\gr^{-1})^{-1}.
\end{equation}
We have $t_{\rm evol}\simeq t_{\rm h}$ at the non-hard and the hard stages and
$t_{\rm evol}\simeq t\gr$ at the gravitational radiation stage.

\subsection{Dynamical friction stage}
\label{sec:dynamics::dynfri}

In the dynamical friction stage, each of the BHs sinks into the common center
of the merger remnant independently.  We carry out a simple analysis of the
evolution timescale at the dynamical friction stage. Assume that a BH of mass
$M_1$ is located at the center of a spherical galaxy with stellar mass density
$\rho(r)$ and one-dimensional velocity dispersion $\sigma(r)$. Suppose that a BH
with mass $M_2$ that used to be at the center of a galaxy is orbiting with a
velocity $\bf{v}$ and spiraling into the center of the parent galaxy by
dynamical friction. Because the in-spiraling BH $M_2$ is accompanied by the
stars bound to it with total mass $M_{2,\ast}$, dynamical friction brings the
two BHs together much more rapidly than if $M_2$ is naked
\citep{Milosavljevic01, Yu02}. The dynamical friction exerts on $M_2$ and its
accompanying stars $M_{2,\ast}$ and decelerate them at a rate given by
\begin{eqnarray}
\frac{d\mathbf{v}}{dt} = & &
\frac{4\pi G^2( M_2 +M_{2,\ast})\rho(r)\ln\Lambda}{v^3} \nonumber\\
& \times &
\left[{\rm erf}(X)-\frac{2X}{\sqrt{\pi}}{\rm e}^{-X^2}\right]\mathbf{v},
\label{eq:dynfri}
\end{eqnarray}
where $v=\vert\mathbf{v}\vert $, $X=v/(\sqrt{2}\sigma) $, erf is the error
function, and the logarithm of the ratio of the maximum and the minimum impact
parameter $\ln\Lambda$ is set to unity.

The mass of the stars bound to the secondary BH $ M_2 $ can be estimated as
follows. In the potential of the spherical galaxy with central BH $ M_1 $, the
stars around the BH $M_2$ is tidally truncated at the radius
\begin{eqnarray}
r_{\rm t} \simeq
\left[\frac{G( M_2 +M_{2,\ast})}{4\Omega^2-\kappa^2}\right]^{1/3} & \simeq &
\left[\frac{GM_{2,\ast}}{4\Omega^2-\kappa^2}\right]^{1/3}, \nonumber\\
& & M_{2,\ast} \gg  M_2 
\label{eq:rt}
\end{eqnarray}
where $ \Omega $ and $ \kappa $ are the circular frequency and epicycle
frequency in the parent galaxy. If the stars surrounding each of the BHs
follow a singular isothermal distribution, and their one-dimensional velocity
dispersions $ \sigma_1 $ and $ \sigma_2 $, respectively, we have $
\Omega^2=2\sigma_1^2/r^2 $, $ \kappa^2=4\sigma_1^2/r^2 $ and $
M_{2,\ast}=2\sigma_2^2 r_{\rm t}/G $. By applying them into Equation
(\ref{eq:rt}), we have $ r_{\rm t}=\frac{\sigma_2}{\sqrt{2}\sigma_1}r $, and the
total mass of the stars bound to the BH $ M_2 $ is given by
\begin{equation}
M_{2,\ast}(r) =
\frac{2\sigma_2^2r_{\rm t}}{G} \simeq
\frac{\sqrt{2}r\sigma_2^3}{G\sigma_1}.
\label{eq:m2star}
\end{equation}

We follow equations (40)-(42) in \citet{Yu02} to obtain the BBH evolution
timescale at the dynamical friction stage $t_{\rm df}$.

\subsection{Non-hard binary stage}
\label{sec:dynamics:nonhard}

The BBH orbital decay in the non-hard binary stage comes from both dynamical
friction from distant stars and three-body interactions between the BBH and the
stars passing the BBH vicinity.  As done in \citet{Yu02}, we apply the dynamical
friction timescale $t_{\rm df}$ to estimate the BBH hardening timescales $t_{\rm h}$
at the non-hard binary stage, because scattering experiments with the restricted
three-body approximation basically give a hardening timescale similar to the
application of the dynamical friction timescale \citep{Quinlan96}.

Stars may gain energy and be removed from the galactic core by interactions with
the BBH. Before the BBH becomes bound, the stellar mass removed from the
galactic core due to the BH orbit decay can be ignored.  During the non-hard
binary stage, the depletion of the stars can be significant in some galaxies. Before
the BBH becomes hard, if the mass of the initial stellar population in the loss
region is not sufficient, the BBH would lose energy mainly by dynamical friction
with distant stars (which becomes more and more inefficient as the BBH hardens)
and by three-body interactions with the low-angular-momentum stars diffused from
high-angular-momentum orbits by two-body relaxation (which will dominate the BBH
hardening timescales in the hard binary stage).  Thus the BBH hardening
timescale should be higher than those estimated from $t_{\rm df}$ and smoothly
increase to connect the hardening timescale in the hard binary stage.  We use
the linear approximation to obtain the BBH hardening timescale at $a\rmh\le a
\le a_{\rm dp}$: 
\begin{equation}
\ln[t\rmh(a)]=
\ln[t\rmh(a\rmh)]+\ln[t_{\rm df}(a_{\rm dp})
/t\rmh(a\rmh)]/\ln(a_{\rm dp}/a\rmh),
\end{equation}
where $a_{\rm dp}$ is the semimajor axis where the initial population of stars
in the loss region are all removed from the loss region, $t_{\rm df}$ is
obtained from Section 2.1, and $t\rmh(a\rmh)$ is obtained from Section 2.3.

\subsection{Hard binary stage}
\label{sec:dynamics:hard}

In the hard binary stage, the orbital decay of a BBH is mainly driven by
three-body interactions with low-angular-momentum stars passing its vicinity.
The BBH orbital decay timescale is mainly determined by how many stars are
available to move onto low-angular-momentum orbits and how fast they can move
onto the orbits to strongly interact with the BBH. After such strong
interactions, the stars escape with energy gain and are considered to be cleared
from the loss region. We denote the orbital binding energy of a star per unit
mass by $\calE$. Assuming that $ F(\calE,a)d\calE $ is the number clearing rate
from the loss region for stars with energy $ \calE\rightarrow\calE+d\calE $ and
$ \Delta\calE $ is the corresponding average change of the specific energy of
the stars escaping from the BBH during an interaction with the BBH, the orbital
energy loss rate of the BBH is given by
\begin{equation}
\left\vert\frac{dE}{dt}(a)\right\vert =
-m_\ast\int\Delta\calE F(\calE,a)d\calE,
\label{eq:ener_loss_rate}
\end{equation}
where $ E=-GM\bh\mu_{12}/2a $ is the orbital energy of the BBH, $\mu_{12}= M_1
M_2 /M\bh $ is the reduced mass of the BBH, $m_\ast$ is the mass of a single
star, and all stars are assumed to have the same mass.  The quantity $
\Delta\calE $ is only weakly dependent on $ \calE $ for hard binaries. Hence we
assume $ \Delta\calE=-KG\mu_{12}/a $ where $ K $ is a constant which is
determined from the numerical experiment of \citet{Quinlan96}.  Therefore, after
the BBH becomes hard, its hardening timescale due to stellar interaction is
given by
\begin{equation}
t_{\rm h}(a) =
\left\vert\frac{a}{\dot{a}}\right\vert =
\left\vert\frac{E}{\dot{E}}\right\vert =
\frac{M\bh}{2Km_\ast}\frac{1}{\int F(\calE,a)d\calE},
\label{eq:hard_time}
\end{equation}
where $m_*$ is the mass of a star and is assumed to be a solar mass in this
work.
We obtain the clearing rate $F(\calE,a)$ by following the analysis in section
3.2 in \citet{Yu02}, which is affected by the size of the loss region and the
two-body relaxation processes. 

\subsection{Gravitational radiation stage}
\label{sec:dynamics:trace:gw}

In the gravitational radiation stage, the evolution timescale of the BBH is
described by
\begin{equation}
t\gr(a,e) =
\frac{5}{64}\frac{c^5a^4(1-e^2)^{7/2}}
{G^3\mu_{12}M^2\bh(1+73e^2/24+37e^4/96)},
\label{eq:gw_timescale}
\end{equation}
where $ e $ is the orbital eccentricity of the binary \citep{Peters64} is set to
zero in this work as argued above.  The transition from the evolution stages of
the BBH interacting with its surrounding environments (the hard binary and
non-hard binary stages) to the gravitational radiation stage can be marked by
$t\gr(a=a\gr)=t_{\rm h}(a=a\gr)$. At $a=a\gr$, the BBH GW frequency and the
orbital period are labeled by $f\gr$ and $P\gr$, respectively.

Note that a large eccentricity can not only make the BBH enter the gravitational
radiation earlier, but also accelerate the orbital decay, and the gravitational
radiation energy
spectrum can be correspondingly changed.

\subsection{Loss region} 
\label{sec:dynamics:hybrid}

The size of the loss region in the phase space depends on galaxy shapes.  The
loss region in a spherical stellar system is a `loss cone' given by
\begin{equation}
J \le J_{\rm lc}\simeq \sqrt{2GM\bh f_a a}
\label{eq:Jlc}
\end{equation}
($f_a$ is a dimensionless factor $\sim 1$).
In an axisymmetric or triaxial system, there exist centrophilic orbits such as
box orbits, which pass arbitrarily close to the center and have low angular
momentum, as well as centrophobic orbits such as loop or tube orbits, which
avoid the center and have high angular momentum.  We use a characteristic
specific angular momentum $J_{\rm s}$ to mark the transition from centrophilic
($J\la J_{\rm s}$) to centrophobic ($J\ga J_{\rm s}$) orbits.  The loss region
in an axisymmetric system can be approximated as a `loss wedge' where stars on
centrophilic orbits with $|J_z|<J_{\rm lc}$ can precess into the loss cone,
while those on centrophobic orbits with $J>J_{\rm lc}$ cannot.  In a triaxial
system, the loss region can be approximated by $J<J_{\rm s}$ in which most of
the stars can precess into the loss cone.

In this work, we obtain $J_{\rm s}(\calE)$ by numerically simulating the motion
of a star under the combined gravitational potential of a central point mass
$M\bh$ and the galactic stars (see eq.~7 in \citealt{Cui14}). The galactic
potential is spherical if it can be described in the form $\Phi(r)$ $(r = |{\bf
r}|)$ and triaxial if it can be described in the form $\Phi(x^2 +y^2/\xi^2
+z^2/\zeta^2)$ $(\zeta<\xi<1)$. If the shape distribution of host galaxies is
described in mass density, we use the following equation (which is
derived for logarithmic potentials) to approximately
convert it into the shape distribution in gravitational potentials:
\begin{equation}
q_{\Phi} = \sqrt{\frac{1+\sqrt{1+8q_{\rho}^2}}{4}},
\end{equation}
where $q_{\rho}$ represents the medium-to-major or the minor-to-major axis ratio
in the density shape, and $q_{\Phi}$ represents the corresponding ratio in the
potential shape (see Eq.~2.72b in \citealt{BT08}). If $1-q_\rho\ll 1$, we have
$1-q_{\phi}\simeq (1-q_{\rho})/3$.

Noting that it is time-consuming to obtain $J_{\rm s}(\calE)$ in a large number
of galaxies by numerically tracing the motion of each star in a galaxy for a
sufficiently long time to judge whether it can move to the center on a
centrophilic orbit, we improve the judging efficiency for a centrophilic orbit
and describe the method to obtain $J_{\rm s}$ below.

In a general triaxial system, for a centrophilic orbit,
the motion of the stars can pass arbitrarily close to the center, and the
stellar orbital angular momentum will flip its direction during the orbital
motion. Numerically we count a star as one in the loss region, if all the three
$x$, $y$, and $z$ components of its orbital angular momentum flipped their
directions.
This criterion is effective and efficient enough in practice, although in principle
the stars that change the signs of their orbital angular momentum components are
not exactly the same population of the stars that can move to the center (on
centrophilic orbits).
Given the potential of a merger remnant, we obtain the characteristic angular
momentum $J_{\rm s}$ as a function of the binding energy $\calE$. In practice,
only stars in a certain range of $\calE$ have considerable importance to the
three-body interactions with a BBH.  We set the boundaries of $\calE$ as
$[-\Phi(r_{\rm max}),\,-\Phi(r_{\rm min})]$, where $r_{\rm min}= 0.1R\infl$ and
$r_{\rm max}=100 R\infl$.  For each system, we obtain $J_s(\calE)$ for $401$
values of $\calE$ equally spaced within the given range.  For each $\calE$, we
use the Monte Carlo method to generate $2000$ test particles whose squared
angular momenta are uniformly distributed in $[0,\,J^2_{\rm c}(\calE)]$ with
isotropic velocity distributions, where $J_{\rm c}(\cal E)$ is the specific
angular momentum of a circular orbit at $\cal E$.  We follow the evolution of
these test particles within the potential of the host merger remnant, which is
composed of two parts, i.e., the potential from the BBH and the galactic
potential from the host galaxy, where the potential of the BBH is simplified as
a Keplerian potential around a point mass $M\bh$ and the galactic potential is
assumed to be triaxial. We
calculate the fraction $\eta$ of the test particles that flipped the signs of
all the $x$, $y$, and $z$ components of the orbital angular momentum during an
evolution time of $160 P\rmc(\calE)$, where $P\rmc(\calE)$ is
the period of a circular orbit at energy $\calE$. The characteristic angular
momentum $J_{\rm s}(\calE)$ is then obtained by $J_{\rm
s}(\calE)=\sqrt{\eta(\calE)J^2_{\rm c}(\calE)}$.  The above method to obtain
$J_{\rm s}(\calE)$ is efficient enough.

We have done some tests to show that the above method is effective, although the
stars that change the signs of their orbital angular momentum components are not
exactly the same population stars that can move to the center and the evolution
of the orbital angular momentum of a star is traced with an evolution time of
only $160P\rmc(\calE)$. The tests are done for the galaxies (with
a limited number $\sim 800$) randomly selected from the mock galaxy sample (with
the shape distribution in \citealt{Padilla08}) presented in
Section~\ref{sec:data:mock}. In each galaxy, the motions of 500 stars at a given
${\cal E}$ are traced for a sufficiently long time ($8000P\rmc(\calE)$), where
the two values of ${\cal E}=-\Psi(10 R\infl)$ and $-\Psi(100 R\infl)$ are
tested.  Our tests show that the number ratios of the stars that change the
signs of their orbital angular momentum components to the stars that can move
the center is close to $1$ and the extension of the evolution time of a star
does not affect our results significantly.

As mentioned above, in a configuration close to being axisymmetric, the loss region
is approximated as a loss wedge, and the $J\rms(\calE)$ obtained by the above
method may be too low and is not appropriate to be taken as the $J\rms(\calE)$
defined in the loss wedge, as a star is recognized as being in the loss region
only if all the three $x$, $y$, and $z$ components of its angular momentum
flipped their directions during the orbital evolution. We estimate the size of
the loss wedge by the following criterion: a star is recognized as being in the
loss wedge if at the least one (but not all) of the three $x$, $y$, and $z$
components of its angular momentum flipped their directions and the remaining
components possess a combined specific angular momentum smaller than $J_{\rm
lc}$ (where $a$ is set to $GM\bh/44\sigma\rmc^2$ in Eq.~\ref{eq:Jlc}, i.e., the
values of $a\rmh$ with $q\bh=0.1$. We calculate the fraction $\eta$ of the
stars satisfying the above conditions, and the characteristic angular momentum
$J\rms(\calE)$ in the loss wedge is then obtained by $J_{\rm
s}(\calE)=\eta(\calE)J^2_{\rm c}(\calE)/(2J_{\rm lc})$.

Given a shape configuration of a system, we use the above methods to
numerically estimate both the size of the loss region by applying the criterion
for a general triaxial system and the size of the loss wedge by applying the
criterion for an axisymmetric system, and the stars in the loss wedge and in
the loss region both contribute to the stellar reservoir to have strong
interactions with the BBH.

In the model, for simplicity, we do not consider the time evolution of galaxy
shapes (e.g., rotation of a triaxial system or changes in axis ratios)
\footnote{Some possible effects of the changes in axis ratios can be partly
inferred from this work done by adopting three different sets of the galaxy
triaxial shape distributions and comparing their results.  Regarding the
rotation of a triaxial system, the characteristic rotating pattern timescale is
generally much longer than one orbital period of a star with the relevant
energy, but it can be shorter than the timescale of the star on a centrophilic
orbit to precess into the loss cone.  The assumption of ignoring the shape
rotation is plausible and does not affect the above estimate of $J_{\rm s}$
much if the stellar centrophilic orbits in the triaxial system are mostly
stochastic (which generally have relatively large apocenter distances and low
${\cal E}$; cf.\ the right panel in Figure 4 of Magorrian \& Tremaine
1999). The shape rotation may affect the above estimate of $J_{\rm s}$ if the
precessions of the stellar orbits into the loss cone are regular, which needs a
further detailed complement for its statistical effect; nevertheless, an
extreme case of fast rotation of a triaxial system would result in the
dynamical effects partly close to those in axisymmetric systems, and the
effects in axisymmetric systems can be partly seen from the results obtained
from one of the three observational galaxy shape distribution adopted in this
work (i.e., Weijmans et al.\ 2014, which prefers the axisymmetric shape
configurations).}
and the difference of galaxy shapes at different galactic radii.

\section{Modeling of the distributions of merging BBHs and surviving BBHs}
\label{sec:modelBBH}

\subsection{Distributions of surviving BBHs}
\label{subsec:BBHdis}

We define the distribution of surviving BBHs by $\Phi\bbh(M\bh,q\bh,a,z)$
($q\bh\le 1$) so that $\Phi\bbh(M\bh,q\bh,a,z)dM\bh dq\bh da$ is the comoving
number density of BBHs at redshift $z$ with total mass in the range
$M\bh\rightarrow M\bh+dM\bh$, mass ratio in the range $q\bh\rightarrow
q\bh+dq\bh$, and semimajor axis (or separation) in the range $a\rightarrow
a+da$.

We define the galaxy stellar mass function (GSMF) by $n\gal(M\gal,z)$ so that
$n\gal(M\gal,z)dM\gal$ is the comoving number density of galaxies with mass in
the range $M\gal\rightarrow M\gal+dM\gal$ at redshift $z$.  We describe the
galaxy merger rate per galaxy at redshift $z$ by $\calR\gal(q\gal,z|M\gal)$
($q\gal \le 1$) so that $\calR\gal(q\gal,z|M\gal)dq\gal dt$ represents the
average number of galaxy mergers with mass ratio in the range $q\gal\rightarrow
q\gal+dq\gal$ within time $t\rightarrow t+dt$ for a descendant galaxy with mass
$M\gal$, where $t$ is the corresponding cosmic time at redshift $z$.
Similarly as the above definition for galaxies, we also define the corresponding
variables for the spheroidal components of merging remnants.  (which are also
called ``bulge'' for simplicity and denoted by the subscript $\bgal$ below),
i.e, the stellar mass function $n\bgal(M\bgal,z)$, the merger rate per
spheroidal component $\calR\bgal(q\bgal,z|M\bgal)$, and the mass ratio
$q\bgal$.  For major mergers, we assume that galaxy mergers form spheroids
characterized by mass $M\bgal=M\gal$ and merger mass ratio $q\bgal=q\gal$.  For
minor mergers, the secondary galaxy is assumed to be accreted onto the
spheroidal component of the primary galaxy, and the mass of the spheroidal
component of the merging remnant is assumed to be the sum of the stellar masses
in the secondary galaxy and in the spheroidal component of the primary galaxy,
and the merger mass ratio $q\bgal$ is their mass ratio.

We connect the BBH distribution function with the galaxy merger rate by the
following equations:
\begin{align} & \Phi\bbh(M\bh,q\bh,a,z) \nonumber\\
= & \int_0^t dt' \int d M\bgal \int d q\bgal n\bgal(M\bgal,z')\calR\bgal(q\bgal,z'|M\bgal)
\nonumber \\
& \times p\bh(M\bh,q\bh|M\bgal,q\bgal,z') \nonumber \\
& \times p_a(a,t-t'|M\bgal,q\bgal,M\bh,q\bh,z') \nonumber \\
& \times P\intact(z,z'|M\bgal)
\label{eq:phibbh}
\end{align}
and
\begin{align}
& n\bgal(M\bgal,z')\calR\bgal(q\bgal,z'|M\bgal)\nonumber \\
= & \int d M\gal \int d q\gal n\gal(M\gal,z')\calR\gal(q\gal,z'|M\gal) \nonumber \\
& \times p\bgal(M\bgal,q\bgal|M\gal,q\gal,z') 
\end{align}
where $p\bh(M\bh,q\bh|M\bgal,q\bgal,z')$ is a
probability function defined so that $p\bh(M\bh,q\bh|M\bgal,q\bgal,z') dM\bh
dq\bh $ represents the probability that a galaxy merger characterized by the bulge mass
of the merging remnant $M\bgal$
and mass ratio $q\bgal$ at redshift $z'$ leads to a BBH with total mass in the
range $M\bh\rightarrow M\bh+dM\bh$ and mass ratio in the range $q\bh\rightarrow
q\bh+dq\bh$, $p_a(a,\tau|M\bgal,q\bgal,M\bh,q\bh,z')$ is a probability function
defined so that $p_a(a,\tau|M\bgal,q\bgal,M\bh,q\bh,z')da$ represents the
probability that the galaxy merger leads to a BBH with semimajor axis in the
range $a\rightarrow a+da$ at a time $\tau$ after the galaxy merger,
$p\bgal(M\bgal,q\bgal|M\gal,q\gal,z')$ is the probability that a galaxy merger with
total mass $M\gal$ and mass ratio $q\gal$ at redshift $z'$ results in 
the merging remnants characterized by $M\bgal$ and $q\bgal$,
and
$\int\int p\bh(M\bh,q\bh|M\bgal,q\bgal,z') dM\bh dq\bh=1$, $\int
p_a(a,\tau|M\bgal,q\bgal,M\bh,q\bh,z')da=1$, and $\int\int
p\bgal(M\bgal,q\bgal|M\gal,q\gal,z') d M\gal d q\gal =1$.
The distributions of $p\bh$ and $p\bgal$ depend on BH demography and galaxy
demography, and the distribution of $p_a$ depends on the distribution of
the intrinsic structure of galaxies. As to be seen below, the BBH mergers
contributing to the GWB comes mainly from redshift $z\la 2$.  The distribution
of the intrinsic structure of galaxies and BH demography used in this work are
mainly based on observations in the nearby universe, and we assume that they
are independent of redshift at low redshift $z\la 2$ and then distributions of
$p\bh$ and $p_a$ are independent of redshift in this work. 
The function $p_a$ is obtained by assuming that the BBH evolves independently
inside the merging galaxy remnant and that no other later galaxy mergers affect
the BBH evolution.
However, before the BBH reaches the gravitational radiation stage, it is
possible that the host merger remnant undergoes a big impact, that is, a major
merger with another galaxy or being accreted by another bigger galaxy to become
a satellite galaxy in this work; in these cases the interactions of triple or
multiple BHs may lead to quick BBH coalescence or BH kicks out of the host
merger remnant.
The function $P\intact(z,z'|M\bgal)$ ($z<z'$) is introduced in Equation
(\ref{eq:phibbh}) to represent the probability that the host merger remnant
$M\bgal$ does not experience such big impacts from redshift $z'$ to $z$.  The
detailed method to obtain $P\intact(z,z'|M\bgal)$ is described in
Section~\ref{sec:multi}. In this work, the effects of the big impacts
(e.g., on the surviving BBH distributions, BBH coalescence rates, GWBs;
see also Eqs.~\ref{eq:f3} and \ref{eq:hc2} below) are included
by removing a corresponding BBH system from consideration if it experiences
a big impact during its evolution, and we neglect the contribution from
possibly newly formed BBHs caused by triple or multiple BH interactions.

As to be shown below, we use Monte-Carlo simulations to generate galaxy mergers
and calculate the BBH evolution in the galaxy mergers. Suppose that we have $N$
galaxy mergers with parameters $(M\bgal,q\bgal)$ leading to the formation of BBHs
with parameters ($M\bh,q\bh$) and semimajor axis
$a_{i|M\bgal,q\bgal,M\bh,q\bh}(\tau)$ ($i=1,2,...,N$) at a time $\tau$ after the
galaxy mergers.  The probability function $p_a(a,\tau|M\bgal,q\bgal,M\bh,q\bh)$
can be obtained by
\begin{align}
& p_a(a,\tau|M\bgal,q\bgal,M\bh,q\bh) \nonumber \\
=& \frac{1}{N} 
\sum_{i=1}^{N}
\delta[a-a_{i|M\bgal,q\bgal,M\bh,q\bh}(\tau)].
\label{eq:f2}
\end{align}

By applying Equation (\ref{eq:f2}) into Equation (\ref{eq:phibbh}), we have
\begin{align}
& \Phi\bbh(M\bh,q\bh,a,z) \nonumber \\
= & \frac{1}{N} \sum_{i=1}^{N}
\int d M\bgal \int d q\bgal n\bgal(M\bgal,z_i)
\calR\bgal(q\bgal,z_i|M\bgal) \nonumber \\
& \times p\bh(M\bh,q\bh|M\bgal,q\bgal,z_i)
H(t-\tau_{a,i})\frac{t_{{\rm evol},i}}{a} \nonumber \\
& \times P\intact(z,z_i|M\bgal),
\label{eq:phibbhi}
\end{align}
where
\begin{equation}
t_{{\rm evol},i}=a\left| \frac{da_{i|M\bgal,q\bgal,M\bh,q\bh}}{d\tau}\right|^{-1}_{\tau
=\tau_{a,i|M\bgal,q\bgal,M\bh,q\bh}}, 
\end{equation}
is the BBH evolution timescale at $a$ (see Eq.~\ref{eq:tevol}),
$\tau_{a,i|M\bgal,q\bgal,M\bh,q\bh}$ is the solution of equation
$a_{i|M\bgal,q\bgal,M\bh,q\bh}(\tau)=a$, $H(t-\tau_{a,i})$ is a step function
defined by $H(t-\tau_{a,i})=1$ if $t>\tau_{a,i}$ and $H(t-\tau_{a,i})=0$ if
$t\le\tau_{a,i}$, and $z_i$ is the corresponding redshift of cosmic time
$t-\tau_{a,i}$.

In a simplified or extreme case, if
$p\bh(M\bh,q\bh|M\bgal,q\bgal,z)=\delta(M\bh-A\cdot M\bgal)\delta(q\bh-q\bgal)$ with
$A$ being a constant, and the BBH evolution $a_{i}(\tau)$ ($i=1,...,N$) is the
same for the same ($M\bgal,q\bgal$), Equation (\ref{eq:phibbhi}) is simplified as
\begin{align}
 &
\Phi\bbh(M\bh,q\bh,a,z(t)) \nonumber \\
= & A^{-1} H(t-\tau_a)
a^{-1} t_{\rm evol} \cdot n\bgal(M\bgal=A^{-1}M\bh,z(t-\tau_a)) \nonumber \\
& \times \calR\bgal(q\bgal=q\bh,z(t-\tau_a)| M\bgal=A^{-1}M\bh) \nonumber \\
& \times P\intact(z(t),z(t-\tau_a)|M\bgal=A^{-1}M\bh).
\label{eq:sim1}
\end{align}

\subsection{BBH coalescence rates}

We define the BBH coalescence rate $R\bh(M\bh,q\bh,z)$ so that
$R\bh(M\bh,q\bh,z)dM\bh d q\bh dt$ represents the comoving number density of the
BBH coalescence events occurring during cosmic time $t\rightarrow t+dt$ and with
the total mass of the two coalescing BHs in the range $M\bh\rightarrow
M\bh+dM\bh$ and their mass ratio in the range $q\bh\rightarrow q\bh+dq\bh$. The
BBH coalescence rate can be obtained by
\begin{align}
& R\bh(M\bh,q\bh,z(t)) \nonumber \\
= & \frac{d}{dt} \biggl[
\int_0^t dt' \int d M\bgal \int d q\bgal 
n\bgal(M\bgal,z')\calR\bgal(q\bgal,z'|M\bgal) \nonumber \\
& \times p\bh(M\bh,q\bh|M\bgal,q\bgal,z') \nonumber \\
& \times f\coal(t-t'|M\bgal,q\bgal,M\bh,q\bh,z') \biggr],
\label{eq:Rbh}
\end{align}
where $f\coal(\tau|M\bgal,q\bgal,M\bh,q\bh,z')$ represents the fraction of the BBHs
that can coalesce before cosmic time $t'+\tau$ among those BBHs with parameters
$(M\bh,q\bh)$ caused by galaxy mergers with parameters $(M\bgal,q\bgal)$ occurring
at redshift $z'$. The fraction of $f\coal$ can be obtained by
\begin{align}
& f\coal(\tau|M\bgal,q\bgal,M\bh,q\bh,z') \nonumber \\
= & \frac{1}{N}\sum_{i=1}^{N}\biggr[H(\tau-\tau_{a=0,i|M\bgal,q\bgal,M\bh,q\bh}) 
\nonumber \\
& \times P\intact\left(z(t'+\tau_{a=0,i|M\bgal,q\bgal,M\bh,q\bh}),
z'|M\bgal\right)\biggr],
\label{eq:f3}
\end{align}
where the term $P\intact$ removes the contribution of a BBH system if it
experiences a big impact before its expected coalescence time $t'+\tau_{a=0}$
obtained by assuming it evolves independently.

Applying Equation (\ref{eq:f3}) into Equation (\ref{eq:Rbh}) gives
\begin{align}
 & R\bh(M\bh,q\bh,z(t)) \nonumber \\
= & \frac{1}{N}\sum_{i=1}^N \int d M\bgal \int d q\bgal n\bgal(M\bgal,
z_i)\calR\bgal(q\bgal,z_i|M\bgal) \nonumber \\
& \times p\bh(M\bh,q\bh|M\bgal,q\bgal,z_i)H(t-\tau_{a=0,i}) \nonumber \\
& \times P\intact(z(t),z_i|M\bgal),
\label{eq:Rbhi}
\end{align}
where $z_i$ is the redshift at cosmic time $t-\tau_{a=0,i}$.

In the extreme case used for Equation (\ref{eq:sim1}), if $\tau_{a=0}<t$,
Equation (\ref{eq:Rbhi}) is simplified as
\begin{align}
 & R\bh(M\bh,q\bh,z(t)) \nonumber \\
= & A^{-1} n\bgal(M\bgal=A^{-1}M\bh,z(t-\tau_{a=0})) \nonumber \\
& \times \calR\bgal(q\bgal=q\bh,z(t-\tau_{a=0})| M\bgal=A^{-1}M\bh) \nonumber \\
& \times P\intact(z(t),z(t-\tau_{a=0})|M\bgal=A^{-1}M\bh).
\label{eq:Rbhsim}
\end{align}
The above equation can be further simplified in the following two cases.
\begin{itemize}
\item For sufficiently small $a$, the difference of $n\gal$ and
$\calR\gal$ (or $n\bgal$ and $\calR\bgal$) at different redshifts
$z(t-\tau_a)$ and $z(t-\tau_{a=0})$ is
negligible, and applying Equation (\ref{eq:Rbhsim}) into Equation
(\ref{eq:sim1}) gives
\be
\Phi\bbh(M\bh,q\bh,a,z)=R\bh(M\bh,q\bh,z)|da/dt|.
\label{eq:PhiR}
\ee

\item If the BBH coalescence occurs immediately after a galaxy merger (i.e.,
$\tau_a\sim 0$), Equation (\ref{eq:Rbhsim}) is simplified as
\begin{align}
 & R\bh(M\bh,q\bh,z) \nonumber \\
\simeq & A^{-1} n\bgal(M\bgal=A^{-1}M\bh,z) \nonumber \\
& \times \calR\bgal(q\bgal=q\bh,z| M\bgal=A^{-1}M\bh),
\label{eq:Rbhnodelay}
\end{align}
and the BBH coalescence rate can be approximately obtained from the galaxy
merger rate at the same redshift.

\end{itemize}

\subsection{Stochastic gravitational-wave background}
\label{sec:met:gwb}

We denote the total present-day energy density in gravitational radiation by
$\calE\gw$, which can be expressed by 
\be
\calE\gw \equiv
\int\frac{\pi}{4}\frac{c^2}{G}f^2h\rmc^2(f)d\ln f,
\label{eq:gw_ener_dens}
\ee
where $c$ is the speed of light, $G$ is the gravitational constant, $f$ is the
frequency of gravitational waves observed today on earth, and $h\rmc$ is the
characteristic strain amplitude of the GW spectrum over a logarithmic frequency
interval \citep{Phinney01}. 

The present-day GW energy density can be obtained by an integration over the
contribution from the GW radiation sources (BBHs in this work) in the cosmic
history as follows,
\begin{align}
\calE\gw  = & \int dz d M\bh dq\bh da \left |\frac{dt}{dz}\right|
\nonumber \\
& \times \Phi\bbh(M\bh,q\bh,a,z) \cdot \frac{1}{1+z}
\cdot \left|\frac{E}{t\gr}\right|
\label{eq:gw_ener_dens_inte}
\end{align}
where $f_{\rm r}$ is the frequency of the GW signal in the source's cosmic rest
frame and is related with $f$ through $f_{\rm r}=f(1+z)$,
$\left|\frac{E}{t\gr}\right|$ is the GW energy per unit time radiated by an
in-spiraling BBH with parameters $(M\bh,q\bh,a)$ (see the definition of $t\gr$ in
Eq.~\ref{eq:gw_timescale}), and the term $\frac{1}{1+z}$ accounts for the
redshift of the GWs since emission.

During the in-spiral, the frequency of the GW signal is twice the Keplerian
orbital frequency of the BBH and increases monotonically till the final
coalescence and can be estimated by
\be
f \simeq 3.4\times 10^{-9}
\left(\frac{M\bh}{10^8\msun}\right)^{1/2}
\left(\frac{a}{10^{-2}\pc}\right)^{-3/2}\hz.
\label{eq:fmin}
\ee
The upper limit of $f$ when the circular in-spiral assumption in Equation
(\ref{eq:gw_ener_spec}) is valid can be roughly estimated by twice of the
Keplerian orbital frequency at the innermost stable circular orbit (ISCO) for a
Schwarzschild BH with mass the same as the total mass of the BBH, that is,
\be
f\simeq f_{\rm ISCO} \simeq 2.2\times 10^{-5}
\left(\frac{M\bh}{10^8\msun}\right)^{-1}\hz.
\label{eq:fmax}
\ee
The above upper limit for BBHs is safely outside the frequency band probed by
PTAs. 

From Equations (\ref{eq:gw_ener_dens})-(\ref{eq:gw_ener_dens_inte}) and
(\ref{eq:phibbhi}), we can estimate the overall strain amplitude of the expected
stochastic GWB from BBHs by
\begin{align}
& \frac{\pi}{4}\frac{c^2}{G}f^2h\rmc^2(f) \nonumber \\
= & \int dz d M\bh dq\bh \left|\frac{dt}{dz}\right| \cdot \Phi\bbh(M\bh,q\bh,a,z)
\nonumber \\
 &  \times \left|\frac{da}{d\ln f_{\rm r}}\right|
\cdot \frac{1}{1+z} \cdot
\left|\frac{E}{t\gr}\right|  \\
= & \int dz d M\bh dq\bh \left|\frac{dt}{dz}\right| \nonumber \\
 & \times \frac{1}{N} 
\sum_{i=1}^{N}
\int d M\bgal \int d q\bgal n\bgal(M\bgal,z_i)\calR\bgal(q\bgal,z_i|M\bgal) \nonumber \\
& \times p\bh(M\bh,q\bh|M\bgal,q\bgal,z_i)H(t-\tau_{a,i}) \nonumber \\
& \times P\intact(z,z_i|M\bgal) \nonumber \\
 &  \times
\frac{(\pi G)^{2/3}}{3}\calM^{5/3}f^{2/3}
(1+z)^{-1/3} \cdot
\frac{t_{{\rm evol},i}}{t\gr} 
\label{eq:hc2}
\end{align}
where $z_i$ is the corresponding redshift of cosmic time $t-\tau_{a,i}$, $\calM$
is the chirp mass of the BBH defined by $\calM^{5/3}\equiv M_1  M_2 ( M_1 + M_2
)^{-1/3}$, $t^{-1}_{{\rm evol},i}=t^{-1}_{{\rm h},i}+t^{-1}\gr$, and
\be
\left|\frac{E}{a}\cdot\frac{da}{d\ln f_{\rm r}}\right|
= \left|\frac{dE}{d\ln f_{\rm r}}\right|
= \frac{(\pi G)^{2/3}}{3}\calM^{5/3}f_{\rm r}^{2/3}
\label{eq:gw_ener_spec}
\ee
is used.

For sufficiently high $f$, the difference of $n\gal$ and $\calR\gal$
(or $n\bgal$ and $\calR\bgal$) at
different redshifts $z(t-\tau_a)$ and $z(t-\tau_{a=0})$ is negligible; and in
this case, applying Equation (\ref{eq:Rbhi}) into Equation (\ref{eq:hc2})
gives
\begin{align}
 & \frac{\pi}{4}\frac{c^2}{G}f^2h\rmc^2(f) \nonumber \\
\simeq & 
\int dz d M\bh dq\bh \left|\frac{dt}{dz}\right| R\bh(M\bh,q\bh,z)
\nonumber \\
 &  \times 
\frac{(\pi G)^{2/3}}{3}\calM^{5/3}f^{2/3}(1+z)^{-1/3}
\cdot \left\langle \frac{t_{{\rm evol},i}}{t\gr}\right\rangle.
\label{eq:hchighf}
\end{align}
where
\begin{align}
& \left\langle\frac{t_{{\rm evol},i}}{t\gr}\right\rangle \nonumber \\
\equiv & \frac{1}{N}\sum_{i=1}^{N}
\int d M\bgal \int d q\bgal n\bgal(M\bgal,z_i)\calR\bgal(q\bgal,z_i|M\bgal) \nonumber \\
& \times p\bh(M\bh,q\bh|M\bgal,q\bgal,z_i)H(t-\tau_{a,i}) 
\cdot \frac{t_{{\rm evol},i}}{t\gr} \nonumber \\
& \times R^{-1}_{\rm BH}(M\bh,q\bh,z).
\end{align}
If $t\gr\ll t_{{\rm h},i}$ ($i=1,2,...N$), we have ${t_{{\rm
evol},i}}/{t\gr}\simeq 1$, $\langle{t_{{\rm evol},i}}/{t\gr}\rangle\simeq 1$,
and $h\rmc\propto f^{-2/3}$; and if $t\gr\gg t_{{\rm h},i}$ and $t_{{\rm
h},i}\propto a^{\alpha_a}$ (where the power law $\alpha_a$ is a constant,
$i=1,2,...N$), we have $\langle{t_{{\rm evol},i}}/{t\gr}\rangle\propto
f^{2(4-\alpha_a)/3}$ and $h\rmc\propto f^{\alpha_h}$ with
$\alpha_h=(2-\alpha_a)/3$.

\subsection{Multiple galaxy mergers during the evolution of a BBH} 
\label{sec:multi}

In this subsection, we present a rough way on how to obtain the function 
$P\intact(z,z'|M\bgal)$ shown in Equation (\ref{eq:phibbh}), the probability
that the host merger remnant $M\bgal$ does not experience big impacts (including
major mergers with other galaxies and mergers with bigger galaxies) from
redshift $z'$ to $z$.

We define the progenitor galaxy merger rate per galaxy at redshift $z$ by
$\calR\prog(\msec,z|\mpri)$ so that $\calR\prog(\msec,z|\mpri) d\msec dt$
represents the average number of mergers of a galaxy with mass $\mpri$ undergoing
with a second galaxy with mass within $\msec\rightarrow \msec+d\msec$ during
time $t\rightarrow t+dt$.  We define $\mugal\equiv\msec/\mpri$, and thus
$\qgal(\le 1) =\min(\mugal, 1/\mugal)$. The function
$\calR\prog(\msec,z|\mpri)$ can be obtained from the GSMF
$n\gal(M\gal,z)$ and the descendant galaxy merger rate
$\calR\gal(\qgal,z|\mgal)$ (see their definitions in
Section~\ref{subsec:BBHdis}) through the following equation on the number of
the galaxy merger events per unit comoving volume:
\begin{align} &\calR\prog(\msec,z|\mpri) d\msec dt \cdot n\gal(\mpri,z) d\mpri
\nonumber\\
= &\frac{1}{2}\calR\gal(\qgal,z|\mgal) d\qgal dt \cdot n\gal(\mgal,z) d\mgal,
\label{eq:calRrela} \end{align} where $\mgal=\mpri+\msec$, and the factor of
$1/2$ accounts for the symmetry between $\mpri$ and $\msec$ in the definition
of the progenitor merger rate; and thus we have
\begin{eqnarray}
& & \calR\prog(\msec,z|\mpri) \nonumber \\
& = & \frac{1}{2}
\frac{n\gal(\mgal,z)}{n\gal(\mpri,z)}\calR\gal(\qgal,z|\mgal)
\left|\frac{\partial(M\gal,q\gal)}{\partial(\mpri,\msec)}\right|, \nonumber \\
\label{eq:calRprog}
\end{eqnarray} where
$|\partial(M\gal,q\gal)/\partial(\mpri,\msec)|=(1+\qgal)^2/\mgal$.
The average number of big impacts that a host merger remnant with mass $\mpri$
is expected to go through between redshifts $z$ and $z'$ ($z'>z$)is given by
\begin{eqnarray} \calI(z,z'|\mpri)
& = & \int_{z}^{z'}\int_{q_{\rm major}}^{\infty} \calR\prog(\msec,z''|\mpri)
\nonumber \\
& & \times
\left|\frac{d\msec}{d\mugal}\right| \cdot \left|\frac{dt(z'')}{dz''}\right| d\mugal
dz'', \end{eqnarray} where $q_{\rm major}$ is the lower limit of the mass ratios
for major mergers, and $|d\msec/d\mugal|=\mpri$.
The probability that the host merger remnant keeps intact between redshifts $z$
and $z'$, $P\intact(z,z'|\mpri)$ can be given by
\begin{eqnarray}
& & P\intact(z,z'|\mpri) \nonumber \\
&= & [1-\calI(z,z'|\mpri)]H(1-\calI(z,z'|\mpri)),
\label{eq:pintact}
\end{eqnarray} which is reduced to 1 if $\calI(z,z'|\mpri)=0$
and 0 if $\calI(z,z'|\mpri)>1$.

For major mergers of galaxies, whose merging remnants are spheroids, the term
$P\intact(z,z'|M\bgal)$ in Equation (\ref{eq:phibbh}) can be obtained by
replacing $\mpri$ directly with $M\bgal$ in the function $P\intact(z,z'|\mpri)$
obtained in Equation (\ref{eq:phibbh}).  For minor mergers of galaxies, we also
do this replacement, which is a rough approximation and deserves further
improvements.

\section{Galaxy distributions, SMBH demography, and galaxy merger rates} \label{sec:dynamics:sample}

In this section, we present the galaxy samples, the realistic galaxy
distributions, the galaxy merger rates, and the BH-host galaxy relations used
in the model. We also describe the BBH evolution tracks obtained with those
distributions.

We generally use a $\Lambda CDM$ cosmology model with parameters $\Omega_{\rm
m}=0.307$, $\Omega_{\rm \Lambda}=0.693$, $h=0.678$, $\sigma_8=0.823$, and
$n_S=0.96$ \citep{Planck16}.  The results of some previous works adopted by our
model may have some differences in their used cosmological models.  For
example, the GSMF adopted from \citet{Behroozi19} is obtained with a cosmology
model with parameters as mentioned above, and the galaxy merger rates adopted
from \citet{Rodriguez-Gomez15} is obtained with a cosmology model with
parameters $\Omega_{\rm m}=0.2726$, $\Omega_{\rm \Lambda}=0.7274$, $h=0.704$,
$\sigma_8=0.809$, and $n_S=0.963$ \citep{Hinshawetal13}. In this work, it is
plausible to assume the differences in our conclusions caused by the different
versions of the cosmological models are negligible. We use the results from
the previous works directly without doing conversion between the
different cosmological models, except that they are converted to the values
with $h=0.678$.

\subsection{Observational galaxy surface brightness and triaxial shape
distributions} \label{sec:met:data}

To model the dynamics of BBHs inside galaxy merger remnants, the stellar density
distributions of the host remnants are needed. To construct a sample of host
galaxy merger remnants with a realistic stellar distribution, we use the
early-type galaxies obtained in the $\atlas$ survey \citep{Cappellari11} and in
\citet{Lauer07a}, both of which have high spatial resolution $\hst$ (Hubble
Space Telescope) imaging to
detect the inner part of the stellar surface brightness distributions, as
described in Section~\ref{subsubsec:galsb}. Regarding the intrinsic triaxial
shape distributions of galaxies, we use the observational results summarized
from the $\atlas$ survey and the Sloan Digital Sky Survey (SDSS), as described
in Section~\ref{subsubsec:galtri}. 

\subsubsection{Observational galaxy surface brightness distributions}
\label{subsubsec:galsb}

The $\atlas$ survey is based on a volume-limited sample of nearby ($\la
42\mpc$) early-type galaxies, with multi-band imaging as well as
two-dimensional bulk kinematic information as derived from the optical
integral-field spectroscopy. Among these early-type galaxies, $135$ were found
to have $\hst$ archive data \citep{Krajnovic13}. In the compilation of
\citet{Krajnovic13}, $13$ galaxies show strong dust features in the nuclei and
are thus excluded from the analysis. The surface brightness of the remaining
$122$ galaxies, which are observed by $\hst$ surveys are all converted to
the $\rm F555W$ band (broad-band $V$), in order to make comparison with previous
studies. The ``Nuker law'' profile \citep{Lauer95} is used to fit the surface
brightness of these galaxies, i.e., 
\begin{equation}
I(R)= 
2^{\frac{\beta-\gamma}{\alpha}}I\rmb
\left(\frac{R}{R\rmb}\right)^{-\gamma} 
\left[1+\left(\frac{R}{R\rmb}\right)^{\alpha}\right]^{\frac{\gamma-\beta}{\alpha}}, 
\label{eq:sb}
\end{equation}
where $I\rmb$ is the surface brightness at the break radius $R\rmb$, $\gamma$
and $\beta$ are the asymptotic inner and outer slopes of the surface brightness
profile, respectively, and $\alpha$ parameterizes the sharpness of the break.
\citet{Cappellari13} provides
the mass-to-light ratio $M/L$ in the $r$ band for the $\atlas$ galaxies.
which is obtained by the best-fitting self-consistent
Jeans Anisotropic Multi-Gaussian Expansion model.
Note that the five Nuker-law parameters are obtained in the $V$ band, whereas
the above mass-to-light ratios are obtained in the SDSS $r$ band. We convert
the $M/L$ from the $r$ band to the $V$ band as follows, based on the magnitude
differences in the two bands. \citet{Cappellari13} provides the total SDSS
$r$-band luminosities for these galaxies.
Among the 122 $\atlas$ galaxies adopted in our study, 53
have the $V$-band apparent magnitudes and all of them have the $B$-band
apparent magnitudes. For the 66 galaxies without the $V$-band magnitudes but
with the SDSS $ugriz$ ones, we convert their $B$-band apparent magnitudes to
the $V$-band ones, based on one empirical equation, $B-V=0.90(g-r)+0.21$
listed in \url{http://www.sdss3.org/dr8/algorithms/sdssUBVRITransform.php}
(see also Jester et al.\ 2005). For the remaining 3 galaxies without the
$V$-band and the SDSS $g$- and $r$-band magnitudes, we do the conversion by
$B-V=0.95$ (e.g., \citealt{Lauer07a}). The absolute magnitudes $M_V$
can be converted from their apparent magnitudes, given the distances and the
extinctions of these galaxies (listed in Table~3 of \citealt{Cappellari11}),
by assuming the extinction ratio $A_B/A_V=1.324$ (Table 3.21 in \citealt{BM98}).

\citet{Lauer07a} provide the ``Nuker law'' parameters $R\rmb$, $\mu\rmb$,
$\alpha$, $\beta$, and $\gamma$ together with some other parameters for a sample
of $219$ galaxies, which are compiled from several $\hst$ investigations of the
central structure of early-type galaxies. The mass-to-light ratio of these
galaxies can be estimated according to the following empirical galaxy scaling
relation in the V-band \citep{Magorrian98}: 
\begin{equation}
\frac{(M/L)_V}{(M/L)_{\odot}}=
4.9\left(\frac{L_V}{10^{10}L_{\odot}}\right)^{0.18},\label{eq:m2l}
\end{equation}
where $L_V$ is the galaxy luminosity in the V-band.

In Figure~\ref{fig:gala_samp}, we show the $\atlas$ galaxy sample (blue) and
the \citet{Lauer07a} sample (red) in the six-dimensional parameter space
composed of the five ``Nuker law'' parameters (i.e., break radius $R\rmb$,
surface brightness at break radius $\mu\rmb$, $\alpha$, $\beta$, $\gamma$) and
the mass-to-light ratio $M/L$ in the $V$ band. In the figure, the mass-to-light
ratios of the \citet{Lauer07a} sample are estimated by Equation (\ref{eq:m2l}).
There are 61 common galaxies in the two samples, and the fit
parameters of the same galaxies may be different in the two samples, which are
all shown in Figure~\ref{fig:gala_samp}.

\subsubsection{Observational galaxy triaxial shape distributions}
\label{subsubsec:galtri}

The triaxial shape of a galaxy can be characterized by the two parameters,
i.e., the medium-to-major axis ratio $b/a$ and the minor-to-major axis ratio
$c/a$, where $a$, $b$, and $c$ represent the major, the medium, and the minor
axis of an iso-density layer of the galaxy, respectively.  The triaxial shape
can also be described by an equivalent set of parameters i.e., the triaxiality
$T\equiv(a^2-b^2)/(a^2-c^2)$ and the minor-to-major axis ratio $c/a$. The
general triaxial description can be reduced to the following special cases:
spherical shape with $b/a=c/a=1$, oblate spheroid with $c/a<b/a=1$, and prolate
spheroid with $c/a=b/a<1$.

In the left panel of Figure~\ref{fig:tsc_contour}, we show the shape distributions of early-type
galaxies obtained from observations.  The results obtained from the photometric
study in the SDSS indicate that early-type galaxies are better described by
triaxial shapes than by purely spheroidal shapes.  As shown in the figure, the
green plus symbol marks the possible range of the mean medium-to-major and
minor-to-major axis ratios of the early-type galaxies in the SDSS DR3 data set
\citep{Vincent05}.
The blue and red contours indicate the two-dimensional axis
ratio distribution of early-type galaxies in the SDSS DR6 data set
\citep{Padilla08} and the SDSS DR8 data set \citep{Rodriguez13}, respectively.
The results obtained from the $\atlas$ galaxy sample, where both the
photometric and the bulk kinematic information are used, show that early-type
galaxies can be classified by fast- and slow-rotators \citep{Emsellem07,
Krajnovic11}.  The magenta color-bar marks the mean and the standard deviation
of the minor-to-major axis ratios for a nearly oblate fast-rotator galaxy
sample from the $\atlas$ galaxy survey, obtained through a triaxial
configuration fitting \citep{Weijmans14} (where the mean values of $c/a$ and
$\ln(1-b/a)$ are $0.33$ and $-5.0$, respectively, and their standard deviations
are $0.11$ and $0.08$, respectively). 
The shape parameters cannot be well constrained for the slow-rotator galaxies,
due to the limited sample size and the lack of a clearly defined projected
rotation axis in many of the systems.

We will apply the different shape distributions to our study and see how
different the results are affected by them.

\subsection{Mock galaxy sample}
\label{sec:data:mock}

\begin{figure*}
\centering
\includegraphics[width=0.8\textwidth]{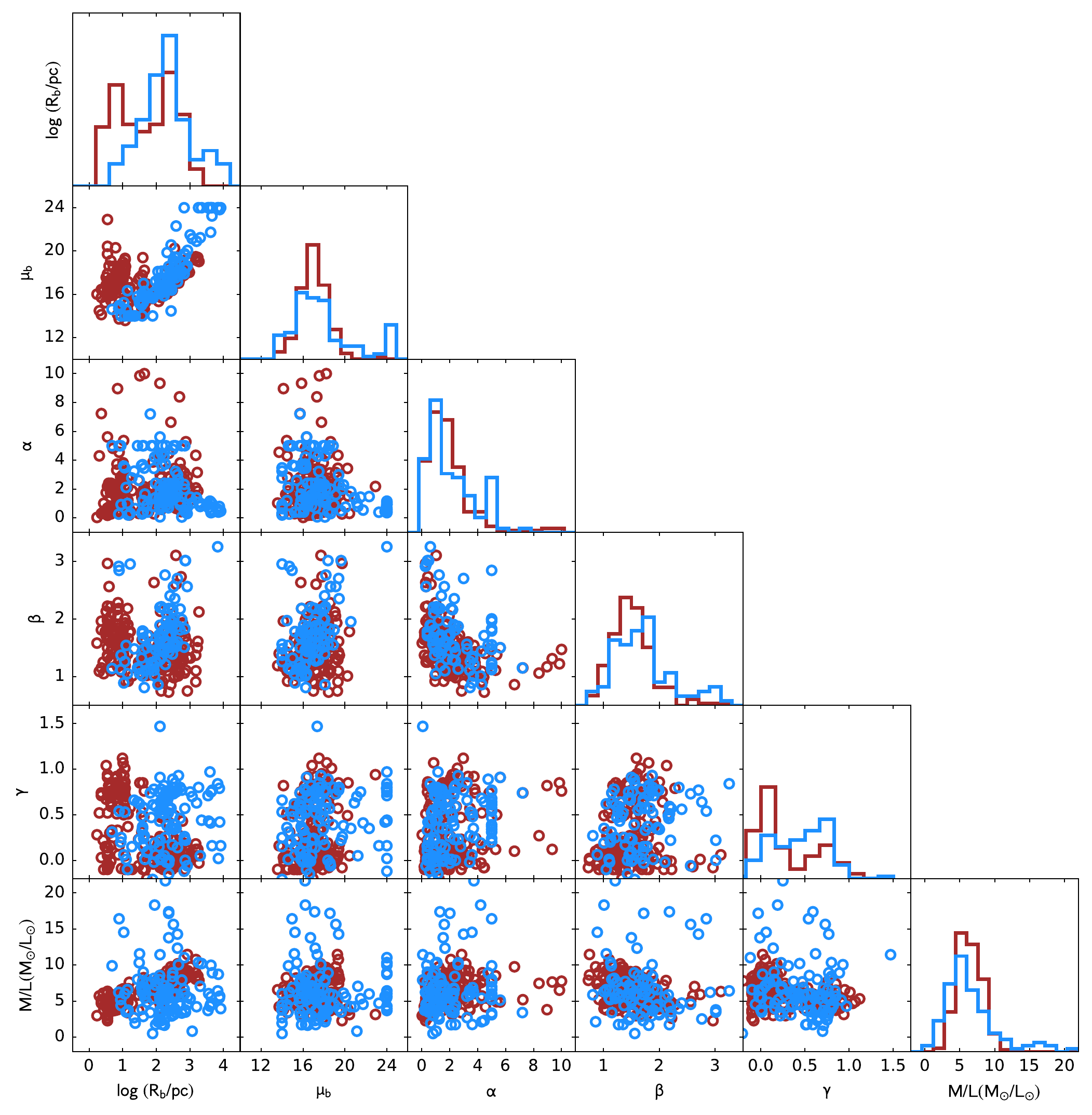}
\caption{Distributions of the galaxy properties for the galaxies from the
$\atlas$ \citep{Cappellari11} sample (blue) and \citet{Lauer07a} sample (red).
The parameters shown in the figure include the five ``Nuker law'' parameters
(i.e., break radius $R\rmb$, surface brightness at break radius $\mu\rmb$,
$\alpha$, $\beta$, $\gamma$; see Eq.~\ref{eq:sb}) and mass-to-light ratio $M/L$
in the $V$ band.  Each circle represents one galaxy, and the solid lines
represent the histograms of the corresponding parameters. See Section~\ref{subsubsec:galsb}.  }
\label{fig:gala_samp}
\end{figure*}

As the observational sample size shown in Figure~\ref{fig:gala_samp} is small,
we construct a mock galaxy sample with a larger size for the statistical study
of the BBH evolution in galaxies with realistic property distributions.  The
mock galaxy sample is generated based on all the galaxies shown in
Figure~\ref{fig:gala_samp} and follow the statistical distributions of their
properties. For each galaxy in the mock galaxy sample, we first select a
galaxy, e.g., galaxy $i$ with parameters $\{\log R\rmb^i, \mu\rmb^i, \alpha^i,
\beta^i, \gamma^i, M/L^i\}$, randomly from Figure~\ref{fig:gala_samp}, then add
small scatters $\{\delta \log R\rmb^i, \delta\mu\rmb^i, \delta\alpha^i,
\delta\beta^i, \delta\gamma^i, \delta M/L^i\}$ to the corresponding parameters
and assign the new parameter set to the galaxy. All of the scatters are assumed
to follow Gaussian distributions with zero means, and their standard deviations
are assumed to be $\sigma_{\log R\rmb}=0.2$, $\sigma_{\mu\rmb}=0.4$,
$\sigma_{\alpha}=1.0$, $\sigma_{\beta}=0.2$, $\sigma_{\gamma}=0.2$, and
$\sigma_{M/L}=2.0$. None of those assumed standard deviations are larger than
the standard deviations of the corresponding parameters of all the
observational points shown in Figure~\ref{fig:gala_samp}. We delete those
systems generated with unphysical $\beta<0$ or $M/L<0$. It is likely that some
generated parameters lead to
extreme systems, but they are rare and do not affect our results statistically.
 
Our calculations show that the detailed results are not sensitive
to the exact values of the boundaries.
Thus, a mock sample with a size of $\sim 10^7$ galaxies is generated in our
study (see details in Section~\ref{subsec:mockBBH}).

The generation of the scatters in
observational parameters and the large number of the mock galaxy sample will
make the statistical results obtained from our Monte Carlo study appear smooth.
If the standard deviations assigned to the parameter set are too small, the
parameters of the mock galaxy sample will cluster around the points shown in
Figure~\ref{fig:gala_samp} and the application of the mock sample will result
in significant discreteness or discontinuity in the statistical results (e.g.,
the curves of the surviving BBH distributions shown below). If the standard
deviations are too large, some intrinsic cross correlations among the
parameters of the observational galaxies shown in Figure~\ref{fig:gala_samp}
will be lost significantly in the mock sample. The values of the
standard deviations in our mock sample are roughly chosen to be as small as
possible to make our statistical results smooth and keep the correlations
originally existing in the observational sample as much as possible.

Given a galaxy in the Mock galaxy sample with parameters $\{\log R\rmb,
\mu\rmb, \alpha, \beta, \gamma, M/L\}$, the stellar density distribution within
the galaxy can be determined as done by Equations (48)-(49) in Section~5.1 in
\citet{Yu02}.

The triaxial shape of a galaxy in the mock sample is generated randomly from
the observational galaxy triaxial shape distribution described in
Section~\ref{subsubsec:galtri}, without considering the possible dependence of
the galaxy shapes on galaxy properties, such as galaxy masses.  In this work,
we explore three sets of galaxy shape distributions presented in \citet{Padilla08},
\citet{Rodriguez13}, and \citet{Weijmans14}, respectively. 

\subsection{GSMFs and galaxy merger rates}
\label{sec:data:gmfmr}

As shown by Equation (\ref{eq:phibbh}) in Section~\ref{sec:modelBBH}, the
GSMFs, and the galaxy merger rates are needed to model the BBH
populations. We present the GSMFs and the galaxy merger rates
in Section~\ref{subsubsec:galaxymass} and \ref{subsubsec:galaxymerger},
respectively.

\subsubsection{Galaxy stellar mass functions}\label{subsubsec:galaxymass}

The GSMFs can be obtained from either observations (e.g., see
\citealt{Bernardi13, Muzzin13, Ilbert13, Tomczak14, Kelvin14}) or
hydrodynamical simulations (e.g., see \citealt{Torrey15, Lim17}).

In the left panel of Figure~\ref{fig:mass_func}, we show the observational GSMFs
obtained at several discrete redshifts.  In the right panel of
Figure~\ref{fig:mass_func}, we show the GSMF used in this work, which is
obtained from \citet{Behroozi19} and derived by matching their model to a
compilation of different observational results and constraining the galaxy
growth history inside DM haloes. The GSMF in \citet{Behroozi19} extends from
$z\sim 0$ to high redshifts up to $z\sim 10$ in 
sufficiently dense redshift space, and we illustrate them only at several
redshifts in the figure.  The observational results obtained from
\citet{Ilbert13} are also shown in this panel for comparison. We also illustrate
the GSMF evaluated from the Illustris simulation \citep{Vogelsberger14}, by
adopting the fitting formula and the best-fitting parameters from
\citet{Torrey15}, which appears to be relatively higher at both the low-mass
(i.e., $\la 10^{10}\msun$) and the high-mass (i.e., $\ga 10^{11}\msun$) ends
than the observational ones. The differences in the results caused by the
difference in the GSMFs will be discussed in Section~\ref{sec:discussion}.

\subsubsection{Galaxy merger rates}\label{subsubsec:galaxymerger}

The galaxy merger rates can be evaluated from different ways including
observations, numerical simulations, semi-analytical models, and semi-empirical
methods \citep{Hopkins10a}.

Observationally, it is usually determined by dividing the fraction of merging
galaxies by a characteristic merger timescale, i.e., the fractional merger
rate. The fraction of merging galaxies is usually determined through either
close-pair counts or galaxy morphological disturbance. The characteristic
merger timescale, which is usually called merger-averaged observability time,
can be calibrated by hydrodynamic simulations of galaxy mergers \citep{Lotz08,
Lotz10, Lotz11, Snyder17} or cosmological dark matter halo $N$-body simulations
\citep{Kitzbichler08}. For example, the fractional major merger rate of galaxies with
total mass greater than $M\gal$ at redshift $z$ can be defined by
\begin{equation}
\calR(>M\gal,z)\equiv 
\frac{f\mrg(>M\gal,z)}{\langle T\mrg\rangle} =
\frac{C\mrg f\pair (>M\gal,z)}{\langle T\mrg\rangle}.
\label{eq:calR}
\end{equation}
Here major mergers mean that the mass ratio of the two galaxies $q\gal$ is
greater than a value $q_{\rm major}$, such as $1/3$ or $1/4$. In
Equation~(\ref{eq:calR}), $f\mrg$ represents the fraction of galaxies undergoing
merging processes, which is the fraction of galaxies in close pairs $f\pair$
multiplied by a factor $C\mrg$ describing the probability of an observed close
pair that will merge within a given time, e.g., $\sim 0.6$ in \citet{Lotz11},
and $\tobs$ is the merger-averaged observability time.

Alternatively, given the GSMF $n\gal(M\gal,z)$ and the galaxy
merger rate per galaxy $\calR\gal(q\gal,z|M\gal)$ as defined in
Section~\ref{subsec:BBHdis}, the fractional major merger rate can be obtained
by
\begin{eqnarray}
& \calR(>M\gal,z) = \nonumber\\
& \frac{\int_{M\gal}^{\infty} dM\gal n\gal(M\gal,z) 
\int_{q_{\rm major}}^1 \calR\gal(q\gal,z|M\gal)dq\gal }
{\int_{M\gal}^{\infty} n\gal(M\gal,z)dM\gal},
\label{eq:calRinte}
\end{eqnarray}
where $q_{\rm major}=1/4$.
In this work, we adopt the galaxy merger rate from the Illustris simulation
\citep{Rodriguez-Gomez15} (see Table 1 therein), where a simple formula is
provided to describe the merger rate as a function of progenitor stellar mass
ratio, descendant total stellar mass, and redshift.

We illustrate the fractional galaxy merger rates obtained from both the 
Illustris simulation and observations in Figure~\ref{fig:merg_rate}. To obtain
galaxy merger rate from observations, the merger timescale $\tobs$ used
in Equation (\ref{eq:calR}) is obtained from numerical simulations.
Below we list several examples.
\begin{itemize}
\item Based on virtual galaxy catalogues derived from the Millennium Simulation
\citep{Springel05}, \citet{Kitzbichler08} calibrate the average observability
timescale for galaxies in close pairs within a projected separation of $r\proj$
as
\begin{eqnarray}
& \tobs = 2.2\gyr\left(\frac{r\proj}{50\kpc}\right) \nonumber \\
& \times \left(\frac{M\gal}{4\times 10^{10}h^{-1}\msun}\right)^{-0.3}
\left(1+\frac{z}{8}\right),
\label{eq:tKitzbichler08}
\end{eqnarray}
where this timescale is for close pairs with relative radial velocity
$<300\kms$. For those close pairs without the radial velocity difference
constraint, the normalization of the observability timescale is increased by a
factor of $\sim 1.5$.  In Equation (\ref{eq:tKitzbichler08}), the dependences of
$\tobs$ on the total mass $M\gal$ and redshift $z$ are both relatively weak.
\item
\citet{Lotz08} and \citet{Lotz10} conduct a series of hydrodynamic simulations
of galaxy mergers with different mass ratios and orbital parameters to
determine their average observability timescale. Based on these results, in
\citet{Lotz11}, the distributions of baryonic gas fraction and mass ratio, as
well as their redshift evolutions from cosmological-scale galaxy evolution
models are adopted to obtain the average observability timescale for candidate
galaxy mergers identified based on either close-pair counts or morphology
disturbance. By fitting the results of \citet{Lotz11}, $\tobs$ is estimated to
be $0.3\gyr$ for a projected separation of $<20\kpc$; 
given that the candidates are selected by their masses
($\ga 10^{10}\msun$, according to \citealt{Xu12}), and thus we have
\begin{eqnarray}
& \tobs = 0.3\gyr\left(\frac{r\proj}{20\kpc}\right) \nonumber \\
& \left(\frac{M\gal}{4\times 10^{10}h^{-1}\msun}\right)^{-0.3}
\left(1+\frac{z}{8}\right),
\label{eq:Xu12}
\end{eqnarray}
where the dependences on the total mass $M\gal$ and redshift $z$ are adopted as
those in \citet{Kitzbichler08} because the study in \citet{Lotz11} does not provide
the dependence.
The timescale shown in Equation (\ref{eq:Xu12}) is a factor of ${\sim}3$
smaller than that of Equation (\ref{eq:tKitzbichler08}).
\item By comparing the synthetic close-pair catalogue
with the intrinsic galaxy merger rate in the Illustris Simulations
(up to redshift $z\sim 3$),
\citet{Snyder17} conclude that the averaged observability time
incorporating $C\mrg$ (i.e., $\tobs/C\mrg$) is a strong function of
redshift, that is
\begin{equation}
\tobs/C\mrg\simeq 2.4\times(1+z)^{-2}\gyr.
\label{eq:Snyder17}
\end{equation}
\end{itemize}
In the upper panel of Figure~\ref{fig:merg_rate}, we show the observational
close-pair fractions at different redshifts and different galaxy mass ranges.
In the lower panel of Figure~\ref{fig:merg_rate}, we show the fractional galaxy
merger rates obtained from the numerical simulations in
\citet{Rodriguez-Gomez15}, and the observational ones converted by using
Equation (\ref{eq:calR}) and
Equation (\ref{eq:Snyder17}).  We also convert the fractional galaxy merger
rates obtained from the numerical simulations in \citet{Rodriguez-Gomez15} to
the close-pair fractions by using Equation (\ref{eq:calR}) and
Equation (\ref{eq:Snyder17}) and show them in the upper panel.  As seen from
the figure, the fractional galaxy merger rates obtained from the Illustris
simulation in \citet{Rodriguez-Gomez15} is close to the observational one
converted by using Equation (\ref{eq:Snyder17}) (which evolves with redshift
strongly). However, if the merger timescale depends weakly on redshifts as
shown in Equation (\ref{eq:Xu12}) or (\ref{eq:tKitzbichler08}), the galaxy
merger rates of \citet{Rodriguez-Gomez15} will be much higher than those
converted from observations at high redshifts $z\ga 2$.

\subsection{BH--host galaxy relations}
\label{sec:data:msigma}

As shown by the modeling of the BBH population in Equation (\ref{eq:phibbh}) in
Section~\ref{sec:modelBBH}, besides the GSMFs and the galaxy
merger rates, the BH--host galaxy relation is also needed to determine
the probability function $p\bh(M\bh,q\bh\vert M\gal,q\gal,z)$
and $p\bgal(M\bgal,q\bgal|M\gal,q\gal,z)$.

The determination of the BH--host galaxy relation is composed of the two parts:
the relation between the MBH mass and the physical property of the hot
component of a galaxy (i.e., either an elliptical galaxy or the bulge of a
spiral or S0 galaxy), and the relation (e.g., mass ratio) between the physical
properties of the hot component and the total component of a galaxy.  Here the
two parts are simply called as the BH--bulge relation and the bulge--galaxy
relation.

In Figure~\ref{fig:scal_rela}, we show the BH--bulge relations obtained from
different works in the literature (many of which can also be seen in
\citealt{Sesana13}).
The BH--bulge relations are parameterized in the
equation below and the related parameters are listed in Table~\ref{tab:scal}.
\begin{equation}
\log M_{\rm BH,1} = 
\tilde{\gamma}+
\tilde{\beta}\log\sigma_{200}+
\tilde{\alpha}\log M_{*,11},
\label{eq:scal} 
\end{equation}
where $M_{\rm BH,1}$, $\sigma_{200}$, and $M_{*,11}$ correspond to the
MBH mass in units of $1\msun$, the one-dimensional velocity
dispersion in units of $200\kms$, and the bulge mass in units of $10^{11}\msun$,
respectively. The intrinsic scatters of the different BH--bulge relations are
quantified by $\tilde{\epsilon}$ in the table. We call the relation as
the $M\bh$-$\sigma$ relation if $\tilde{\alpha}=0$,  the $M\bh$--$M\bgal$
relation if $\tilde{\beta}=0$, and the $M\bh$-$M\bgal$-$\sigma$ relation
if none of $\tilde{\alpha}$ and $\tilde{\beta}$ is zero.
For those $M\bh{-}\sigma$ relations and $M\bh$-$M\bgal$-$\sigma$
relations
with $\tilde{\beta}\ne 0$, we apply the relation between the velocity dispersion
$\sigma$ and the bulge mass $M\bgal$ from \citet{Gallazzi06},
$\log(\sigma/\kms) = 0.286\log(M\bgal/\msun)-0.895$ (with an intrinsic scatter
of $0.071{\rm dex}$) and convert them to the $M\bh{-}M\bgal$ ones to be shown
in Figure~\ref{fig:scal_rela}, and the corresponding scatters in the
converted $M\bh{-}M\bgal$ relations are set to be
$\sqrt{\tilde{\epsilon}^2+0.071^2 \tilde{\beta}^2}$.

The bulge--galaxy relation is often characterized by the bulge-to-total stellar
mass ratio $B/T$.  The bulge-to-total ratio depends on the properties of the
host galaxies, such as the morphology or the galaxy mass. In this work, we
follow the prescription used by \citet{Ravi15}, where early-type galaxies are
divided into ellipticals and S0s.  The fractions of S0s and ellipticals are
$75\%$ and $25\%$, respectively, for $10\leq\log(M\gal/\msun)\leq 11.25$, and
$55\%$ and $45\%$, respectively, for more massive galaxies. For ellipticals,
the bulge-to-total mass ratio is $B/T=1$; whereas for S0s, $B/T=0.7$. For
late-type galaxies, $90\%$ of them have the bulge-to-total ratio $B/T=0.2$, and
the remaining $10\%$ have no bulge component, i.e., $B/T=0$. The scatter in
$B/T$ is assumed to be $0.1~{\rm dex}$ for both early- and late-type galaxies.
The fraction of early-type galaxies as a function of redshift and galaxy mass
are approximated by the fraction of quenched galaxies presented in
\citet{Behroozi19}.

\subsection{Mock BBH evolution track library}
\label{subsec:mockBBH}

In this subsection, we describe the construction of the mock BBH evolution
library, based on the mock host galaxy sample described in
Section~\ref{sec:data:mock}, which will be used to determine the statistical
distributions of surviving BBHs, BBH coalescence rates, and GWB etc.\ in
Section~\ref{sec:res}.

The BBH evolution track library contains the BBH evolution tracks, with
different parameters $\{M\bgal, q\bgal, M\bh, q\bh\}$. The parameter grids are
set in the following way.  We set the BBH properties $(M\bh,q\bh)$ so that
$\log(M\bh/\msun)$ is in the range from 5 to 10 with an interval of 0.1 and
$\log q\bh$ is in the range from -3 to 0 with an interval of 0.1.  Given each
$(M\bh, q\bh)$, we select 25,000 galaxies
randomly from the mock galaxy sample constructed in Section~\ref{sec:data:mock}.
The mass of the spheroidal component of each galaxy can be evaluated based on
its associated Nuker-law parameters and mass-to-light ratio.  We use the method
presented in Section~\ref{sec:dynamics} to calculate the BBH evolution track in
each (triaxial) galaxy.  For each host galaxy, the pre-merger mass ratio
$q\bgal$ is set so that $\log q\bgal$ is in the range from -3 to 0 with an
interval of 0.1. In the BBH evolution library, the BBH evolution tracks do not
depend on $q\bgal$. The number of the parameter grids is 
$3.75\times 10^7$.  Note that the different systems in
the mock library do not appear in the universe with the same probability, their
existence probabilities in the universe are modulated by galaxy merger rates and
the correlations between the MBHs and their host galaxies etc.\ (see more in
Section~\ref{subsubsec:lib_app} below).

We have tested that the above grids are sufficient for obtaining the GWB
results, but more are needed to obtain the statistical distributions of
surviving BBHs. For the systems with $\log(M\bh/\msun)$ being in the range from
5 to 10 with an interval of 1.0 and $\log q\bh$ being in the range from -2 to 0
with an interval of 0.5 (see the illustrated example systems in
Figs.~\ref{fig:trac_tau0}--\ref{fig:calc_vphi}), we increase the selected galaxy
number from 25,000 to 100,000 for each
($M\bh,q\bh$), which adds another $1.5\times 10^6$ evolution tracks to the
library. 

The mock BBH evolution track library is applied in the following way.
Given a system with
parameters $(M\bgal, q\bgal, M\bh, q\bh)$, its BBH evolution track is obtained
by matching to the appropriate BBH track in the BBH evolution library. The
match is done by finding out the one in the BBH evolution library, e.g., system
$k$, which has the minimum value of $\sum_j(\log X^k_j-\log X_j)^2$
($j=1,2,...,4$), where $X_j\in\{M\bgal, q\bgal, M\bh, q\bh\}$.

\subsubsection{Monte Carlo integrations: application of the BBH evolution library}
\label{subsubsec:lib_app}

Below we describe how to obtain the distributions of the surviving BBHs
$\Phi\bbh(M\bh,q\bh,a,z)$ (see Eq.~\ref{eq:phibbhi}) by applying the above BBH
evolution library in a Monte-Carlo method.  The similar method is also applied
to obtain the coalescence rate $R\bh(M\bh,q\bh,z)$ and the stochastic GWB
$h\rmc(f)$ etc.\ in this work. 

We evaluate the distribution function of $\Phi\bbh(M\bh,q\bh,a,z)$ at the mesh
points $(M\bh,q\bh)$ of the BBH evolution track library, with
$\log(a/\pc)$ being in the range from -6 to 1 with an interval of 0.1 and $z$
being in the range of
0--3 with an interval of 0.1. As seen from Equation (\ref{eq:phibbhi}), given
each point of $(M\bh, q\bh, a, z)$, the integration of Equation
(\ref{eq:phibbhi}) is done over the $(\log M\bgal, \log q\bgal)$ space.  We set
the integration range of $\log (M\bgal/\msun)$ to be within $[8,13]$ and that of
$\log q\bgal$ to be within $[-3,0]$. We divide the space of $(\log M\bgal, \log
q\bgal)$ into 1500 small grids with
intervals $\Delta\log M\bgal=0.1$ and $\Delta\log q\bgal=0.1$, do a Monte Carlo
integration within each small grid, and then add the integration results of all
the grids up. 

The part of the function $\Phi\bbh(M\bh,q\bh,a,z)$ contributed by the integrand
within each small grid
(see Eq.~\ref{eq:phibbhi}) 
can be obtained by
\begin{align}
& (\ln 10)^2 \Delta\log M\bgal \Delta\log q\bgal \nonumber \\
\times & \frac{1}{N}\sum_{i=1}^{N} 
n\bgal(M\bgal,z_i)\calR\bgal(q\bgal,z_i|M\bgal) \nonumber \\
\times & p\bh(M\bh,q\bh|M\bgal,q\bgal,z_i)H(t-\tau_{a,i})\frac{t_{{\rm evol},i}}{a},
\label{eq:phibbhiN}
\end{align}
where we set the number of the randomly generated galaxy merger events within a
grid $N=1000$, and the function
$\Phi\bbh(M\bh,q\bh,a,z)$ can be obtained by the sum over all the grids. The
value of $n\bgal(M\bgal,z)R\bgal(q\bgal,z|M\bgal)$ in Equation (\ref{eq:phibbhiN})
can be obtained (from some analytical formula) as described in
Section~\ref{sec:data:gmfmr}.  The function $p\bh(M\bh,q\bh|M\bgal,q\bgal,z)$
can also be described in an analytical way (see Section~\ref{sec:data:msigma}).
The variables of $\tau_{a,i}$ and $t_{{\rm evol},i}$ in the integrand of
Equation (\ref{eq:phibbhiN}) represent the BBH evolution in each generated
galaxy merger event, which are evaluated by matching to the appropriate BBH
track in the BBH evolution library as described in Section~\ref{subsec:mockBBH}.

Note that the integration is dependent on the following physical ingredients,
including the GSMF ($n\gal$), the galaxy merger rate per
galaxy, the BH-host galaxy relation ($p\bh$), as well as the shape distribution
of the host galaxies. In this paper, we consider how the results are affected
with different inputs of those ingredients in our results.

\section{Results} 
\label{sec:res}

In this section, we present the main results of this study.
In Section~\ref{sec:res:trac}, we illustrate some examples of BBH evolution
tracks and the dependence of BBH evolution timescales on
host galaxy triaxial shapes.
In Section~\ref{sec:res:prop}, we present the statistical properties of BBH
evolution tracks expected in the realistic universe, e.g., the distribution of
their bottleneck evolution timescales and the distribution of the slopes in
their evolution timescale curves during the hard binary stages.
In Section~\ref{sec:res:surv}, we present statistical distributions of
surviving BBHs, which are the very target of many observational surveys. 
In Section~\ref{sec:res:cols}, we present the BBH coalescence rates.
In Section~\ref{sec:res:gwb}, we present the strain amplitude of the stochastic
GWB produced by the cosmic population of BBHs.
In Section~\ref{sec:res:indv}, we discuss some prospects for detecting
individual sources at the PTA bands.
In Section~\ref{sec:res:lisa}, we extend the study to investigate the
contribution of the BBH mergers to the LISA detection bands.

\subsection{BBH evolution tracks and timescales}
\label{sec:res:trac}

In this subsection, we illustrate the examples for the BBH evolution tracks and
the dependence of the BBH evolution timescales on galaxy triaxial shapes.

Figure~\ref{fig:evlv_trac} shows an example for the evolution tracks of a BBH
inside a galaxy merger remnant. The example host galaxy (i.e., the merger
remnant containing the BBH) has the typical properties, with surface brightness
following a Nuker-law profile characterized by parameters $\alpha=3.97$,
$\beta=1.26$, $\gamma=0.72$, and $R\rmb=466\pc$. The system has a total stellar
mass of $\log(M\bgal/\msun)=12$ and contains a BBH at the center with total mass
satisfying $\log(M\bh/\msun)=8.6$.  Two sets of BH mass ratios are assumed,
i.e., $1/1$ and $1/10$, as shown by solid and dashed curves in the figure.
Given a BBH mass ratio, the five sets of intrinsic triaxial shapes are assumed
for the host merger remnant, i.e., $(b/a,c/a)=(1.0,1.0)$ (red), $(1.0,0.8)$
(green), $(1.0,0.6)$ (blue), $(0.9,0.8)$ (cyan) and $(0.8,0.6)$ (magenta).

As described in Section~\ref{sec:dynamics}, the evolution of a BBH goes through
several stages driven by different physical mechanisms.  In
Figure~\ref{fig:evlv_trac}, we mark the boundaries between these different
stages, according to the definitions in Section~\ref{sec:dynamics}.  Along the
direction of decreasing $a$, the black arrow labeled with ``bound'' marks the
formation of a bound BBH, the black arrow labeled with ``hard'' marks the
formation of a hard BBH, and the color arrows mark the transition to the
gravitational radiation stage for the BBH evolution curves with the
corresponding colors and line styles.  We use the horizontal arrows labeled with
$t\peak$ 
to mark the maximum BBH evolution timescales after the
BBHs become bound. As seen from the figure, the galaxy
triaxiality can decrease the maximum BBH evolution timescale significantly, as
shown by Figures~7-9 in \citet{Yu02}.

For the example BBH system with $q\bh=1$
shown in Figure~\ref{fig:evlv_trac}, we also calculate their evolution with
other galaxy host galaxy shapes, and show the dependence of their maximum
evolution timescales on the shapes in
Figures~\ref{fig:tsc_contour}--\ref{fig:tpeak_rely}.

The right panel of Figure~\ref{fig:tsc_contour} shows the contours of $\log t\peak$, as a
function of the two axis ratios of the mass density distribution (i.e., the
medium-to-major axis ratio $b/a$ and the minor-to-major axis ratio $c/a$).
As seen from the figure, the contours of $\log t\peak$ are quite
dense at the boundaries of $b=c$ and $a=c$, which implies that even a minor
triaxiality can decrease $t\peak$ significantly. As shown by the label on the
contours, the values of $t\peak$ decrease to significantly short timescales
(shorter than the Hubble timescale) in a wide parameter space of triaxiality.

As an alternative way to illustrate the results in
Figure~\ref{fig:tsc_contour}, Figure~\ref{fig:tpeak_rely} shows $t\peak$ of
the example BBH system as a function of the triaxiality parameter
$T=(a^2-b^2)/(a^2-c^2)$ for the five sets of different minor-to-major axis
ratios $c/a$. Except for the curves in nearly spherical system ($c/a=0.99$ or
0.95), $t\peak$ decreases sharply when the triaxiality deviates from $T=0$
($b/a=1$) and $T=1$ ($b=c$), which is consistent with the result shown
in Figure~\ref{fig:tsc_contour}.

\subsection{Statistical properties of BBH evolution tracks}
\label{sec:res:prop}

In this section, the statistical distributions of some properties describing
the BBH dynamical evolution tracks are investigated, including the
coalescence timescale $\tau_0\equiv\tau_{a=0}$ (e.g., see Eq.~\ref{eq:f3})
defined by the total time to be taken by a BBH to
the final coalescence from the mergers of their host galaxies, the maximum
timescale $t\peak$ defined by the maximum evolution timescale of a BBH after the
formation of the binary, the slope of the evolution curve of a BBH in the $\log
t_{\rm evol}$--$\log a$ space before its gravitational radiation stage, and
some other orbital properties of a BBH when it enters the gravitational
radiation stage.

Consider that a population of galaxy mergers occur at redshift $z'$. Below we
describe the statistical distributions of the evolution tracks of the BBHs
formed by those galaxy mergers (note that these distributions are different
from the statistical distributions of the surviving BBHs at a given redshift to
be described in Section~\ref{sec:res:surv}). Given a physical property $x$
describing the BBH evolution tracks, we define its probability distribution
function $\psi_x(x|M\bh,q\bh,z')$ by
\begin{equation}
\psi_x(x|M\bh,q\bh,z')\equiv
\frac{\Psi_x(x,M\bh,q\bh,z')}{\Psi(M\bh,q\bh,z')},
\label{eq:pdfx}
\end{equation}
where
\begin{align}
&\Psi(M\bh,q\bh,z') \nonumber\\ \equiv
&\int dM\bgal\int dq\bgal 
 n\bgal(M\bgal,z')\calR\bgal(q\bgal,z'|M\bgal) \nonumber\\
&\times p\bh(M\bh,q\bh|M\bgal,q\bgal,z'),
\label{eq:phi}
\end{align}
and
\begin{align}
&\Psi_x(x,M\bh,q\bh,z') \nonumber\\ \equiv
&\int dM\bgal\int dq\bgal 
 n\bgal(M\bgal,z')\calR\bgal(q\bgal,z'|M\bgal) \nonumber\\
&\times p\bh(M\bh,q\bh|M\bgal,q\bgal,z') \nonumber\\
&\times p_x(x|M\bgal,q\bgal,M\bh,q\bh,z'),
\label{eq:phix}
\end{align}
In Equation (\ref{eq:phix}), $p_x(x|M\bgal,q\bgal,M\bh,q\bh,z')$ is the
probability function defined so that $p_x(x|M\bgal,q\bgal,M\bh,q\bh,z')dx$
represents the probability that the evolution tracks of the BBHs characterized
by parameters $(M\bh,q\bh|M\bgal,q\bgal,z')$ has the corresponding physical
property in the range $x\rightarrow x+dx$, where the BH mass growth is ignored
before the BBH coalescence. We have $\int dx
p_x(x|M\bgal,q\bgal,M\bh,q\bh,z')=1$.
The probability function $p_x(x|M\bgal,q\bgal,M\bh,q\bh,z')$ can be obtained
by using Monte-Carlo simulations to generate $N(=1000)$ galaxy mergers with
the grid of the parameters $(M\bgal,q\bgal)$ at $z'$ leading to the formation of
BBHs with parameters ($M\bh,q\bh$) and calculating the physical property $x_i$
($i=1,2,...,N$) of the corresponding BBH evolution tracks. Thus, we have 
\begin{align}
&p_x(x|M\bgal,q\bgal,M\bh,q\bh,z') \nonumber\\=
&\frac{1}{N}\sum_{i=1}^{N}\delta(x-x_i|_{M\bgal,q\bgal,M\bh,q\bh}) \nonumber\\=
&\frac{1}{N}\frac{d}{dx}\sum_{i=1}^{N}H(x-x_i|_{M\bgal,q\bgal,M\bh,q\bh}).
\end{align}
In a similar way to obtain $\Phi\bbh(M\bh,q\bh,a,z)$ as described in
Section~\ref{subsubsec:lib_app},
the part of the function $\Psi_x(x,M\bh,q\bh,z')$ contributed by the
integrand within each small grid of $(M\bgal,q\bgal)$ (see Eq.~\ref{eq:phix})
can be obtained by
\begin{align}
& (\ln 10)^2 \Delta\log M\bgal \Delta\log q\bgal \nonumber \\
\times & \frac{1}{N}\sum_{i=1}^{N}
 n\bgal(M\bgal,z')\calR\bgal(q\bgal,z'|M\bgal) \nonumber\\
\times & p\bh(M\bh,q\bh|M\bgal,q\bgal,z') H(x-x_i|_{M\bgal,q\bgal,M\bh,q\bh}),
\label{eq:phix1}
\end{align}
and $\Psi_x(x,M\bh,q\bh,z')$ can be obtained by the sum over all the grids.

In Figures~\ref{fig:trac_tau0}--\ref{fig:trac_turn}, we show the statistical
results for the evolution tracks of BBHs caused by galaxy mergers occurring at
$z'=0$, by using Equation (\ref{eq:pdfx}) and the method of the Monte Carlo
integration as described above, where the galaxy merger rate is determined by
the GSMF $n\gal$ of \citet{Behroozi19} and the galaxy merger
rate per galaxy $\calR\gal$ of \citet{Rodriguez-Gomez15}.  The BH mass is set,
according to the $M\bh{-}M\bgal$ relation of \citet{Kormendy13} (KH13b in
Table~\ref{tab:scal}).  The mock BBH
evolution tracks obtained with the different intrinsic shape distributions of
the early-type galaxies (see Figure~\ref{fig:tsc_contour}) are adopted, and
their results are compared.

Figure~\ref{fig:trac_tau0} shows the probability distribution of the logarithm
of the BBH coalescence timescale, $\log\tau_0$. The left panels show the
results for different $M\bh$, and the right panels show the results for
different $q\bh$. From top to bottom panels, the results are obtained by
choosing the different sets of the shape distribution of the host galaxy merger
remnants presented in \citet{Weijmans14}, \citet{Rodriguez13}, and
\citet{Padilla08}, respectively. As seen from the figure, we find the following
points.
\begin{itemize}
\item The distribution of the coalescence timescale $\tau_0$ spreads widely,
from $\sim 10^8$ to $\sim 10^{12}\yr$. 
\item The left panels of the figure show that the distribution depends on the
BBH total mass. The BBHs with high total masses (e.g., $\sim
10^9$-$10^{10}\msun$) have a relatively large fraction of long $\tau_0$ (e.g.,
longer than the Hubble timescale $\sim 10^{10}\yr$), as shown by that their
distribution peaks shift rightwards, and those with low total masses have
lower probabilities to have $\tau_0$ longer than the Hubble timescale.
\item The right panels of the figure show that the distribution depends on the
BH mass ratios. The BBHs with smaller BH mass ratios (e.g., $q\bh=0.01$) have a
relatively larger fraction of long $\tau_0$, as shown by that the left sides of
the distribution curves shift rightwards, which is mainly caused by the
increasing dynamical friction timescales with smaller mass ratios, in
consistency with the conclusions in \citet{Yu02}.  As shown in \citet{Yu02}, for
BBHs with mass ratio $\la 10^{-3}$, the dynamical friction timescale is
generally so long that they are hardly able to finish the in-spiral within a
Hubble timescale.
\item As seen from the different rows of the figure, the different choice of
the shape distributions does not affect the above summary qualitatively.
Quantitatively, the coalescence timescale distributions in the bottom panels,
obtained by a nearly oblate spheroidal shape distribution of \citet{Weijmans14}
are relatively broadened towards longer $\tau_0$,, compared with those in the
top and the middle panels.  The coalescence timescale distributions in the top
and the middle panels obtained from the triaxial shape distributions of
\citet{Rodriguez13} and \citet{Padilla08}, are close to each other.
\end{itemize}

Figure~\ref{fig:trac_tmax} shows the probability distribution of the logarithm
of the maximum BBH evolution timescale $t\peak$ after the BBHs become bound. 
As done in
Figure~\ref{fig:trac_tau0}, the results for the different shape distributions of
the host merger remnants are adopted, the different BBH total masses, the
different BBH mass ratios are shown in this figure.  As seen from the left
panels in Figures~\ref{fig:trac_tau0} and \ref{fig:trac_tmax}, the distributions
of the peak evolution timescales 
are quite similar to those of the coalescence timescales, which suggests that
the coalescence timescales $\tau_0$ of BBHs with large mass ratios are dominated
by their peak evolution timescale $t\peak$ after the BBH becomes
bound. However, different from the distributions shown in
the right panels of Figure~\ref{fig:trac_tau0}, the distributions of the peak
evolution timescale after the BBH becomes bound is
less sensitive to the BBH mass ratios.

Figure~\ref{fig:trac_slop} shows the probability distribution of the slopes of
the BBH evolution curves in the $\log t_{\rm evol}$--$\log a$ space, during the
hard binary stage, defined by $\alpha\bbh\equiv \frac{\log t_{\rm
evol}(a\gr)-\log t_{\rm evol}[\max(a\rmh,a_{\rm dp})]}{\log a\gr-\log
[\max(a\rmh,a_{\rm dp})]}$. The slope is an indicator of whether the loss cone
is full or not.  If the loss cone remains full at the hard binary stage, the
slope is expected to be $-1$.  However, as the stars in the loss cone can be
gradually removed due to the interaction with the BBH, the slope can deviate
from $-1$ and be steeper significantly, if the stellar refilling rates into the
loss cone are not high enough. 
The upper panels show the results obtained by using the intrinsic shape
distribution of early-type galaxies from \citet{Padilla08}, and the lower panels
show the results obtained by using the shape distribution from
\citet{Weijmans14}. The results obtained by using the shape distribution from
\citet{Rodriguez13} are not shown here, as they are quite similar to those in
the upper panels.
As seen from the figure, the distribution of the slopes tends to shift leftwards
(with decreasing or steeper values), as the BBH total mass becomes larger in the
left panels or the BBH mass ratio becomes smaller in the right panels.
Those trends are relatively stronger in the lower panels.
Overall, the slopes are in the range from -4 to -1 for BBH systems with mass
ratio greater than 0.1, not deviating from the value $-1$ by orders of
magnitude, which indicates
that in most of those cases the stellar refilling rates to the loss cone are not
low, compared to the rates in the full loss cone case.

Figure~\ref{fig:trac_turn} shows the probability distributions of the physical
properties marking the transition to the gravitational radiation stage, i.e.,
the semimajor axis $a\gr$, the GW frequency $f\gr$, and the orbital period
$P\gr$ in the top, middle, and bottom panels, respectively.  
As expected, the figure shows that larger BBHs tend to evolve into the
gravitational radiation stage with larger semi-major axes and orbital periods
and with smaller GW frequencies. The typical ranges of these properties shown
in the figure can help to determine the chance of a given BBH system being at
the gravitational radiation stage.  In addition, the transitional orbital
frequency is closely related with the bending frequency of the strain spectrum
of the GWB from the cosmic population of BBHs, as will be seen in
Section~\ref{sec:res:gwb}.

\subsection{Distributions of the surviving BBHs} 
\label{sec:res:surv}

In this subsection, we present the results on the distributions of the surviving
BBHs (e.g., see $\Phi\bbh(M\bh,q\bh,a,z)$ in Equation~\ref{eq:phibbh}), by using
the Monte Carlo integration described in Section~\ref{subsubsec:lib_app}. We
show the distributions for the following physical variables: the BBH semi-major
axis $a$, the orbital period $P$, the frequency $f$ of their emitted GWs, and
the relative circular velocity $v$ of the BBHs. Given the total BBH mass, those
variables are mutually convertible after the binary becomes bound (i.e.,
$a\lesssim R\infl$), since the gravitational force between the binary dominates
the orbital motion of the BBH at such scales. For example, the orbital period
distribution can be expressed through the semi-major axis distribution, i.e.,
$\Phi\bbh(M\bh,q\bh,a,z)\left\vert\frac{da}{dP}\right\vert$. The distributions
of these variables may provide clues to discover surviving BBHs.

Figures~\ref{fig:calc_aphi}--\ref{fig:calc_vphi} present the distributions of
the BBH semi-major axis $a$, the orbital period $P$, the GW frequency $f$, and
the relative circular orbital velocity $v$ of the surviving BBHs, respectively.
In each of these figures, the upper left panel shows the dependence of the
distribution functions on the total mass of the BBHs at $z=0$ with $q\bh=1.0$
and $\log(M\bh/\msun)=5.0$, $6.0$, $7.0$, $8.0$, and $9.0$; the upper right
panel shows the dependence on the BBH mass ratio for BBHs at $z=0$ with
$\log(M\bh/\msun)=8.0$ and $\log q\bh=0.0$, $-0.5$, $-1.0$, $-1.5$, and $-2.0$.
Most of those surviving BBHs with properties significantly deviated from the
corresponding peak locations have the evolution timescales significantly shorter
than the Hubble timescale.
The lower panels show the cumulative distributions of the corresponding upper
panels. The figures show the medians of the results obtained by the different
BH--host galaxy relations (similarly for Figs.~\ref{fig:surv_frac} and
\ref{fig:cols_rate} below). Regarding the orbital period distribution shown in
the top left panel of Figure~\ref{fig:calc_pphi}, we note that the
observationally constrained orbital periods of some individual BBH candidates,
e.g., $\sim 12\yr$ for OJ 287 with
$M\bh\simeq 1.83\times 10^{10}\msun$ and $q\bh\simeq 0.01$ \citep{Valtonen08},
$\sim 5.16\pm 0.24\yr$ for PG 1302 with
$\log(M\bh/\msun)\simeq 8.3$--$9.4$ \citep{Graham15}, or $\sim 1.2\yr$ for Mrk
231 with $M\bh\simeq 1.5\times 10^8 \msun$
and $q\bh\simeq 0.026$ \citep{Yan15}, are not located at the peak locations of
the distributions corresponding to their BBH masses, which would imply a
relatively higher probability that more surviving BBH candidates are likely to
be detected in future. 
In these figures, we also show the distributions contributed only by those BBHs
with coalescence timescale $\tau_0$ shorter than the Hubble timescale (by the
dot-dashed curves). The differences between the solid and the dot-dashed curves
in the figure indicate that the distributions of the surviving BBHs with large
semimajor axes/long orbital periods/low GW frequency/low relative velocities (at
the binary stages) are mainly contributed by those not being able to finish
their final coalescences within the Hubble timescale, while the distributions of
the BBHs with low semimajor axes (at the gravitational radiation stage) are
those that can coalesce within the Hubble timescale.

Figure~\ref{fig:surv_frac} plots the fraction of MBHs in galaxies that host
surviving BBHs, as a function of redshift. In the top left panel, only the BBHs
with separation $0<a\leq 10\pc$ and mass ratio $q\bh\geq 1/3$ are included in
estimating the BBH fraction, i.e., $\int_0^{10\pc} d a \int_{1/3}^1 dq\bh
\Phi\bbh(M\bh,q\bh,a,z) /n\bh(M\bh,z)$, where $n\bh(M\bh,z)$ is defined so that
$n\bh(M\bh,z)dM\bh$ represents the comoving number density of MBHs with mass
within the range $M\bh\rightarrow M\bh+dM\bh$ at redshift $z$. As seen from the
top left panel, the surviving BBH fractions are around $\sim$1\%--3\% (solid
curves) for different BH masses at $z\la 3$, and the BBH surviving fractions
increase with increasing BH masses for $M\bh\la 10^7\msun$ and decrease with
increasing BH masses for $M\bh\ga 10^7\msun$.  The surviving fractions change
mildly with redshifts for BBHs mass in the range of $10^6\msun\la M\bh \la
10^9\msun$. 
Among the surviving BBHs, more than about a half have total evolution
timescales shorter than the Hubble timescale, according to the comparison
between the dot-dashed curves and the solid curves.
The surviving BBH fractions with $q\bh\ge 1/100$
increase to $\sim 10\%$ as shown in the bottom left panel.
In the right panels, only
the BBHs with orbital periods $1\yr<P<10\yr$ or $1\yr<P<30\yr$ are included in
estimating the BBH fraction. As seen from the top right panel, at $z=0$, the fractions of
surviving BBHs with $q\bh\ge 1/3$ and orbital periods $1\yr<P<10\yr$ are expected to be
$\sim$0.5\%--1\% in MBHs with $M\bh\sim 10^{6}$-$10^7\msun$, the fractions of
those with $1\yr<P<30\yr$ are expected to be $\sim$2\% in MBHs with $M\bh\sim
10^{6}$-$10^7\msun$, and the fractions decrease in MBHs with other masses,
down to $10^{-4}$ for $M\bh\sim 10^9\msun$. 
The fractions are not sensitive to redshifts at $z\la 3$.
The fractions of surviving BBHs with $q\bh\ge 1/100$ and $1\yr<P<30\yr$
increase to $\sim 10\%$ for $M\bh\sim 10^{6}$-$10^7\msun$ and $\sim 6\times
10^{-3}$ for $M\bh\sim 10^9\msun$, as shown in the bottom right panel.

\subsection{BBH coalescence rates}
\label{sec:res:cols}

In this subsection, we present the results on the BBH coalescence rates (see
Eq.~\ref{eq:Rbh}), by using the Monte Carlo method described in
Sections~\ref{sec:modelBBH}-\ref{sec:dynamics:sample}.  The BBH coalescence rate
traces the merger rate of their host galaxies, but with a BBH evolution time
delay. This time delay is obtained based on the BBH dynamical evolution model as
described in Section~\ref{sec:dynamics}.

Figure~\ref{fig:cols_rate} shows the BBH coalescence rates ($R\bh$; see
Eq.~\ref{eq:Rbh}) as a function of redshift (thick solid curves), and their
dependence on BBH total masses (left panels) and mass ratios (right panels). In
Equation (\ref{eq:Rbh}), the BBH coalescence rates are obtained with including
the BBH dynamical evolution after their host galaxy mergers, where we adopt the
galaxy shape distribution of \citet{Padilla08}, the galaxy merger rate based on
the galaxy stellar mass function of Behroozi et al.\ (2019), the merger rate per
galaxy of \citet{Rodriguez-Gomez15}, and all the different BH--host galaxy
relations shown in Table~\ref{tab:scal}.  We show the median values of the
results obtained with the different BH--host galaxy relations as the thick solid
curves in the figure. In the lower panels, we show the corresponding
volume-integrated BBH coalescence rates as a function of redshift.  If the BBH
evolution is ignored and the BBH coalescence is assumed to occur immediately
after the host galaxy merger, we use Equation (\ref{eq:Rbhnodelay}) to obtain
the BBH coalescence rates and show them as dashed lines as a reference.  As seen
from the figure, for those BBHs with smaller mass ratios which have relatively
larger coalescence timescales, the results obtained with the inclusion of the
BBH evolution time delay (solid curves) have relatively lower rates at high
redshifts for low $q\bh$, compared to the results obtained without including the
BBH evolution time delay (dashed curves). The total BBH coalescence rate at
$z\la 3$ can be up to ${\sim}1\pyr$.

To illustrate the uncertainty in the BBH coalescence rate estimation, we use
Figure~\ref{fig:cols_rate_scale} to show the BBH coalescence rates obtained by
the different sets of the BH--host galaxy relations in Table~\ref{tab:scal}. The
left panels show the coalescence rates of BBHs with total mass $M\bh=10^5\msun$
and mass ratio $q\bh=1$, and the right panels show the coalescence rates of BBHs
with total mass $M\bh=10^8\msun$ and mass ratio $q\bh=0.01$. The figure shows
that the BBH coalescence rates estimated with the different choices of the
BH--host galaxy relations can differ by two orders of magnitude. 

\subsection{Stochastic GWB at the PTA bands}
\label{sec:res:gwb}

In this subsection, we obtain the stochastic GWB spectra radiated from the
cosmic population of BBHs, by using Equation (\ref{eq:hc2}) and the Monte Carlo
integration method described in Section~\ref{subsubsec:lib_app}, where the
characteristic strain amplitude of the background $h\rmc$ is determined by
several astrophysical ingredients, including the GSMF, the merger rate per
galaxy, the BH--host galaxy relation, and the BBH evolution timescale after
the host galaxy mergers.  We try the different sets of BH--host galaxy
relations in the estimation; with the results being treated with equal
weight.  their median and the standard deviation are then used to characterize
the expectation and uncertainty of the stochastic GWB.

We show the obtained characteristic strain spectra of the stochastic GWB across
the PTA bands, produced by the modeled cosmic population of BBHs, in
Figure~\ref{fig:strn_line}. In the estimation, we adopt the GSMF of
\citet{Behroozi19} and the galaxy merger rate given by
\citet{Rodriguez-Gomez15}, and we use the different color curves to show the
results obtained by the different BH--host galaxy relations listed in
Table~\ref{tab:scal}.  By treating all the results with equal weight, we
show their median as the black solid curve and their
standard deviation around the median as the shaded region in the lower panel of
Figure~\ref{fig:strn_fill}. In Figure~\ref{fig:strn_line}, at the relatively
high-frequency bands ($f\ga 10^{-9}\hz$), the black dashed line represents the
median strain spectrum
obtained by assuming that the BBH coalescence occurs immediately after the
galaxy merger (i.e., applying Eq.~\ref{eq:Rbhnodelay} and setting
$\left\langle \frac{t_{{\rm evol},i}}{t\gr}\right\rangle=1$ in
Eq.~\ref{eq:hchighf}), which follows the
$-2/3$ power law as a function of the observational frequency. 
The bending of the strain spectrum shown at the low-frequency end ($\la
10^{-9}\hz$) is caused by that a significant part of the BBH energy loss is
driven by three-body interactions with surrounding stars, and the spectrum
corresponds to the BBH evolution tracks at $a\ga a\gr$.
As seen from
Figure~\ref{fig:strn_line}, the largest strain amplitude at $f=1\pyr$, denoted
by $\ayr$, is $\ayr\simeq 5.4\times 10^{-16}$, obtained by adopting the $M\bh{-}M\bulge$ relation of
\citet{Kormendy13}; whereas the lowest strain amplitude is $A_{\yr}\simeq
6.1\times 10^{-17}$, obtained by
adopting the $M\bh{-}M\bulge{-}\sigma$ relation of \citet{Shankar16}. The
difference in the GWB estimates can be as large as $\sim 10$ due to different
choices of the BH--host galaxy relationships.
As seen from Figure~\ref{fig:strn_fill}, the median
value (black solid curve) at $f=1\pyr$ is
$\ayr\simeq 2.0\times 10^{-16}$, which is lower than that of
the black dashed line by a factor of $\sim 1.5$, with a standard deviation 
of $^{+1.4}_{-0.8}\times 10^{-16}$ around the median. 
As shown by the red and the yellow points in the figure, the current 95\% upper
limits on the stochastic GWB from the NANOGrav and the IPTA at $f=1\pyr$ are
$1.45\times 10^{-15}$ \citep{Arzoumanian18} and $1.7\times 10^{-15}$
\citep{Verbiest16}, respectively. The 95\% upper limits from the PPTA at 
$f=1\pyr$ is $1.0\times 10^{-15}$ \citep{Shannon15}.
The largest strain amplitude at $f=1\pyr$ shown in Figure~\ref{fig:strn_line}
is below those currently constrained upper limit by a factor of $\sim 2$,
while the median of the predicted values is smaller than the current limit
by a factor of $\sim 5$.
As shown in the figure, the expected strain amplitudes at a little lower
frequency $f\sim 10^{-8}$~Hz are well within the expected detection ability of
the on-going PTA experiments in the future [cyan curve estimated for
the Square Kilometer
Array (SKA) PTA formed by monitoring 50 pulsars at 100ns precision for 15
years \citep{Bonetti18}], except for the lowest strain amplitudes estimated
by using the BH-host relation provided by \citet{Shankar16}.

To quantify the bending trends of the strain spectrum, we fit each estimated
strain spectrum shown in Figure~\ref{fig:strn_line}, by using the
following equation:
\begin{equation}
h\rmc(f)=A\frac{(f/\pyr)^{-2/3}}{[1+(f_{\rm bend}/f)^{\kappa\gw\gamma\gw}]^{1/(2\gamma\gw)}},
\label{eq:fbendkappa}
\end{equation}
where the parameters $f_{\rm bend}$, $\kappa\gw$, and $\gamma\gw$ encode the
information about
the physical processes driving the evolution of the BBHs before they enter the
gravitational radiation stage and the distributions of the BBHs.
Equation (\ref{eq:fbendkappa}) is reduced
to Equation~(24) of \citet{Arzoumanian16} if $\gamma\gw=1$.
The $\kappa\gw$ is the low-frequency spectral index
of the broken power-law GWB strain spectrum, related to the slope of the BBH
evolution track during the hard binary stage, i.e., $\kappa\gw=8/3-2/3(d\log
t_{\rm evol}/d\log a)$. The turnover frequency $f_{\rm turn}$, defined by the
frequency at which the strain spectrum reaches its maximum, is a function of
$f_{\rm bend}$, $\kappa\gw$, and $\gamma\gw$ as follows (see Equation~30 in \citealt{Arzoumanian16}).
\begin{equation}
f_{\rm turn}=f_{\rm bend}\left(\frac{3\kappa\gw}{4}-1\right)^{1/(\gamma\gw \kappa\gw)}.
\label{eq:fturn}
\end{equation}

In Table~\ref{tab:scal}, we list the estimated GWB strain amplitudes at the
observational GW frequency $1\pyr$, $\ayr$, and the fitting parameters $f_{\rm
turn}$ and $\kappa\gw$ for each BH--host galaxy relation used in this study.
As shown in Figure~\ref{fig:strn_fill}, the assumption of instantaneous BBH
mergers leads to an overestimate of the GWB strain amplitudes, compared to the
case with considering the BBH evolution processes. Our calculations show that
the overestimates are in the range of $\sim$0.10--0.20~dex for those
corresponding cases listed in Table~\ref{tab:scal} (see the ``Drop'' column).
In the model, we use the shape distribution of \citet{Padilla08} for the BBH
evolution. We have checked that the adoption of galaxy shape distribution of
\citet{Rodriguez13} gives similar strain amplitudes, and the adoption of galaxy
shape distribution of \citet{Weijmans14} leads to a further decrease by ${\sim}0.15$
dex.

In Table~\ref{tab:scal}, the turnover frequencies $f_{\rm turn}$ are all within
the range $[10^{-10},10^{-9}]\hz$, with a median of $0.25\nhz$,
which is beyond but close to the lower boundary of
current PTA frequency band. If the evolution of the in-spiraling BBHs during the
hard binary stage is always in the full loss-cone regime, we have $d\log t_{\rm
evol}/d\log a=-1$ and $\kappa\gw=10/3$.  In realistic cases, however, the
evolution of the BBHs gradually departs from the full loss-cone regime and
correspondingly $\kappa\gw$ (in the range of $\sim$3.3--4.6) 
increases a little
compared to the full loss-cone case.

At the high-frequency end of the PTA band, the integrated number of
BBHs emitting GWs at these frequencies is too small for Equation~(\ref{eq:hc2})
to act as a good approximation \citep{Roebber16, Kelley17}. In addition to
showing the results of the Monte Carlo integration, we evaluate the strain
spectrum from a realization of the cosmic population of BBHs and show the
results in Figure~\ref{fig:strn_indv}. The realization of the cosmic BBH
population is based on the number distribution function of these objects in the
parameter space $(z, M\bh, q\bh, f\obs)$ (in a rejection method; e.g., see
\citealt{BR03}) and the total number of the sources is
the integration of the number distribution over the 4-dimensional parameter
space.  We obtain ten realizations of the cosmic BBH population by using the
method.
With the realizations, we can also select out the individual sources loud enough
to stand out of the average background, which will be presented in the next
subsection.

Figure~\ref{fig:strn_indv} shows that the GWB strain spectrum obtained from the
ten realizations of the cosmic BBH populations (grey curves) which deviates from
the $-2/3$ power law (purple curve) and fluctuates due to the small number of
BBH sources at the high-frequency end.  Here the strain spectrum is obtained by
adopting the $M\bh{-}M\bulge$ relation of \citet{Kormendy13}, which gives the
largest estimate of the GWB amplitude in Figure~\ref{fig:strn_line} (see the
purple curve). 
In Figure~\ref{fig:strn_indv}, each grey jagged curve show the strain spectrum
estimated from one Monte Carlo realizations of the cosmic BBH sources. We
present the details on how to divide frequency bins and obtain this curve in
Section~\ref{sec:res:indv} below.
Note that in each frequency bin, the source with the strongest strain
($h_{\rm c}$; see Eq.~\ref{eq:hcindividual}) in each realization is
excluded to obtain the grey curve, and it is counted as an individual source
with its characteristic strain (cf.\ Eq.~\ref{eq:hc}) being larger than the
background, and the grey curve (background) is contributed by the remaining
sources in the same frequency bin.
As seen from the figure, the strain spectra based on Equation (\ref{eq:hc})
below for the Monte Carlo realizations of the cosmic BBH sources are well
described by the $-2/3$ power law at $f\la 10^{-8}\hz$. At higher frequencies,
the strain spectra based on the Monte Carlo realizations tend to fall below the
power-law expectation, but with some spikes being above the power law
expectation (see also \citealt{Roebber16}). This is the frequency range where
individual sources emerge from the background, and the detection of the
individual sources will be discussed in the next subsection.

To study which parts of galaxy merger populations contribute significantly to
the stochastic GWB, in Figure~\ref{fig:gwe_contr} we show
the fractions of the total GW energy density (blue dashed) and the total
characteristic strain amplitude (blue solid) contributed by galaxy mergers with
merger redshift $<z$ (left), total galaxy mass $<M\gal$ (middle), and galaxy
mass ratio $>q\gal$ (right), at the observational frequency $f=1\pyr$.
Similarly, to tell the contribution of different parts of BBH populations to
the stochastic GWB,
in Figure~\ref{fig:gwe_contr} we also show the
fractions of the total GW energy density (red dashed) and the total characteristic
strain amplitude (red solid) contributed by BBHs with coalescence redshift $<z$
(left), total BBH mass $<M\bh$ (middle), and BBH mass ratio $>q\bh$ (right), at
the observational frequency $f=1\pyr$.
As seen from Figure~\ref{fig:gwe_contr},
most of the GWB energy density is contributed by galaxy or BBH mergers within
redshift lower than 2, with total galaxy mass within $\sim
10^{10}$--$10^{12}\msun$ or BBH mass within $\sim 10^{8}$--$10^{10}\msun$, and
with merging galaxy or BH mass ratio greater than $\sim 0.01$.

\subsection{Individual sources}
\label{sec:res:indv}

Detection of signals from individual BBH systems, though beyond the current
capability of PTAs \citep{Leeetal11,Arzoumanian14, Zhu14, Babak16}, would be also of great
interest and provide insights into the BBH population in a different way from
the global constraint from the stochastic GWB. Individual BBH systems can be
possible targets of telescopes detecting electromagnetic signals so that the
physical properties of some sources can be further investigated individually.
Detection of a large number of individual BBHs systems will also enable a
statistical study of the formation and evolution of the BBH population
\citep{Rosado15, Kelley18}.

The prospects of detecting GW signals from individual BBHs by current and future
PTAs have been investigated in the previous works (e.g., see \citealt{Sesana09,
Ravi15, Rosado15, Kelley18}). For example, based on the halo merger rate from
the Millennium simulation, together with different BH--host halo relations,
\citet{Sesana09} assemble a cosmic BBH sample and estimate that at least one
resolvable source might be detected by future PTAs involving with SKA
observations.  \citet{Ravi15} find that though beyond the capability of current
PTAs, signals from up to ${\sim}3$ BBHs could be detected by a PTA consisting of
${\sim}100$ pulsars with ${\sim}100{\rm\,ns}$ timing precision for an
observation period of $5\yr$ with a cadence of two months \citep{Ellis12},
which might be achievable with the SKA. Despite this, bursts from coalescing
BBHs are still not likely to be detectable. On the other hand, \citet{Rosado15}
try to answer if the detection of signals from individual sources comes before
that from the stochastic background.
A broad set of simulations covering a large parameter space of the BBH
populations are employed. And appropriate detection statistics are used to
evaluate the detection probability for both kinds of signals. They find that
though the stochastic GWB is more likely to be detected first, the detection
probability for individual sources is not negligible. Recently, based on mock
PTAs, \citet{Kelley18} estimate that the individual sources are likely to be
detected with ${\sim}20\yr$ total observing baselines of PTAs. However, this
detection time may double if several red noises hide in the PTA data.  

As mentioned in Section~\ref{sec:res:gwb}, we use the Monte Carlo method to
generate the cosmic BBH sources; and then we obtain their GW strain spectrum in
the following way.
As done in \citet{Roebber16}, we assume a $T\obs=25\yr$ PTA program, with an
observational cadence of 6 weeks. The width of each frequency bin is
$\Delta f = 1/T\obs \simeq 1.3\nhz$. And the GWs at the frequencies between a
minimum value $f_{\min}\simeq 1.3\nhz$ and a maximum value $f_{\max}\simeq
100\nhz$ is assumed to be explored by the PTA program. Within each frequency bin
$[f_k,f_k+\Delta f]$ ($k=1,2, ...$), the characteristic strain amplitude 
is the incoherent sum of the GW signal of each BBH with GW frequency
$f_i\in [f_k,f_k+\Delta f]$ and given by
\begin{equation}
h\rmc(f_k) = \sqrt{\sum_i h_i^2(f_{{\rm r},i})\min({\cal N}_i,f_i T\obs)},
\label{eq:hc}
\end{equation}
where $h_i(f_{{\rm r},i})$ is the sky- and polarization-averaged
strain amplitude of GW produced by the BBH
system $i$ at the source-rest GW frequency $f_{{\rm r},i}=f_i(1+z_i)$ and given
by the following formula (Thorne 1987)
\begin{equation}
h_i(f\rmr) = 
\sqrt{\frac{32}{5}}\frac{(G\calM)^{5/3}}{c^4R\rmc(z)}
(\pi f\rmr)^{2/3},
\label{eq:hi}
\end{equation}
$R\rmc(z)$ is the comoving distance of the GW source at redshift $z$ and
lower than its luminosity distance by a factor of $1/(1+z)$,
${\cal N}=f\rmr^2/\dot{f\rmr}$,
and
\begin{equation}
\dot{f\rmr} = 
\frac{96\pi^{8/3}G^{5/3}}{5c^5}\calM^{5/3}f\rmr^{11/3}.
\label{eq:dfdt}
\end{equation}
Within the given frequency bin, there are two types of sources, i.e., those
continuous sources satisfying ${\cal N}\ge f T\obs$ and those in-spiraling
sources satisfying ${\cal N}<f T\obs$, and $\min({\cal N}_i,f_i
T\obs)$ is roughly the number of cycles that a binary spends within the given
frequency bin during the observational time $T\obs$.

In Figure~\ref{fig:strn_indv} we show the characteristic strain amplitude
$h\rmc$ of the individual sources of the ten realizations of the cosmic BBH
populations. Note that in our model, given one realization of the cosmic BBH
population, only the loudest source in each frequency bin is picked out as a
candidate of an individual source and shown as a color circle
in Figure~\ref{fig:strn_indv}, where different colors represent
different BBH masses. Those individual sources are excluded when estimating the
backgrounds whose strain spectra are shown by the grey curves in the figure.
However, in principle, multiple loud sources from the same frequency bin could
be extracted simultaneously, since signals may come from different directions of
the sky and could be resolved by PTAs spatially, as well as chromatically
\citep{Ravi12, Babak12, Boyle12}. In this work, we do not consider
this situation, since its probability is small when the detector sensitivity
is not high enough; (e.g., see also \citealt{Kelley18}).
As seen from
Figure~\ref{fig:strn_indv}, the majority of the sources are located at
frequencies $f\ga 10^{-8}\hz$, and some of the loudest individual sources have
the characteristic strain larger than that given by the purple curve following
the $-2/3$ power law. The strongest individual sources are below the current
upper limits from the PTA analysis (see red, green, and blue curves) by one
order of magnitude.

In Figure~\ref{fig:strn_indv}, for reference,  we also illustrate the
characteristic strain amplitudes as a function of frequency of BBHs at redshift
$z=0.1$, 0.2, and 1.0, with BH mass ratio $q\bh=0.1$, and with total masses
$M\bh=10^{10}\msun$, $10^9\msun$, and $10^8\msun$, respectively, which is given by
\begin{equation}
h\rmc(f)=h(f\rmr)\sqrt{\min({\cal{N}},f T\obs)}
\label{eq:hcindividual}
\end{equation}
(see the black lines).
 
Figure~\ref{fig:strn_indv} can provide some prospects for detecting individual
BBH systems in the GW domain, though a quantitative investigation on the
detection time estimation for these sources is beyond the scope of this work,
which involves signal searching strategies, noise models and detector
conditions.  Firstly, the best constraints of current PTAs on continuous
individual sources are at frequencies $f\sim 5\times 10^{-9}$--$10^{-8}\hz$.
However, the majority of the high-signal sources located above the stochastic
background are at larger frequencies.  Secondly, at frequencies $f\ga
10^{-8}\hz$, some of the loudest individual sources have characteristic strain
larger than that given by the $-2/3$ power law. In contrast, the stochastic
background from realizations of the cosmic BBH population has rms characteristic
strain below the power law.  Thirdly, those individual BBHs that have the
highest signals generally have total masses $M\bh\sim10^9\msun$ and low redshift
$z\la 1$, which should be among the first detection of the individual sources.
The average numbers of the individual sources with signals higher
than the background (grey curve) for one realization of the BBH population
shown in the figure are $\sim 1.0$ 
at $h\rmc \ga 10^{-15}$ and $\sim 8.7$ at $h\rmc
\ga 3\times 10^{-16}$, potentially achievable by the PTA experiments involving
next-generation radio telescopes, such as SKA and ngVLA. 

\subsection{Contribution to the LISA detection bands}
\label{sec:res:lisa}

In the above subsection, we investigate the GW radiation from the cosmic
population of the BBHs at the frequency band of PTAs, and in this subsection we
extend the investigation to higher GW frequencies ($\sim 0.1\mhz$--$0.1\hz$;
LISA/Taiji/Tianqin detection bands) and study the prospects of detecting GW
signals from less massive BHs.

By using one Monte Carlo realization of the BBH populations generated in
Section~\ref{sec:res:gwb} and assuming a three-year or ten-year observational
time of the LISA mission, we plot the characteristic strain of the BBH systems
covering the whole mission period in Figure~\ref{fig:strn_lisa} (shown as the
color lines, with different colors representing different total BBH masses, mass
ratios, and redshifts in the top, middle, and bottom panels, respectively), and
their starting values at the beginning of the observation are shown as the
points.
As seen from the figure, some of the sources have significant frequency shifts
during the observation periods. 

At frequencies $f\ga 10^{-4}\hz$, the number of the sources is not significant
to form a background.  However, as a reference, in Figure~\ref{fig:strn_lisa} we
also plot the strain spectrum of the possible stochastic background estimated
from the integration of Equation~(\ref{eq:hc2}) as the red dashed line,
with the integration limits $M\bh>10^5\msun$, $q\bh>10^{-3}$.

For each generated source within the LISA band, we can calculate the
signal-to-noise ratio (SNR) of the source for a given observational duration by
using Equation (19) of
\citet{Moore15}, i.e.,
\begin{equation}
{\rm SNR} = \sqrt{\int_{-\infty}^{\infty} d\ln\,f
\left[\frac{h_c(f)}{h_n(f)}\right]^2},
\label{eq:snr}
\end{equation}
where $h_n(f)$ corresponds to the LISA sensitivity curve shown in
Figure~\ref{fig:strn_lisa}. By simulating ten Monte Carlo realizations of the
BBH samples within the LISA band, the average total number of the events with
SNR above 8 is 2.8 for a 3-year mission and 9.1 for
a 10-year mission, and thus the detection rate for LISA is estimated to be
${\sim}0.9\pyr$ (obtained by using the
$M\bh{-}M\bulge$ relation of \citealt{Kormendy13}). 

The above detection rate of BBHs by LISA is consistent with our previous results
shown in Section~\ref{sec:res:cols}.  For the BBH coalescence rates shown
in Figure~\ref{fig:cols_rate}, the integrated coalescence rate of the BBHs over
the same ranges of total masses $M\bh>10^5\msun$ and mass ratios $q\bh>10^{-3}$
is ${\sim}0.63\pyr$, which is the median value of the results obtained by
adopting all the BH--host galaxy relations listed in Table~\ref{tab:scal}. If
we adopt the $M\bh{-}M\bulge$ relation of \citet{Kormendy13} (see the bottom
left panel of Figure~\ref{fig:cols_rate_scale}), the integrated coalescence
rate over the above same integration limits is $\sim 0.86\pyr$, which is close
to the median value.
The BBH coalescence rate is consistent with the estimated LISA detection rate,
since most of these sources have signals at their final chirping stages well
above the LISA detection threshold.
As seen from Figure~\ref{fig:cols_rate_scale}, the possible range of the BBH
coalescence rate can span two orders of magnitude, due
to the use of the different BH--host galaxy relations (for example, at redshift
$z\sim 3$, the standard deviation of the logarithms of the cumulative BBH
coalescence rates around the median shown in
Figure~\ref{fig:cols_rate_scale} is $^{+0.46}_{-0.68}$dex for
the case of $\log(M\bh/\msun)=5$ and $\log\,q\bh=0$ and
$^{+0.54}_{-0.46}$dex for the
case of $\log(M\bh/\msun)=8$ and $\log\,q\bh=-2$.). If taking into account this
scatter, the possible range of the LISA detection rate should be 
$\sim 0.1{-}10\pyr$, with a median close to $0.9\pyr$.

The events contributing to the LISA detection in Figure~\ref{fig:strn_lisa} are
mainly from those sources with $M\bh \sim 10^5$--$10^7\msun$, $q\bh\ga 0.01$,
and $z\la 3$.
The modeling of the low-mass BBH population here only represents an
extrapolation of the BBH population obtained partly based on the BH--host
galaxy relations determined from the high-mass BH sample, and MBHs with smaller
masses, e.g., $M\bh\la 10^5\msun$, is not considered here. 
The low-mass BBH population from the MBH seeds formed at high redshifts might
provide an additional contribution to the detection. For example, based on the
cosmological hydrodynamical simulations from the EAGLE project
\citep{Schaye15}, \citet{Salcido16} predict a detection rate of ${\sim 2}$ per
year for the eLISA, and Figures~7-8 in \citet{Salcido16} find two peaks of the
distribution of sources in the strain-merging frequency space, one of which has
a location consistent with ours, while the other of which is caused by BBHs
with smaller masses.
A comparison of the previous results obtained from semi-analytical models
(which predict higher rates by about one order of magnitude) and those from
hydrodynamical simulations can also be found in \citet{Katz19} (see Table 2
therein).

\section{Discussion}
\label{sec:discussion}

Below we discuss and summarize the effects on the estimation of the stochastic
GWB caused by the different factors. \begin{itemize}
\item GSMFs: we also calculate the strain amplitudes by adopting the GSMFs of
\citet{Torrey15}, \citet{Tomczak14}, \citet{Ilbert13}, and \citet{Muzzin13},
and find that the ratios of these results to
the result obtained by adopting the GSMF of \citet{Behroozi19} are
$2.06$, $0.92$, $1.04$, and $1.08$, respectively. The
difference in the estimated strain amplitude caused by the difference in the
GSMFs, except for that from \citet{Torrey15}, is negligible.
\item Galaxy merger rates: the uncertainties in the logarithm of the GW strain
can be contributed roughly by half of the uncertainties in the logarithm
of galaxy merger timescales and in the logarithm of 
close-pair fractions given by observations.
The difference in the galaxy merger timescales of
\citet{Kitzbichler08} and \citet{Lotz11} is a factor of $\sim 3$ (see
Eqs.~\ref{eq:tKitzbichler08} and \ref{eq:Xu12}), and the logarithm of the
strain amplitude
estimated by adopting those of \citet{Kitzbichler08} can be $\sim\lg(\sqrt{3})\simeq$0.24~dex
lower than that obtained by adopting those of \citet{Lotz11}.  Observational
uncertainty in the close-pair fractions give an additional scatter of
$\sim 0.3$~dex in the logarithm of the strain.
\item The different choices of BH--host galaxy relations contribute to a
difference as large as $\sim$1~dex in the logarithm of the estimated strain
amplitude, as shown in Figure~\ref{fig:strn_indv}.
\item The difference in the logarithm of the estimated strain amplitude caused
by the different observational galaxy shape distributions of \citet{Padilla08,
Rodriguez13,Weijmans14} is $\la$0.15~dex.

\item In this work, the effect of the gas is not considered in the BBH evolution.
The gas effect (e.g., \citealt{Haimanetal09,Mayeretal07}) is negligible in the
estimate of the GWB strain at the PTA band, as the GWB at the PTA band is
mainly contributed by BBH mergers at $z<2$ and in high-mass galaxies, which is
generally gas-poor.  In addition, the triaxiality of galaxy shapes has been
shown to be effective enough to decrease the BBH coalescence timescales
significantly.
For smaller BBHs (e.g., $M_{\rm BH} \lesssim 10^7M_\odot$), they have
relatively high probabilities to be hosted in gas-rich mergers. However, more
than half of those BBHs have peak evolution timescales shorter than a few Gyrs,
even without considering an acceleration of the merger processes caused by gas
(see Figure~\ref{fig:trac_tmax}). It is not clear, yet, on how much gas can
sink into the vicinity of BBHs in reality and how significant the gas effect
can be on the merger timescales.  If a significant amount of gas can sink into
the BBH vicinity and accelerate its merging process significantly, the
estimation of the surviving fractions and the coalescence rates of those small
BBHs deserve further improvements in the future.
\end{itemize}

There are some other estimates of the stochastic GWB from the cosmic population
of BBHs in the literature (e.g., \citealt{Kelley17, Roebber16, Sesana16,
Ravi15, Kulier15, McWilliams14, Ravi14}). Below we list some recent estimates
of the strain amplitude in the literature, together with some brief
descriptions of the astrophysical ingredients implemented in each of those
models. The differences in those estimates come from the different uses of the
ingredients (such as GSMFs, galaxy merger rates, BBH populations, BBH
evolution models, and different numerical simulation results) in the model.
\begin{itemize}
\item \citet{Kelley17} obtain the BBH population based on the Illustris
Simulation and implement different binary hardening mechanisms at different BBH
separation scales, including dynamical friction on galaxy scales, loss-cone
stellar scatterings within parsec scales, viscous drag from a circumbinary
disk, and GW emission at smaller scales. In the loss-cone stellar scattering
stage, a ``refilling fraction'' coefficient describing the loss-cone refilling
rate compared to the case of full loss-cone refilling is introduced. Within
their model implementation, the BBH lifetimes are generally multiple $\gyr$. As
a consequence, only a limited fraction of BBHs could finish final coalescence
by $z=0$, e.g., $45\%$ for BBHs with total masses above $10^8\msun$ and mass
ratios above $0.2$.
(Our detailed considerations of the dynamical evolution of BBHs in this work
find that according to Fig.~\ref{fig:trac_tau0}, for galaxy mergers occurring at
$z=0$ and their BBHs with $M\bh=10^8\msun$ and $q\bh=0.01$, the fraction of the
BBHs with coalescence timescales shorter than the Hubble timescale is
${\sim}87\%$ if the shape distribution from \citet{Padilla08} is adopted, and
${\sim}75\%$ if the shape distribution from \citet{Weijmans14} is adopted,
according to Fig.~\ref{fig:trac_tau0}. For galaxy mergers at $z=2$ and their
corresponding BBHs, the fraction is $\sim 60\%$ if the shape distribution from
\citet{Padilla08} is adopted, and ${\sim}38\%$ if the shape distribution from
\citet{Weijmans14} is adopted.) Their model taking into account the above
hardening mechanisms predicts a strain amplitude of $\ayr=3.7\times 10^{-16}$.
\item \citet{Roebber16} also studies the GWB produced by the cosmic population
of BBHs, especially on the high-frequency end strain spectral features
contributed by a limited number of massive BBHs systems.  Their halo merger
sample is based on two recent dark matter simulations, i.e., the Dark Sky
simulation \citep{Warren13} and the MultiDark simulation \citep{Riebe11,
Prada12}. The halos are then filled with galaxies whose masses are determined
based on the galaxy mass--halo mass relation of \citet{Behroozi13}. The GSMFs
for galaxies in the two dark matter simulations are consistent with our model
within the mass range $[2\times 10^{10},\,2\times 10^{11}] \msun$, beyond which
their GSMFs tend to be lower than the one that we adopt.
The MBHs are then assigned to galaxies according to
the $M\bh-M\bulge$ relation of \citet{Kormendy13}.  The BBH coalescence is
assumed to occur instantaneously after halo mergers.  Their estimated strain
amplitude is $\ayr\sim$6--7$\times 10^{-16}$.

\item In the estimate of \citet{Sesana16}, four sets of different GSMFs are
adopted with equal weights, including those of \citet{Bernardi13},
\citet{Ilbert13}, \citet{Muzzin13}, and \citet{Tomczak14}; and four different
sets of close-pair fractions are used, including those of \citet{Bundy09},
\citet{de-Ravel09}, \citet{Lopez12}, and \citet{Xu12}. To convert the close pair
fractions into fractional merger rates of galaxies, two sets of observability
timescale of galaxy mergers are adopted, including those of
\citet{Kitzbichler08} and \citet{Lotz11}.
Three sets of the BH--host galaxy relations are used in their model, including
the $M\bh-M\bulge$ relation and the $M\bh-\sigma$ relation of
\citet{Kormendy13}, and the $M\bh-M\bulge-\sigma$ relation of \citet{Shankar16}.
The BBH coalescence is assumed to occur instantaneously after galaxy mergers.
With the $M\bh-M\bulge$ and $M\bh-\sigma$ relations of \citet{Kormendy13}, their
estimated median strain amplitude is $\ayr{\sim}1.3\times 10^{-15}$, and with
the BH--host galaxy relation of \citet{Shankar16}, the median value drops by a
factor of ${\sim}3$, decreasing to $4\times 10^{-16}$.
Our results obtained by adopting the same $M\bh-M\bulge$ relation of Kormendy
\& Ho (2013) is $\ayr{\sim}5.0\times 10^{-16}$ (see
Table~\ref{tab:scal}) and lower than the above $\ayr{\sim}1.3\times 10^{-15}$
estimated by \citet{Sesana16}; the difference may be partly from the
difference in the used galaxy merger rates and partly from the delayed
BBH coalescence after galaxy merger.

\item \citet{Ravi15} adopts the GSMF of \citet{Muzzin13} in their fiducial
model.  The galaxy merger rate in their model is also based on the
observational galaxy close-pair counts, i.e., those of \citet{Conselice09},
\citet{Xu12}, and \citet{Robotham14}.  Regarding the observability timescale of
galaxy mergers, \citet{Ravi15} adopt the normalization of \citet{Lotz11} plus
the redshift dependence of \citet{Kitzbichler08}.  The resulting 3 sets of
galaxy fractional merger rates adopted by \citet{Ravi15} are a factor of
${\sim} 3$ above that in our model.  To convert the bulge masses to the MBH
masses, the $M\bh-M\bulge$ relations of \citet{Kormendy13} and \citet{Scott13}
are adopted by \citet{Ravi15}.  The BBH coalescence is assumed to occur
instantaneously after galaxy mergers.  Their estimated strain amplitude is
$\ayr{\sim}1.3\times 10^{-15}$ for their fiducial model in which the
$M\bh-M\bulge$ relation of \citet{Kormendy13} is chosen.
The estimated strain amplitude is already quite close to the current upper
limits on the stochastic GWB from the NANOGrav and the IPTA.
\end{itemize}

\section{Conclusions}
\label{sec:conclusion}

\begin{figure*} \centering
\includegraphics[width=0.95\textwidth]{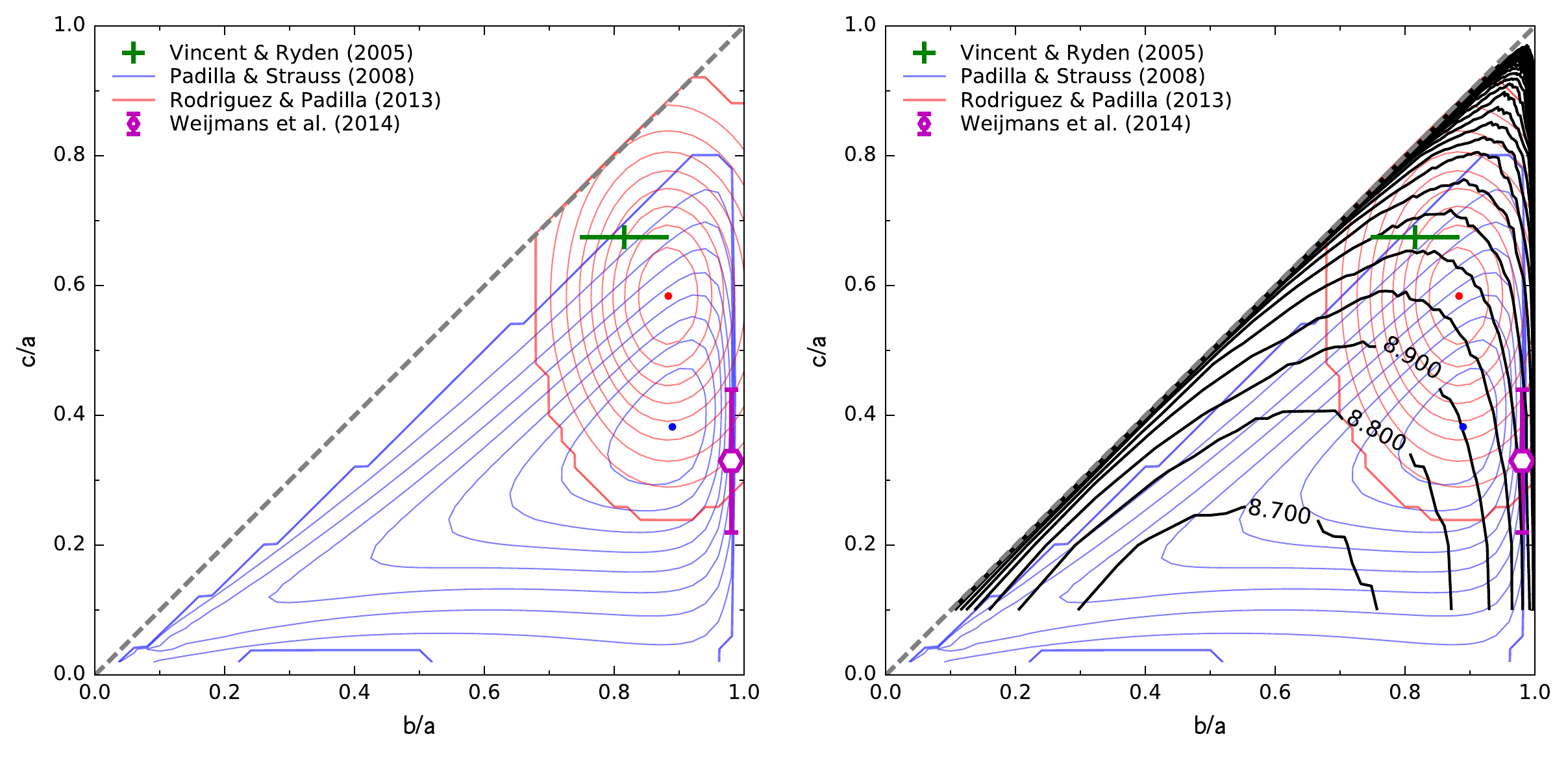} \caption{The left
panel shows the observational shape distributions of early-type galaxies, based
on the different works labeled by the texts.
The blue and red contours show the probability distributions obtained in
\citet{Padilla08} and \citet{Rodriguez13}, respectively, and the contour levels
decrease from the center to the outer parts by an interval of 10\% of
the maximum of the corresponding probability distribution.
The green plus symbol marks the possible ranges of the mean
medium-to-major and minor-to-major axis ratios of the early-type galaxies in
the SDSS DR3 data set, obtained in \citep{Vincent05}, and the ranges
are converted from their constraints on the triaxiality and
the minor-to-major axis ratio of the bright galaxies sample with a
de Vaucouleurs profile therein.
For \citet{Weijmans14}, only the best fitting results for their fast-rotator
galaxies are shown, whose shape distribution is consistent with an almost
oblate sample; and the magenta point and the error bar show the mean value and
the standard deviation of their $c/a$ ratios.  The right panel shows the
dependence of the peak timescale of the BBH evolution on galaxy shapes. The
thick black curves represent the contours of $\log(t\peak/{\rm yr})$ of a BBH
system (with $q\bh=1$) after becoming bound in the example galaxy shown in
Figure~\ref{fig:evlv_trac}.  The values labeled for the corresponding black
curves are
$\log(t\peak/{\rm yr})$ calculated from the BBH evolution model presented in
Section~\ref{sec:dynamics}, with an interval level of 0.1~dex.  This figure
shows that even a mild triaxiality can decrease $t\peak$ significantly (see
also Fig.~\ref{fig:tsc_contour} below).  See details in
Sections~\ref{subsubsec:galtri} and \ref{sec:res:trac}.
}
\label{fig:tsc_contour} \end{figure*}

\begin{figure*}
\centering
\includegraphics[width=0.9\textwidth]{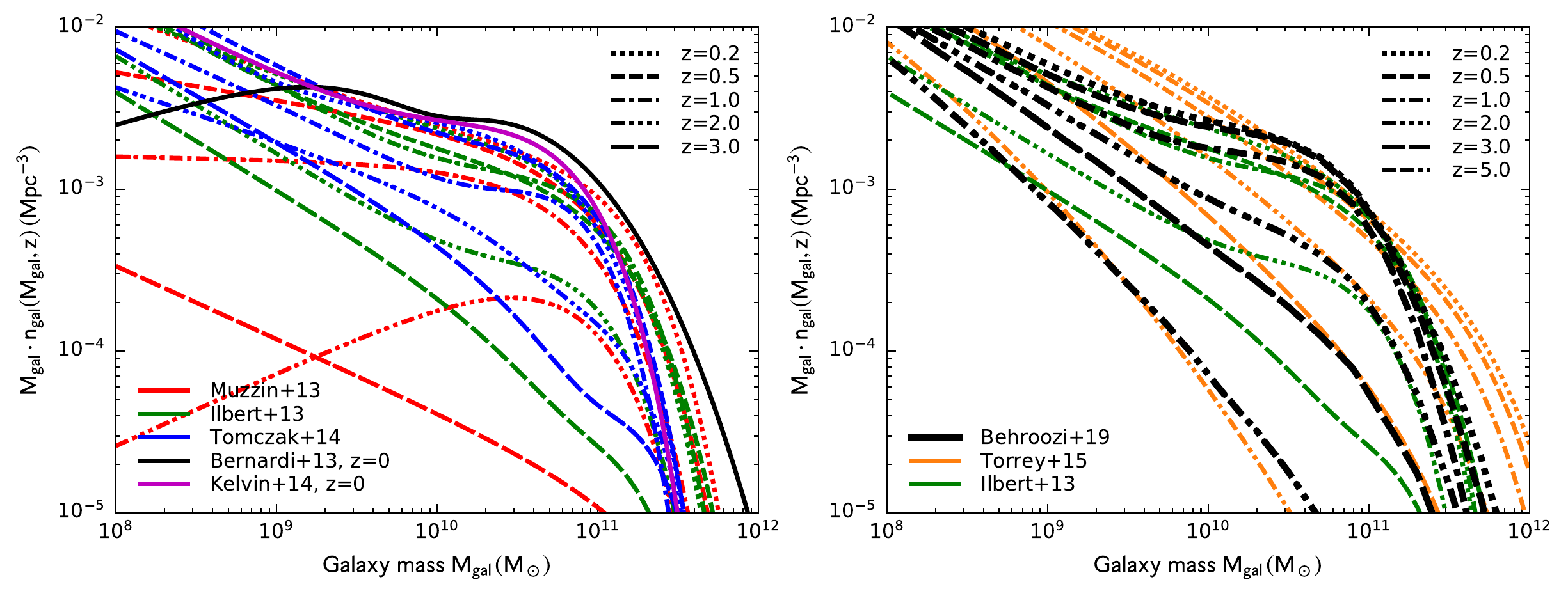}
\caption{Galaxy stellar mass functions $n\gal(M\gal,z)$ obtained from both
observations and hydrodynamical simulations (see
Section~\ref{subsubsec:galaxymass}). The vertical axis represents $M\gal\cdot
n\gal(M\gal,z)$. The left panels show the observational GSMFs. The black and
the magenta curves correspond to the local ($z=0$) GSMFs obtained from
\citet{Bernardi13} and \citet{Kelvin14}, respectively. The red curves
show the GSMFs from \citet{Muzzin13}; and
the different line styles represent the different redshifts ($z=0.2,0.5,1.0,2.0,3.0$), as labeled by the texts.
The green and the blue curves show the
results from \citet{Ilbert13} and \citet{Tomczak14}, respectively.
The right panel shows the GSMFs used in this work, which is from
\citet{Behroozi19} (black).  The GSMFs from the observational results in
\citet{Ilbert13} (green) and the Illustris simulation results in
\citet{Torrey15} (orange)
are also shown in the panel for comparison. The differences in the stochastic
GWB strain amplitude caused by using the different GSMFs are discussed in
Section~\ref{sec:discussion}.
}
\label{fig:mass_func} \label{fig:gsmf_alte} \end{figure*}

\begin{figure} \centering
\includegraphics[width=0.7\textwidth]{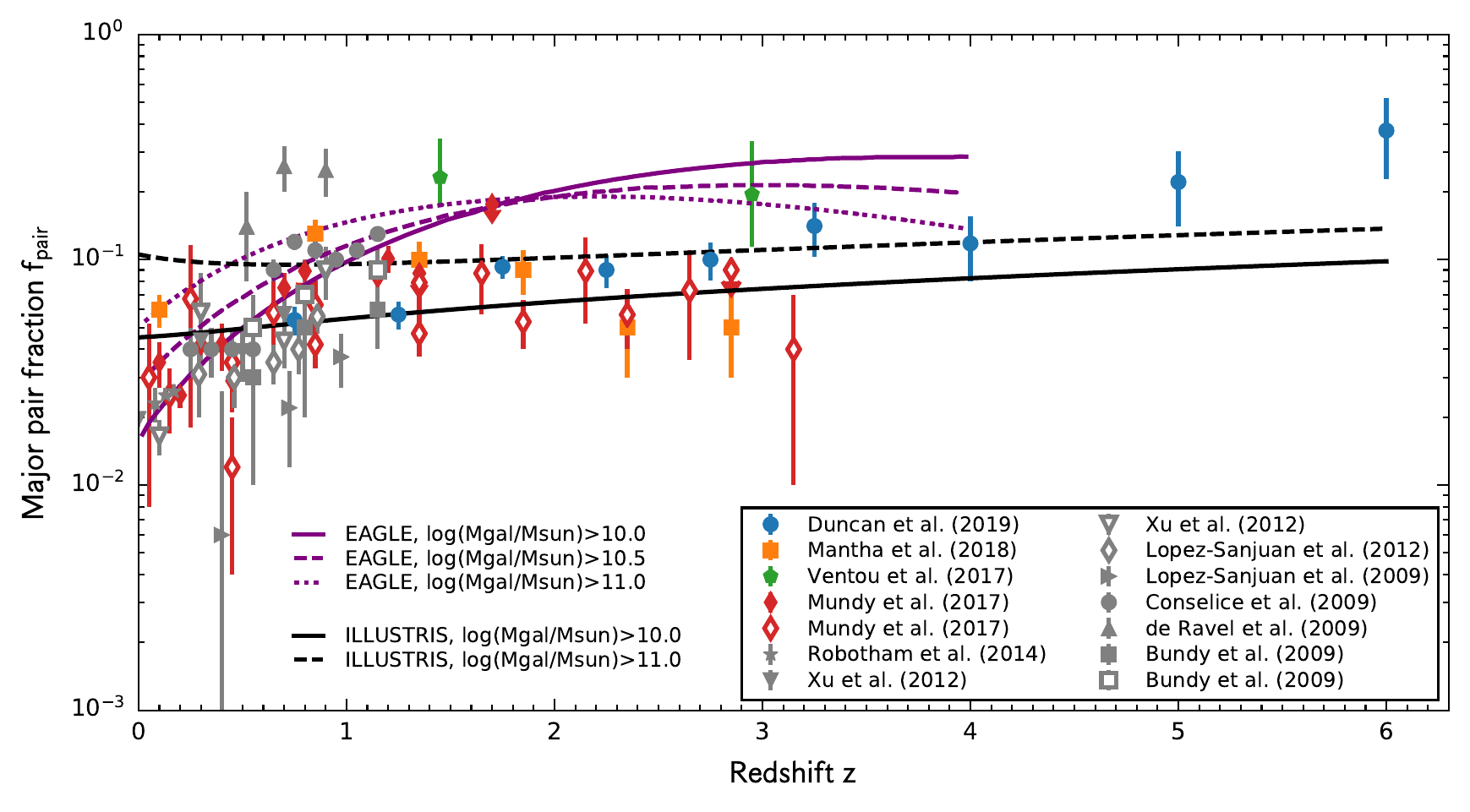}
\includegraphics[width=0.7\textwidth]{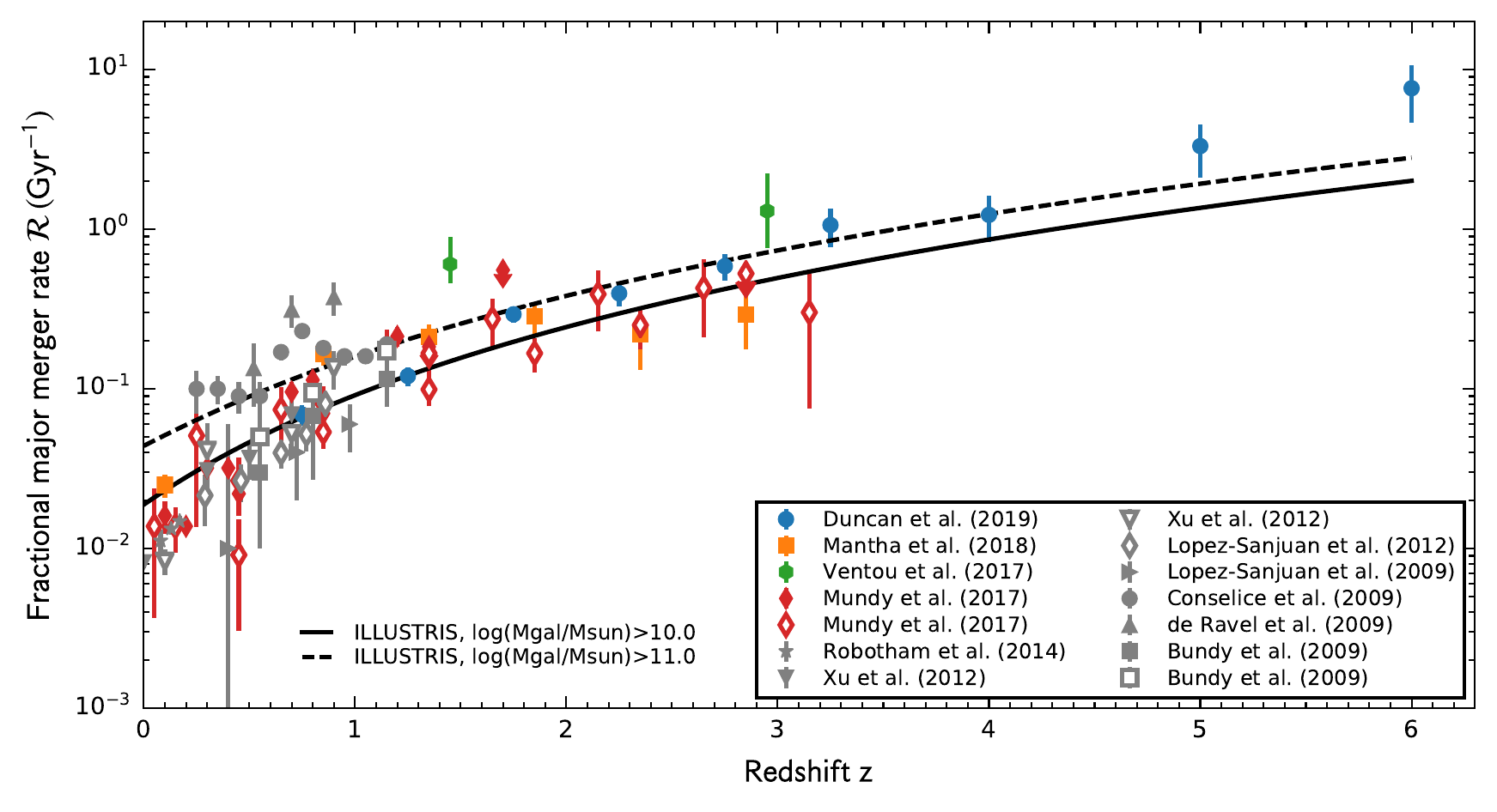} \caption{Galaxy
major pair fractions $f\pair$ (upper panel) and fractional major merger rates
$\calR(>M\gal,z)$ (lower panel) as a function of redshift $z$ (see
Eqs.~\ref{eq:calR} and \ref{eq:calRinte}), which are converted from both
hydrodynamical simulations and observational results.
In the upper panel, the grey symbols show the close pair fraction of galaxies
obtained from the observational samples at $z\la 1$, and the other color
symbols show some recent observational results that extend to high redshift
$z\ga 3$, as labeled by their references.  Those observational results are
based on close pair counts or the morphological disturbance. The filled and the
empty symbols correspond to the samples of galaxies with $M\gal\ga
10^{10}\msun$ and $M\gal\ga 10^{11}\msun$, respectively.
Note that besides the difference in total masses, the different samples have
different cut criteria on the maximum projected separation and velocity
difference between two galaxy members in a pair.
The black curves show the close pair fraction of galaxies with total mass
greater than $10^{10}\msun$ (solid) and $10^{11}\msun$ (dashed), obtained by
the product of the galaxy fractional major merger rate from
\citet{Rodriguez-Gomez15} and the average observability timescale of galaxy
mergers from \citet{Snyder17} (both based on the results of the Illustris
Project described in 
\citealt{Vogelsberger14}). Here $C\mrg=1.0$ is assumed in Equation
(\ref{eq:calR}), and only the major pairs with $q\gal\ge q_{\rm major}=1/4$
(see Eq.~\ref{eq:calRinte}) are considered.
The purple curves correspond to the close pair fraction of the galaxies with
total mass greater than $10^{10}\msun$ (solid), $10^{10.5}\msun$ (dashed) and
$10^{11}\msun$ (dotted), respectively, as shown in \citet{Qu17}, which is based
on the EAGLE simulation \citep{Schaye15}. 
In the lower panel, the black curves show the fractional galaxy merger rates
obtained from the numerical simulations in \citet{Rodriguez-Gomez15}, and the
observational symbols are converted from the corresponding symbols of
close-pair fractions in the upper panel and the average observability timescale
of galaxy mergers obtained from \citet{Snyder17} (see Eq.~\ref{eq:calR}).
This figure illustrates the uncertainty in the close-pair fraction, the
fractional merger rate of galaxies and their dependence on the redshift, from
both observations and hydrodynamical simulations.  This figure shows that the
fractional galaxy merger rates obtained from \citet{Rodriguez-Gomez15} is close
to the observational one converted by using Equation (\ref{eq:Snyder17}) (which
evolves with redshift strongly).  } \label{fig:merg_rate} \end{figure}

\begin{figure}
\centering
\includegraphics[width=0.5\textwidth]{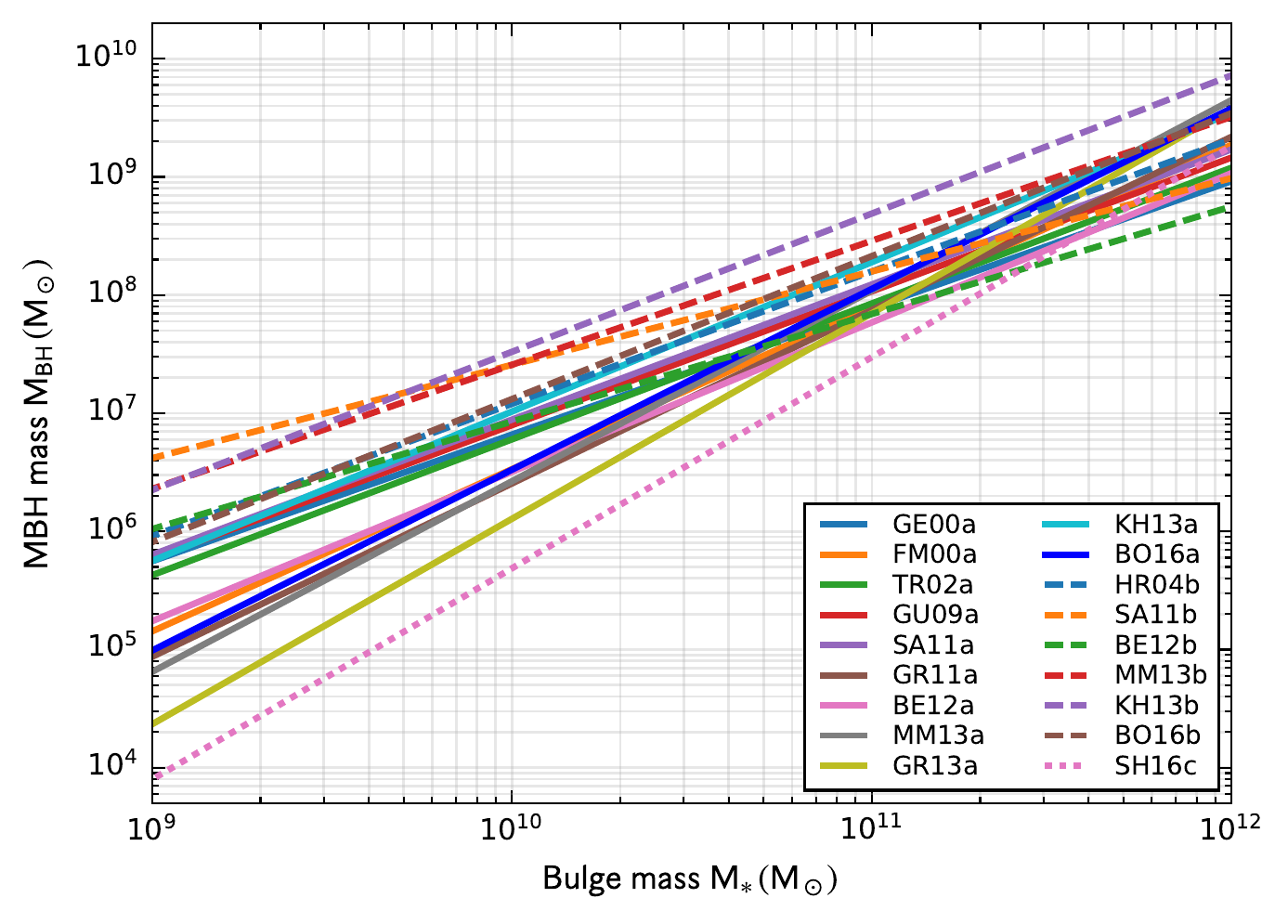}
\caption{Different BH--host galaxy relations used in this study, expressed
through the relations between the MBH masses and the masses of the spheroidal
components of the host galaxies.  The solid, the dashed, and the dotted lines
represent the relations converted from the $M\bh-\sigma$, the
$M\bh-M\bulge$, and the $M\bh-M\bulge-\sigma$ relations shown by Equation
(\ref{eq:scal}), respectively. Different lines represent the different works as
labeled by the texts (see also Table~\ref{tab:scal}).  The $M\bh-M\bulge$ relations
shown by Equation (\ref{eq:scal}) are drawn in this figure directly, and the
$M\bh-\sigma$ and the $M\bh-M\bulge-\sigma$ relations are converted to the
$M\bh-M\bulge$ relation by applying the relation between $\sigma$ and $M\bulge$
obtained in \citet{Gallazzi06}. See Section~\ref{sec:data:msigma}.  }
\label{fig:scal_rela} \end{figure}

\begin{figure*}
\centering
\includegraphics[width=0.7\textwidth]{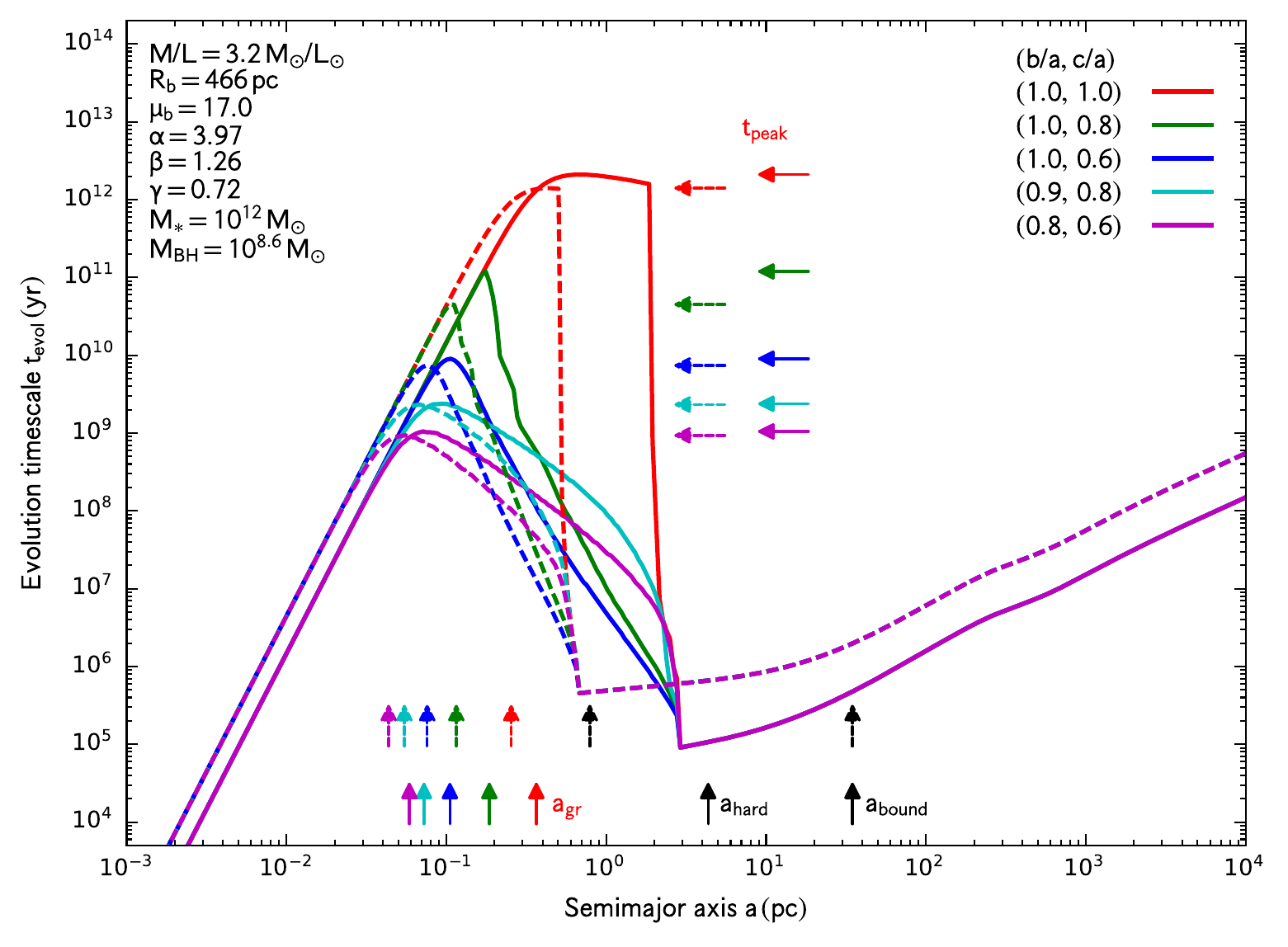}
\caption{Examples of the evolution tracks of BBHs in a galaxy, with different
assumed BBH mass ratios and merger remnant shapes. The radial surface
brightness of the spheroidal component of this galaxy is described by the Nuker-law profile with parameters
$R\rmb=466\pc$, $\mu\rmb=17.0$, $\alpha=3.97$, $\beta=1.26$, $\gamma=0.72$, and
the mass-to-light ratio in the $V$ band $M/L=3.2\,\msun/\lsun$. The total
stellar mass of the merger remnant is $10^{12}\msun$ and the total mass of the
BBH is $10^{8.6}\msun$. The solid and the dashed curves correspond to the cases
with the assumed BBH mass ratio $q\bh=1$ and 0.1, respectively.  The different
colors represent the assumed different shapes of the galaxy merger remnant,
i.e., red for $(b/a,c/a)=(1.0,1.0)$, green for $(1.0,0.8)$, blue for
$(1.0,0.6)$, cyan for $(0.9,0.8)$, and magenta for $(0.8,0.6)$, where $b/a$ and
$c/a$ correspond to the medium-to-major and the minor-to-major axis ratios of
the mass density distribution.  The vertical arrows at the bottom of the figure
mark the semimajor axes at which the BBH enters the gravitational radiation
stage for the corresponding evolution curves with the same colors, and the
horizontal arrows mark the peak evolution timescales after the BBH becomes hard
for the corresponding evolution curves with the same colors. See
Section~\ref{sec:res:trac}.
}
\label{fig:evlv_trac}
\end{figure*}

\begin{figure*}
\centering
\includegraphics[width=0.8\textwidth]{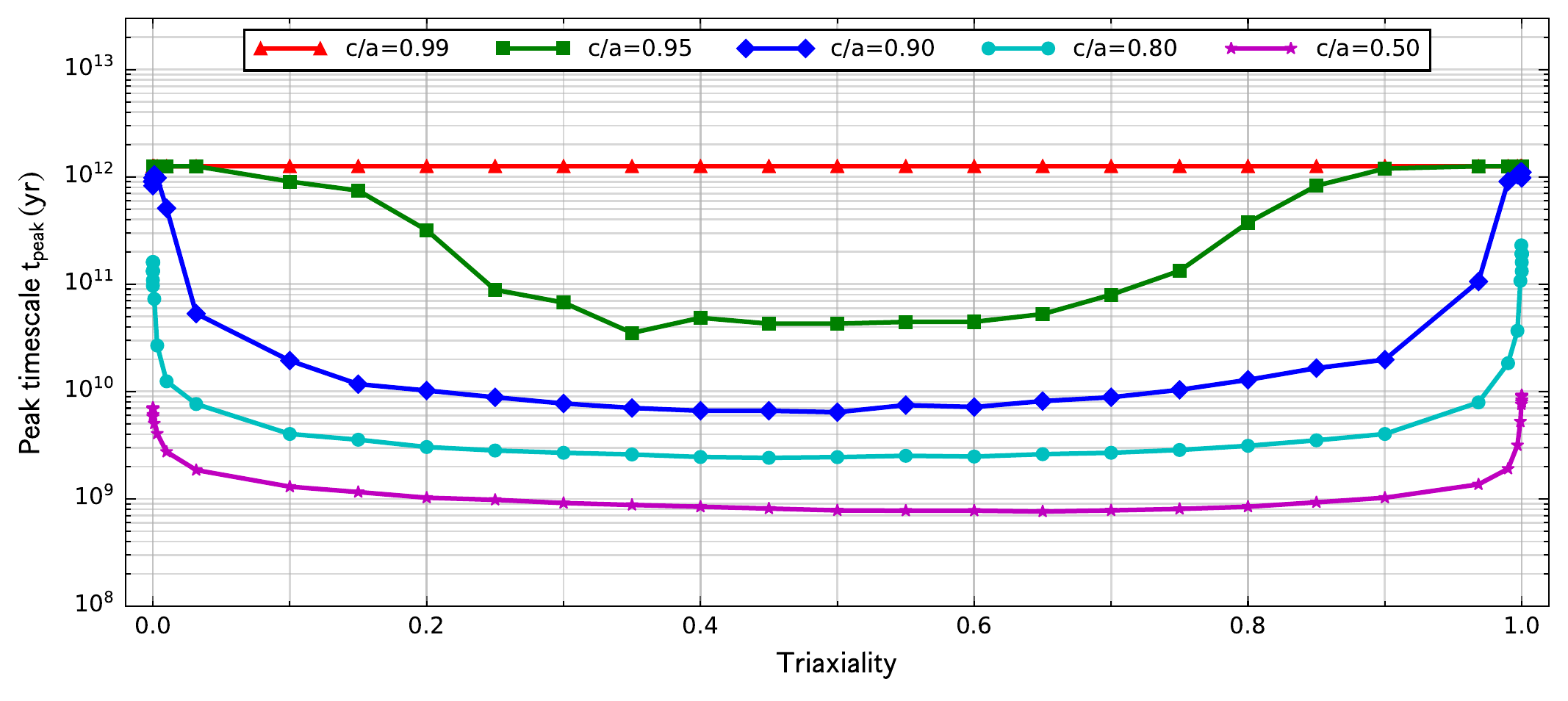}
\caption{
The dependence of the peak timescale $t\peak$
on the triaxiality parameter $T=(a^2-b^2)/(a^2-c^2)$
for the same BBH system shown in Figure~\ref{fig:tsc_contour},
Different curves show the results obtained with the different sets of the
minor-to-major axis ratios $c/a$ of
the mass density distribution of the host merger remnant. 
The sharp decrease of $t\peak$ at $T\sim 0$ and $1$ is consistent with those
shown in Figure~\ref{fig:tsc_contour}.
}
\label{fig:tpeak_rely}
\end{figure*}

\begin{figure*}
\centering
\includegraphics[width=0.8\textwidth]{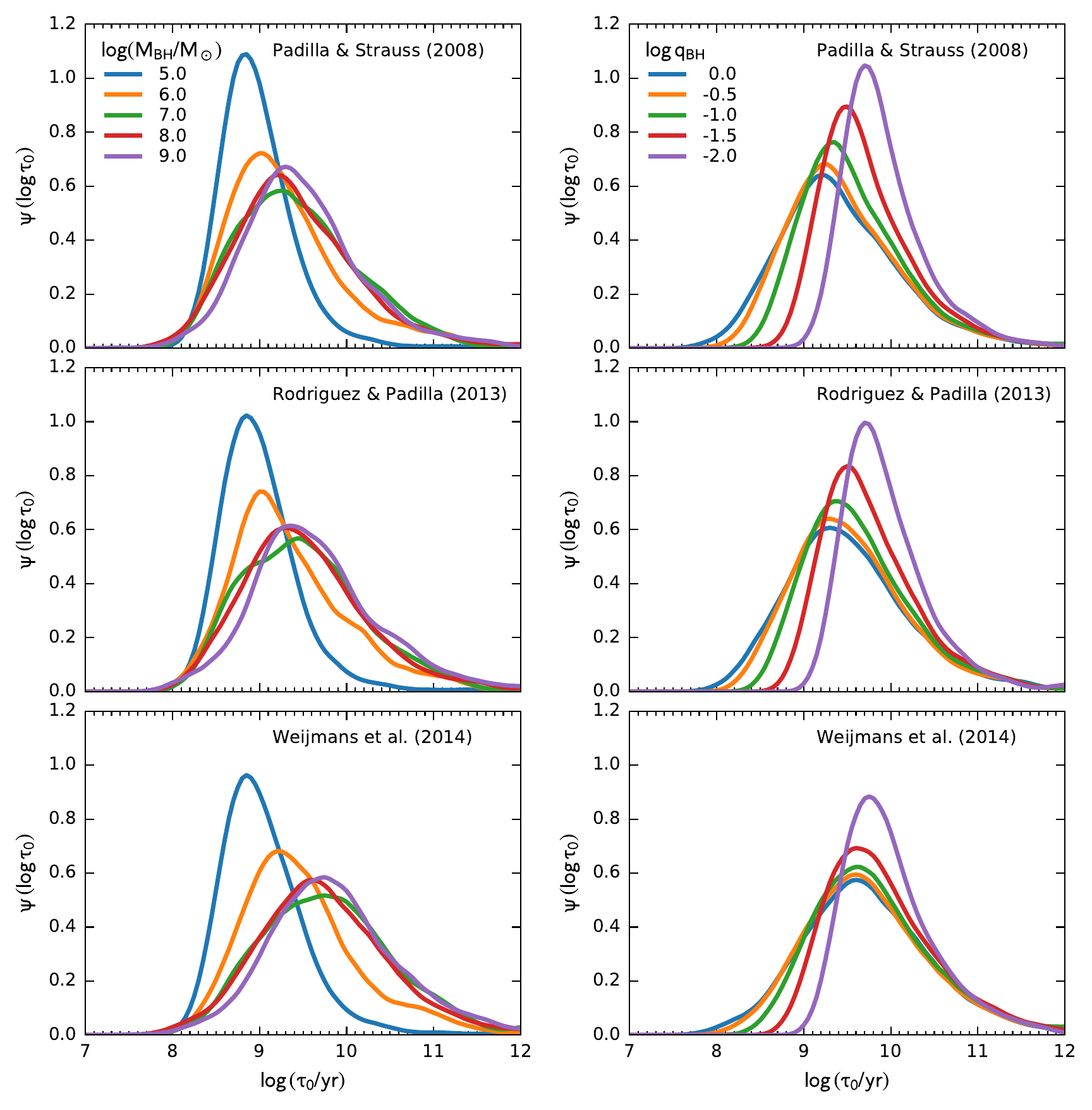}
\caption{
Probability distributions of the logarithm of the coalescence timescale
$\log\tau_0$ (cf. Equation~\ref{eq:pdfx}) of a population of BBHs whose host
galaxies merge at redshift $z'=0$,
$\psi_{\log\tau_0}(\log\tau_0|M\bh,q\bh,z'=0)$, and their dependence on the
properties of the BBHs and their host galaxies.  The left panels show the
distributions of $\log(\tau_0/yr)$ for BBHs with mass ratio $q\bh=1.0$ and with
different total masses $M\bh$ (shown by different colors), which suggest
that the BBHs with high total masses (e.g., $\sim
10^9$-$10^{10}\msun$) have a relatively large fraction of long $\tau_0$ (e.g.,
longer than the Hubble timescale $\sim 10^{10}\yr$.
The right panel shows the distributions of $\log(\tau_0/{\rm yr})$ for BBHs with
total mass $M\bh=10^8\msun$ and with different mass ratios $q\bh$, which
suggest that the BBHs with smaller BH mass ratios (e.g., $q\bh=0.01$) have a
relatively larger fraction of long $\tau_0$. The
different rows represent the results obtained with the shape distributions of
early-type galaxies in the different works, i.e., \citet{Weijmans14},
\citet{Rodriguez13}, and \citet{Padilla08} from top to bottom panels,
respectively. The distributions obtained with triaxial shape distributions
in the top and the middle panels are close to each other, and the distributions
obtained with nearly oblate shape distribution in the bottom panel
are relatively broadened towards longer $\tau_0$.
See Section~\ref{sec:res:prop}. 
} \label{fig:trac_tau0}
\end{figure*}

\begin{figure*}
\centering
\includegraphics[width=0.8\textwidth]{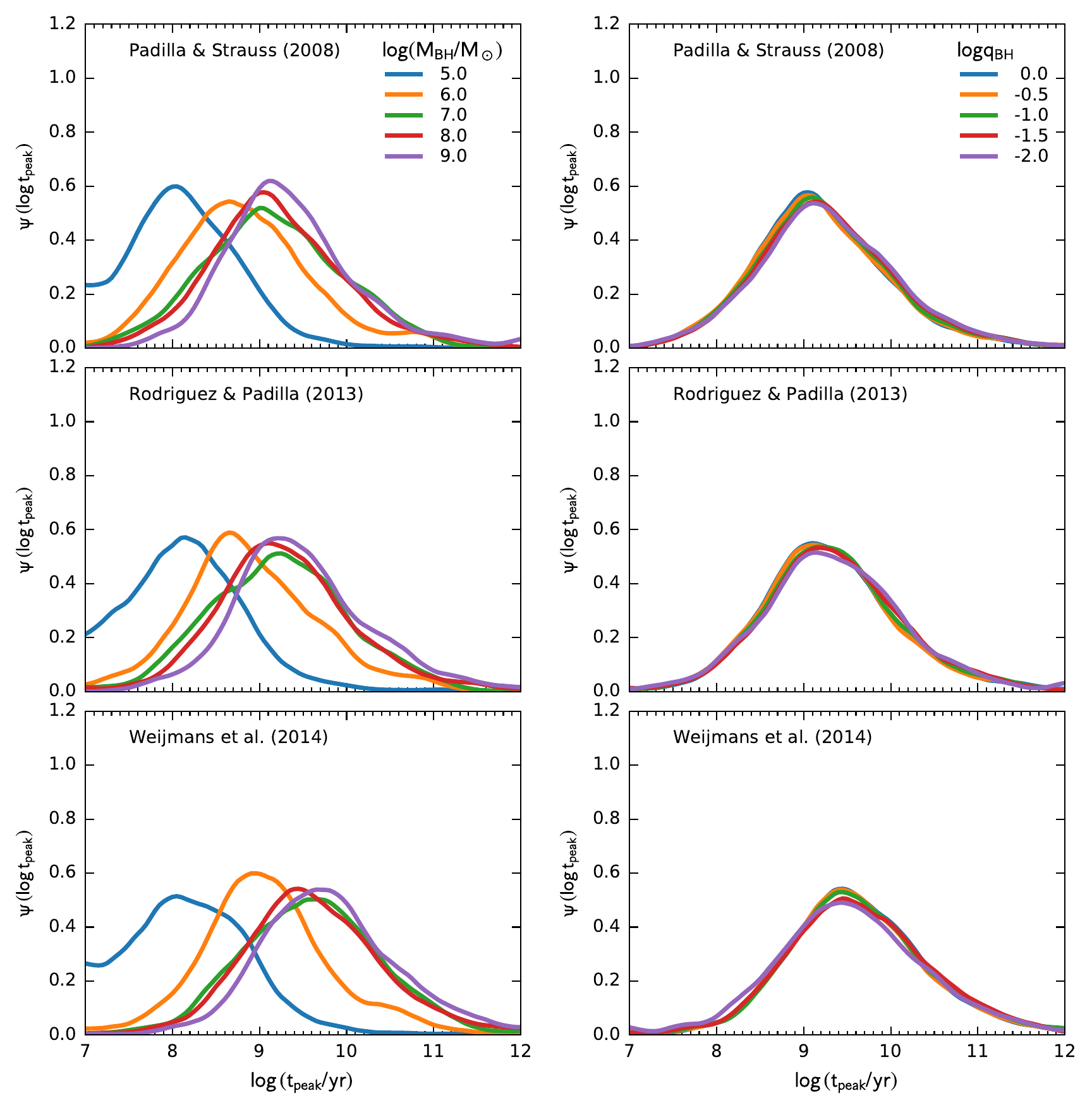}
\caption{
Probability distributions of the logarithm of the timescale 
$\log t\peak$ (cf.\ Equation~\ref{eq:pdfx}) for the same BBH population shown in
Figure~\ref{fig:trac_tau0},
$\psi_{\log t\peak}(\log t\peak|M\bh,q\bh,z'=0)$, and their dependence on the
properties of the BBHs and their host galaxies. The curves with different
colors and
the texts have meanings similar to those in Figure~\ref{fig:trac_tau0}.
Compared to those in Figure~\ref{fig:trac_tau0}, the distributions are less
sensitive to the BBH mass ratios.
}
\label{fig:trac_tmax}
\end{figure*}

\begin{figure*}
\centering
\includegraphics[width=0.8\textwidth]{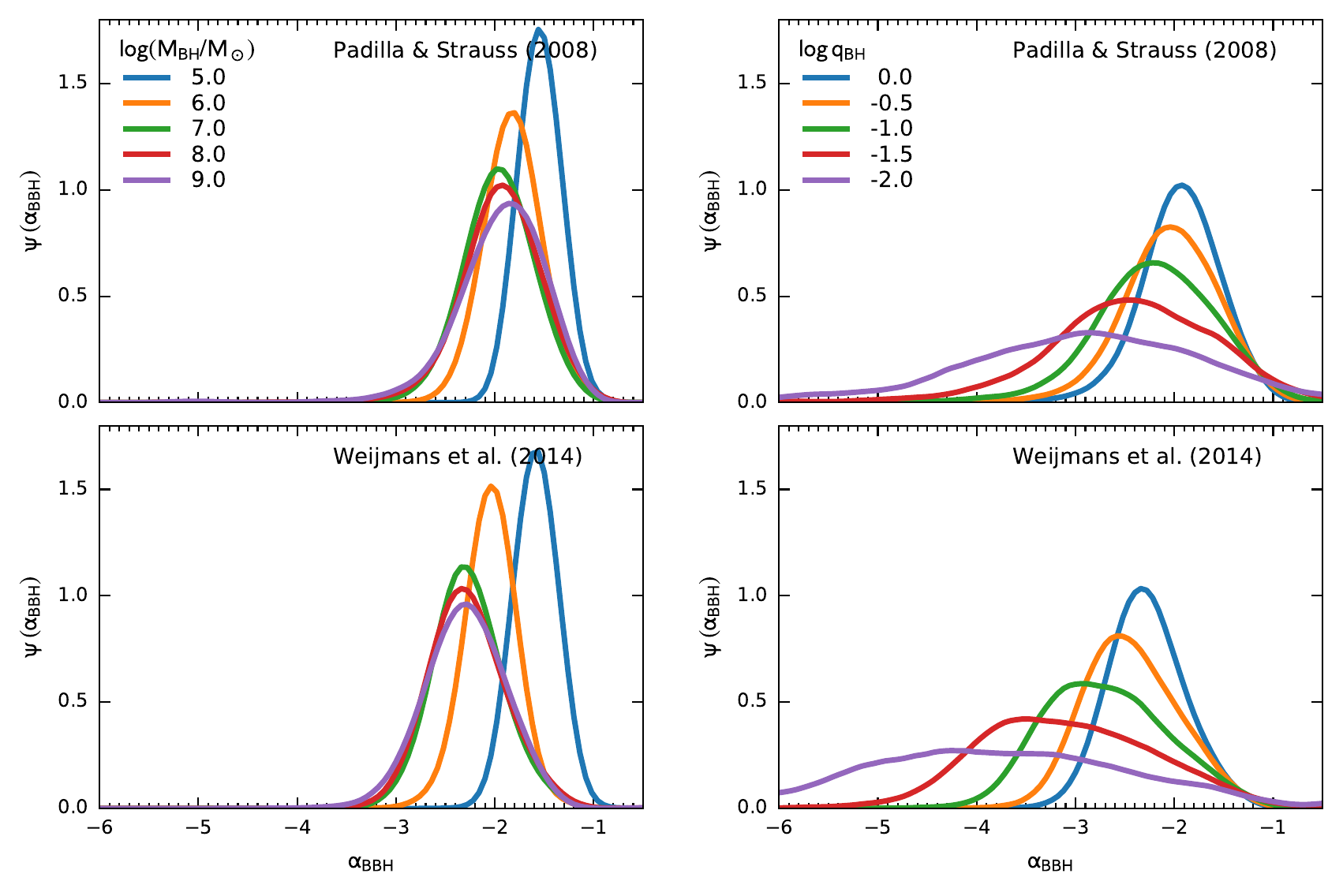}
\caption{Probability distributions of the slope of BBH evolution tracks
$\alpha\bbh$ (cf.\ Equation~\ref{eq:pdfx}) in the logarithmic evolution
timescale $\log t\evol$ versus logarithmic semimajor axis $\log a$ space during
the hard-binary stage of the same BBH population shown in
Figure~\ref{fig:trac_tau0}, $\psi_{\alpha\bbh}(\alpha\bbh|M\bh,q\bh,z'=0)$, and their dependence on the properties of the BBHs
and their host galaxies. The curves, the colors, and the texts have meanings
similar to those in Figure~\ref{fig:trac_tau0}.  In the upper panels, the
galaxy shape distribution of early-type galaxies from \citet{Weijmans14} is
adopted.  Whereas in the lower panels, the shape distribution of
\citet{Padilla08} is adopted. The results obtained by adopting the shape
distribution of \citet{Rodriguez13} are very close to those obtained by
adopting the shape distribution of \citet{Padilla08}, and are not shown here.}
\label{fig:trac_slop} \end{figure*}

\begin{figure*} \centering
\includegraphics[width=0.9\textwidth]{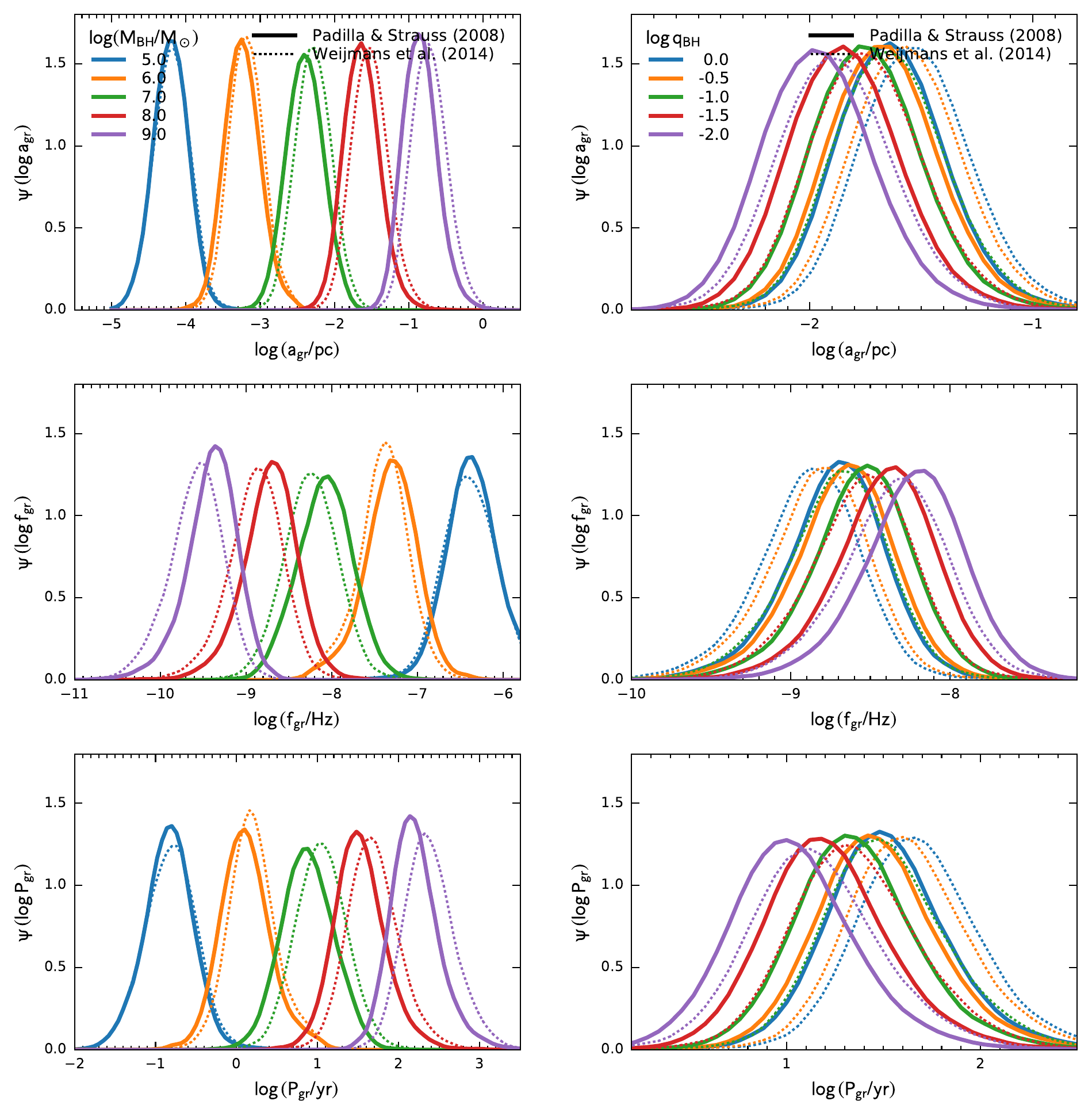}
\caption{Probability distributions of the logarithms of the semimajor axis
$\log\,a\gr$ (top panel), the GW frequency $\log\,f\gr$ (middle panel), and
the orbital period $\log\,P\gr$ (bottom panel) at which the BBHs enter the
gravitational stage for the same population of BBHs shown in
Figure~\ref{fig:trac_tau0}, i.e.,
$\psi_{\log a\gr}(\log a\gr|M\bh,q\bh,z'=0)$,
$\psi_{\log f\gr}(\log f\gr|M\bh,q\bh,z'=0)$,
and $\psi_{\log P\gr}(\log P\gr|M\bh,q\bh,z'=0)$. 
In each panel, the solid and the dotted curves show the results
obtained with adopting the galaxy shape distributions of \citet{Padilla08} and
\citet{Weijmans14}, respectively. The results obtained with adopting the galaxy
shape distribution of \citet{Rodriguez13} are very close to those of
\citet{Padilla08} and are therefore not shown in the figure.
}
\label{fig:trac_turn}
\end{figure*}

\begin{figure*}
\centering
\includegraphics[width=0.8\textwidth]{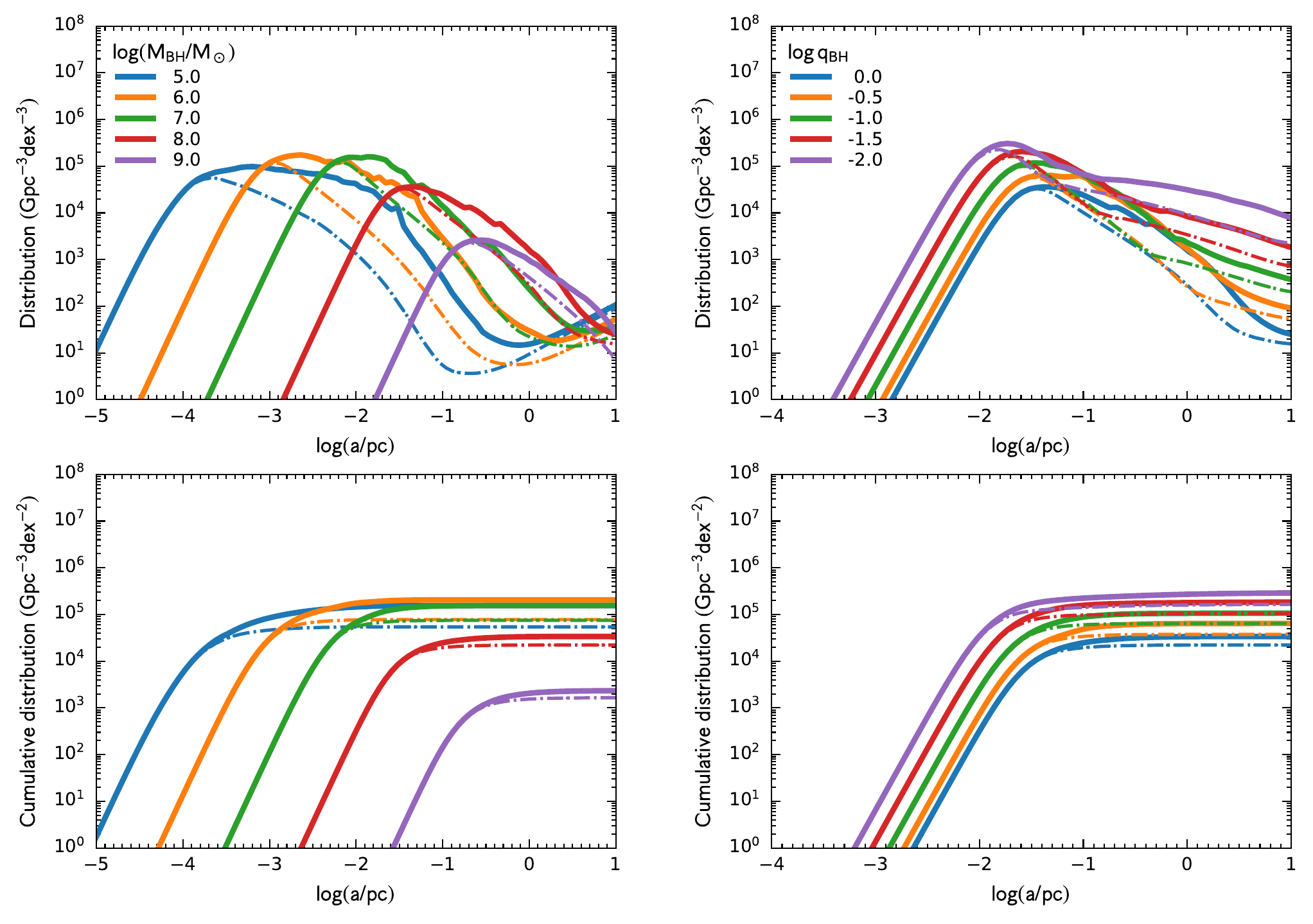}
\caption{
Semimajor axis distributions of the surviving BBHs at redshift $z=0$. In the
upper panels the vertical axis represents the variables $a{M\bh}{q\bh}{(\ln
10)^3}\Phi\bbh(M\bh,q\bh,a,z=0)$. In the lower panels the vertical axis
represents the cumulative semimajor axis distribution over semimajor axis
of the corresponding upper panel, $\int_0^a\,{M\bh}{q\bh}{(\ln
10)^2}\Phi\bbh(M\bh,q\bh,a',z=0)\,da'$.  In the left panels, the BBH
mass ratio is set to $q\bh=1.0$, and the different colors represent the
different BBH total masses $M\bh$,  as labeled by the texts. In the right
panels, the BBH total mass is set to $M\bh=10^8\msun$, and the different colors
represent the different BBH mass ratios $q\bh$, as labeled by the texts.
In the figure, the dot-dashed curves represent the contribution only by those
BBHs whose coalescence timescale $\tau_0$ are shorter than the Hubble
timescale, and the solid curves represent all the contribution.
As seen from the differences between the solid and the dot-dashed curves,
the distributions of the surviving BBHs with large
semimajor axes (at the binary stages) are mainly contributed by
those not being able to finish their final coalescences within the Hubble
timescale, while the distributions of the BBHs with low semimajor axes (at the
gravitational radiation stage) are those that can coalesce within the Hubble
timescale.
In the calculation, we adopt the GSMF from \citet{Behroozi19} and the galaxy
fractional merger rate from \citet{Rodriguez-Gomez15},
and the galaxy shape distribution of \citet{Padilla08} is used in modeling the
BBH evolution (similarly for all the figures below).
All the BH--host galaxy
relations listed in Table~\ref{tab:scal} are used, and the median distributions
obtained with those relations are shown in the figure (similarly for 
Figs.~\ref{fig:calc_pphi}--\ref{fig:cols_rate} below).
See Section~\ref{sec:res:surv}.
} \label{fig:calc_aphi} \end{figure*}

\begin{figure*}
\centering
\includegraphics[width=0.8\textwidth]{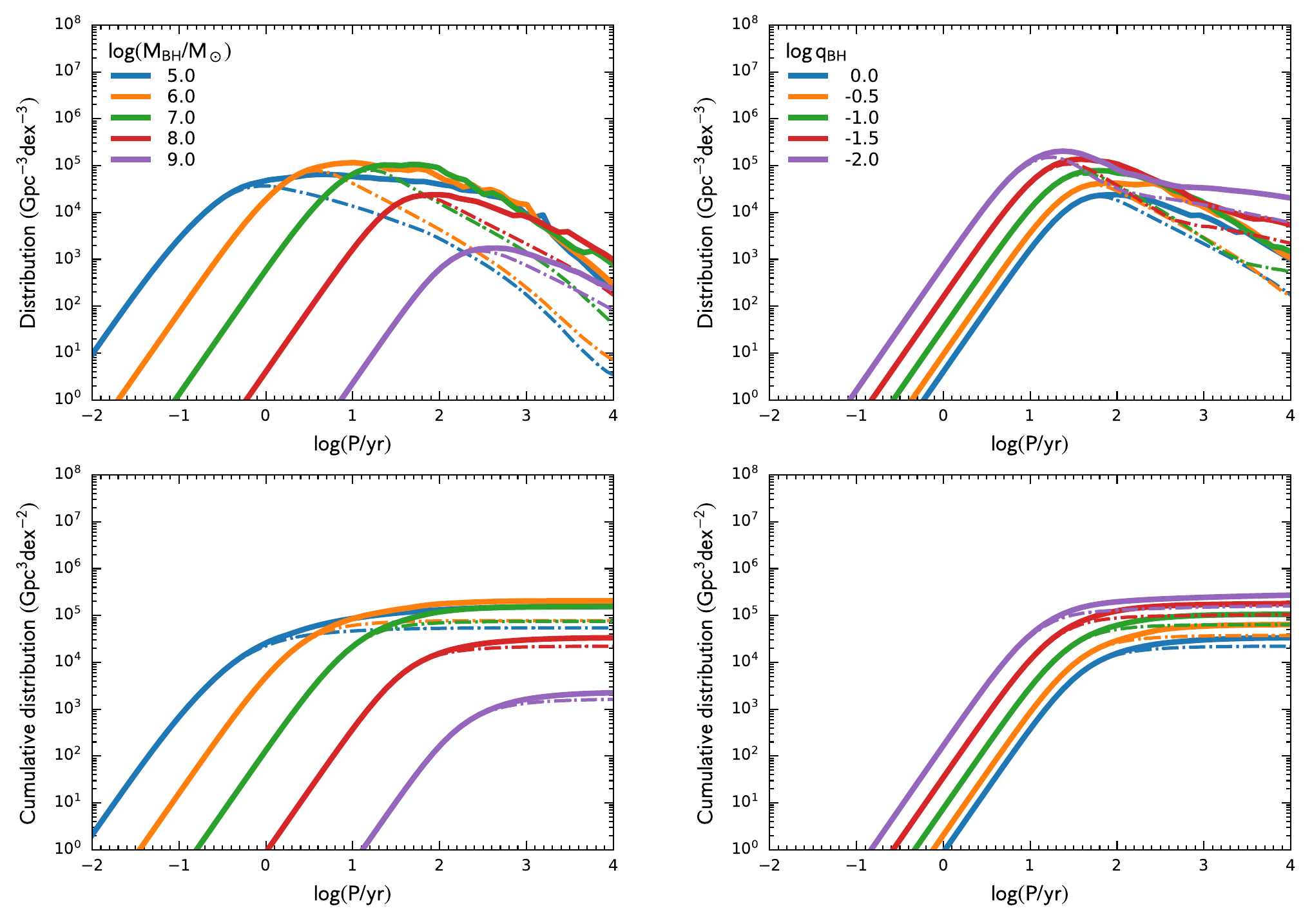}
\caption{The orbital period distributions and
the cumulative orbital period distributions of the same BBH populations
shown in Figure~\ref{fig:calc_aphi}.
The line colors and styles have the same meanings as
those shown in the corresponding panels of Figure~\ref{fig:calc_aphi}.
We set $q\bh=1.0$ in the left panels and $M\bh=10^8\msun$ in the right panels,
as shown in Figure~\ref{fig:calc_aphi}.
See Section~\ref{sec:res:surv}.
}
\label{fig:calc_pphi}
\end{figure*}

\begin{figure*}
\centering
\includegraphics[width=0.8\textwidth]{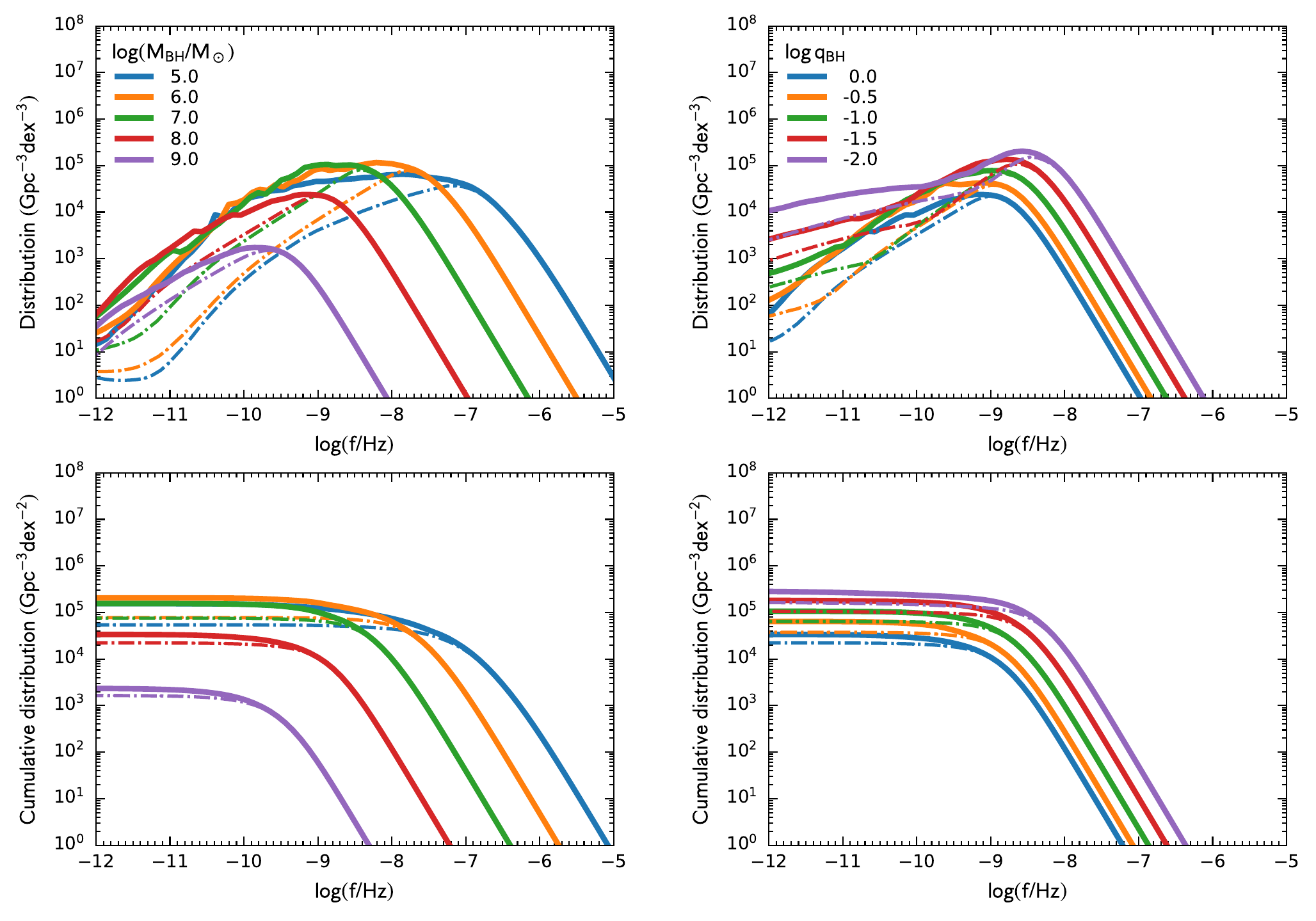}
\caption{ The GW frequency distributions
and the cumulative GW frequency distributions 
of the same BBH populations
shown in Figure~\ref{fig:calc_aphi}.
The line colors and styles have the same meanings as
those shown in the corresponding panels of Figure~\ref{fig:calc_aphi}.
We set $q\bh=1.0$ in the left panels and $M\bh=10^8\msun$ in the right panels,
as shown in Figure~\ref{fig:calc_aphi}.
See Section~\ref{sec:res:surv}.
}
\label{fig:calc_fphi}
\end{figure*}

\begin{figure*}
\centering
\includegraphics[width=0.8\textwidth]{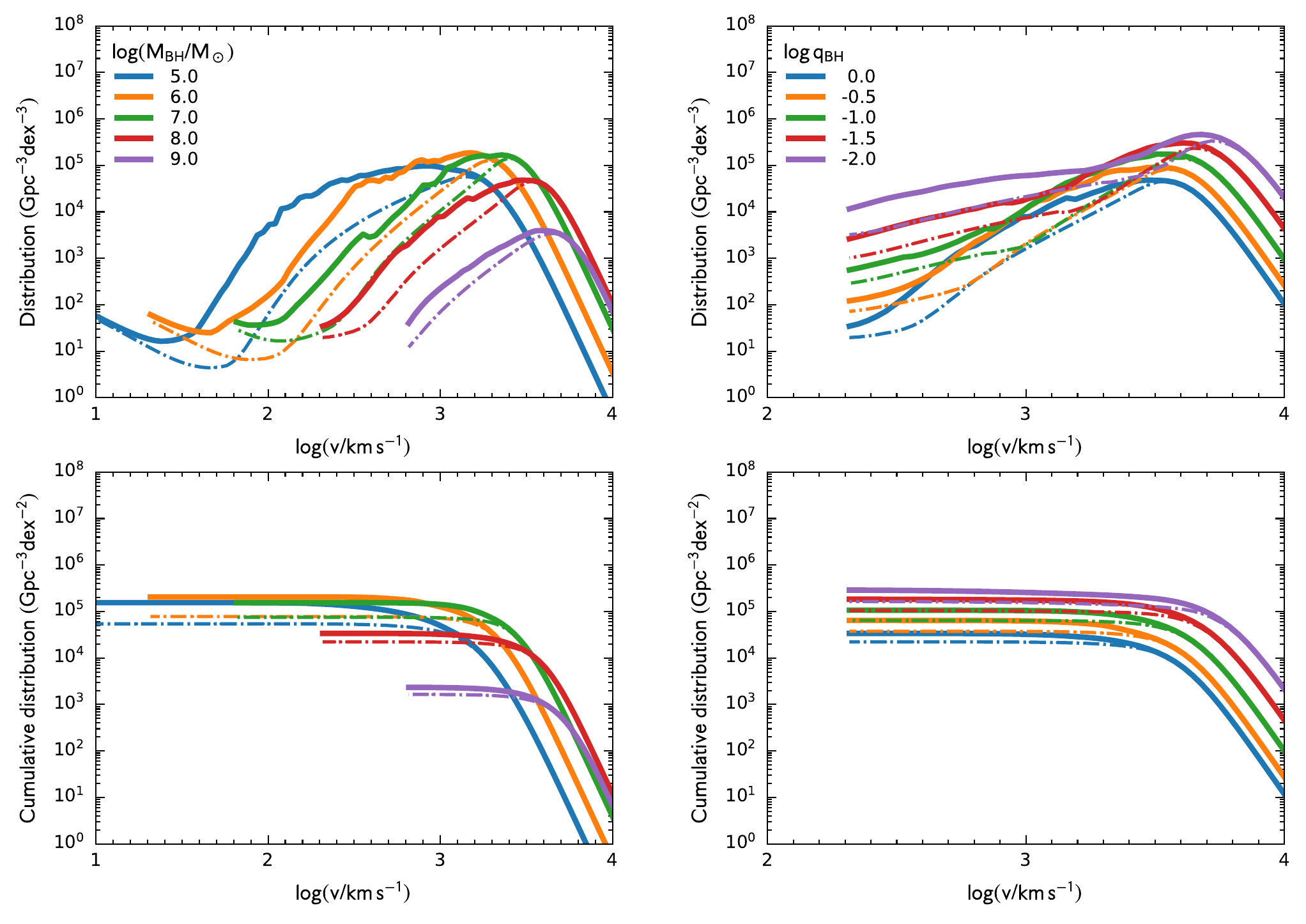}
\caption{The BBH relative velocity distribution and the cumulative distribution
of the same BBH populations shown in Figure~\ref{fig:calc_aphi}, where the
relative velocities are obtained by assuming that BBHs are on circular orbits,
given their masses and separations.
The line colors and styles have the same meanings as those shown in
the corresponding panels of Figure~\ref{fig:calc_aphi}.
We set $q\bh=1.0$ in the left panels and $M\bh=10^8\msun$ in the right panels,
as shown in Figure~\ref{fig:calc_aphi}.
See Section~\ref{sec:res:surv}.
}
\label{fig:calc_vphi}
\end{figure*}

We have investigated the evolution of supermassive BBHs in galaxies with
realistic property distributions, their statistical distributions, and the GW
radiation emitted by their coalescence.  In the model, by adopting the
observational high-resolution surface brightness profiles of some galaxy
samples and the observational galaxy shape distributions, we study the
dynamical interactions of the BBHs with their environments and construct the
BBH evolution tracks in the realistic galaxy distributions.
By incorporating the BBH evolution tracks with the GSMF, galaxy merger rates,
the BH--host galaxy relations, effects of multiple galaxy mergers during the
evolution of a BBH, we obtain the following main results on the evolution of
BBHs, their statistical distributions, and their GW radiation at the PTA and
LISA bands.
\begin{itemize}
\item
Triaxial galaxy shapes, even with a mild triaxiality, can significantly
decrease the BBH peak evolution timescale ($t\peak$) after the BBH becomes
bound, compared to a nearly spherical, or nearly prolate, or nearly oblate
shape (see Figs~\ref{fig:tsc_contour} and \ref{fig:tpeak_rely}), which is
consistent with the results in \citet{Yu02}.  If the galaxy shape is triaxial
and not close to nearly spherical/prolate/oblate, the peak timescale is not
sensitive to the value of the triaxiality.  The statistical distribution of the
BBH coalescence timescales shifts towards longer timescales if the galaxy shape
distribution is close to being axisymmetric, compared to the case in triaxial shape
distributions (see Fig.~\ref{fig:trac_tau0}), and thus the fractions of the
BBHs that can evolve into the
PTA band becomes smaller, leading to a lower stochastic GWB amplitude.
\item The BBHs formed by local galaxy mergers ($z=0$) have their
coalescence timescales distributing widely from ${\sim}10^8$ to
${\sim}10^{12}\yr$ (see Fig.~\ref{fig:trac_tau0}).  High-mass BBHs tend to have
a relatively larger fraction with longer $\tau_0$ than low-mass BBHs. Compared
BBHs with large mass ratios, BBHs with smaller mass ratios tend to have a
larger fraction with longer $\tau_0$, mainly due to the increased dynamical
friction timescales, which is consistent with the study in \citet{Yu02}.
Compared with the distribution of $\log\tau_0$, the distribution of the BBH
peak timescale after the dynamical friction stage, $t\peak$, is less sensitive
to the BBH mass ratios (see Fig.~\ref{fig:trac_tmax}).
For low-mass BBHs, their coalescence timescales are relatively longer than
their peak timescales after they become bound, which is also mainly due to the
significant contribution of the dynamical friction timescale to the coalescence
timescales.
Overall, the slopes of the BBH evolution curves during the hard binary stage are
in the range from -4 to -1 for systems with mass ratio greater than 0.1, not
deviating from the value -1 by orders of magnitude, indicating that in most of
those cases the stellar refilling rates to the loss cone are not low compared to
the rates in the full loss cone case.
\item Given the BBH total mass and its mass ratio, the surviving BBHs with
large semimajor axes are mainly contributed by those systems which have
coalescence timescales longer than the Hubble time and have not reached the
gravitational radiation stage, and the surviving BBHs at the low-semimajor axis
end are those being at the gravitational radiation stage and having coalescence
timescales shorter than the Hubble time (see Figs.~\ref{fig:calc_aphi}).  The
semimajor axis distribution of the surviving BBHs peaks around the transition
to the gravitational radiation stage, where the evolution of most of the BBHs
are located at their ``bottlenecks''.  Below the peak semimajor axis, the
distribution decreases with decreasing semimajor axes sharply, following a
power law.  Above the peak semimajor axis, the distribution decreases with
increasing semimajor axes mildly.  With increasing the total BBH mass or the
BBH mass ratio, the peak of the distribution of surviving BBHs shifts to larger
semimajor axes, longer orbital periods, and lower GW frequencies. The
semimajor axes of some observational BBH candidates do not locate at the peak
locations, which would imply a promising prospect for discovering more BBH
candidates. The distributions of surviving BBHs can serve as a guide for
searching BBH systems at parsec or sub-parsec separation scales.
\item
Among the MBHs at galactic centers in the local universe ($z=0$), the fractions
of surviving BBHs with $q\bh>1/3$ and $a<10\pc$ are expected to $\sim$1\%-3\%.
At $z=0$, the fractions of
surviving BBHs with $q\bh>1/3$ and orbital periods $1\yr<P<10\yr$ are expected
to be $\sim$0.5\%--1\% in MBHs with $M\bh\sim
10^{6}$--$10^7\msun$. The fractions of those with $1\pyr<P<30\pyr$ are expected
to be $\sim$2\% in MBHs with $M\bh\sim 10^{6}$--$10^7\msun$, and the fractions
decrease in MBHs with other masses, down to $10^{-4}$ for $M\bh\sim 10^9\msun$.
The fractions are not sensitive to redshifts at $z\la 3$.
The fractions of surviving BBHs can increase to $\sim 10\%$ for
$q\bh>1/100$.
\item The time delays between host galaxy mergers and embedded BBH coalescences
decrease the BBH coalescence rates at high redshifts while increase them at low
redshifts, compared to the results obtained by assuming instantaneous BBH
coalescences (see Fig.~\ref{fig:cols_rate}). This effect is significant
especially for BBHs with smaller mass ratios, as BBHs with smaller mass ratios
tend to have relatively longer coalescence timescales.  The BBH coalescence
rates obtained by using different sets of the BH--host galaxy relations differ
a lot, up to ${\sim}2$ orders of magnitude (see
Fig.~\ref{fig:cols_rate_scale}).

\item The stochastic GWB strain spectrum around the PTA band contributed by the
cosmic population of BBHs is estimated, and the effects of the physical
ingredients involved in the estimation are discussed. The strain spectrum has a
peak or turnover at the frequency $\sim 0.25\nhz$.
The spectrum follows a power law $\propto f^{-2/3}$ at frequencies higher than
the turnover frequency, unless the frequencies are high enough so that the
number of contributing sources is too small to have a significantly statistical
estimate.  And the spectrum declines with decreasing frequencies with an
exponent of $-\frac{2}{3}+\frac{\kappa\gw}{2}\sim 1.2\pm 0.2$
at frequencies lower than the turnover frequency.
\item Using different BH-host galaxy relations in the model can lead to the
difference in the GWB strain amplitude by one order of magnitude (see
Fig.~\ref{fig:strn_spec_scale}).  At frequency $1\pyr$, the median value of the
characteristic strain amplitudes $\ayr$ obtained with the different BH--host
galaxy relations is $\ayr{\sim} 2.0\times 10^{-16}$, the maximum value is
$\ayr{\sim} 5.4\times 10^{-16}$ obtained with the $M\bh{-}M\bulge$ relation of
\citet{Kormendy13}, and the minimum one is $\ayr{\sim} 6.1\times 10^{-17}$
obtained with the $M\bh{-}M\bulge{-}\sigma$ relation of \citet{Shankar16}.  The
maximum and the median values are below the current upper limit set by the PTA
experiments by a factor of $\sim 2$ and $\sim 5$, respectively. Our estimate is
lower than previous estimates by a factor of $\sim 1.5$--3, partly due to the
different choices of galaxy merger rates, and partly due to the inclusion of the
time delay between BBH coalescences and galaxy mergers. The GWB strain
amplitudes at a little lower frequency $\sim 10$~nHz are expected to be within
the detection ability of future experiments (e.g., SKA, ngVLA).
\item
The above estimates obtained with including the BBH orbital evolution and the
time delays between host galaxy mergers and BBH coalescence has decreased the
estimated strain amplitude by $\sim$0.1--0.2~dex, compared to the estimates
obtained with an instantaneous BBH coalescence assumption.  The above estimates
are obtained by adopting the galaxy shape distributions in \citet{Padilla08} or
\citet{Rodriguez13} is adopted, which prefers triaxial configurations. If the
galaxy shape distribution of \citet{Weijmans14} is adopted, which prefers
axisymmetric configurations, the estimated strain amplitudes will decrease by
another ${\sim}0.15$~dex.
\item At frequencies $f\ga 10^{-8}\hz$, the GWB strain spectrum deviates to
below the $-2/3$ power law and fluctuates due to the small number of BBH sources
at the high-frequency end. The prospect of detecting GW signals from individual
BBH systems around the PTA band is studied, through simulating the cosmic BBH
populations in our model (see Fig.~\ref{fig:strn_indv}). According to the
simulated realizations, it is found that the majority of individual sources with
signals higher than the stochastic backgrounds are located at frequencies $\ga
10^{-8}\hz$. The loudest signals located at $10^{-8}$--$10^{-7}$~Hz have BBH
masses higher than $\sim 10^9\msun$, and redshifts $z\la 1$, and their strain
amplitudes are lower than the best constraints of current PTAs on continuous
individual sources by about one order of magnitude.
\item Finally, we extend the investigation of GW radiation from the cosmic
population of BBHs to the LISA detection bands. Based on our realizations of the
BBH populations, the detection rate of this population is predicted to be
${\sim}0.9\pyr$, which sets the lower limit for the LISA detection rates.
\end{itemize}

We thank Scott Tremaine for helpful discussions.  This work was supported in
part by the National Natural Science Foundation of China under Nos.\ 11673001,
11273004, 10973001, 11690024, 11873056, 11721303, the National Key R\&D
Program of China (Grant Nos.\ 2016YFA0400703, 2016YFA0400704), the Strategic
Priority Program of the Chinese Academy of Sciences (Grant No.\ XDB 23040100),
and National Supercomputer Center in Guangzhou, China.


\clearpage

\begin{figure*} \centering
\includegraphics[width=0.45\textwidth]{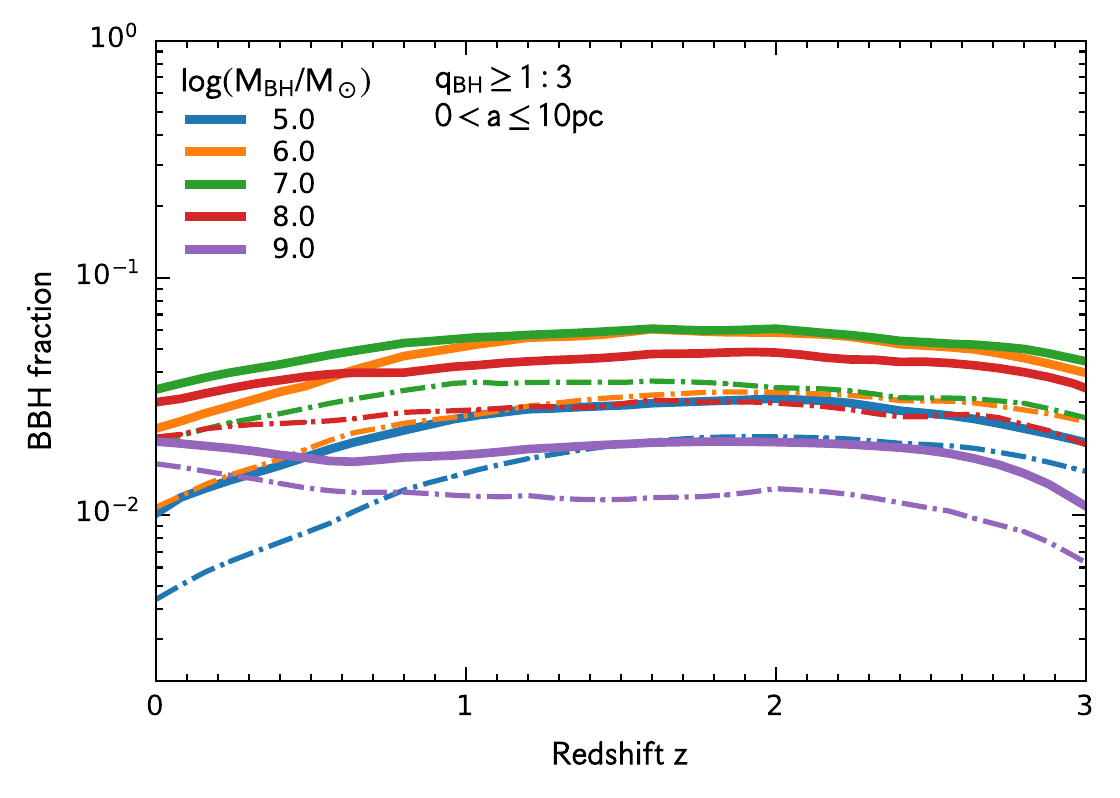}
\includegraphics[width=0.45\textwidth]{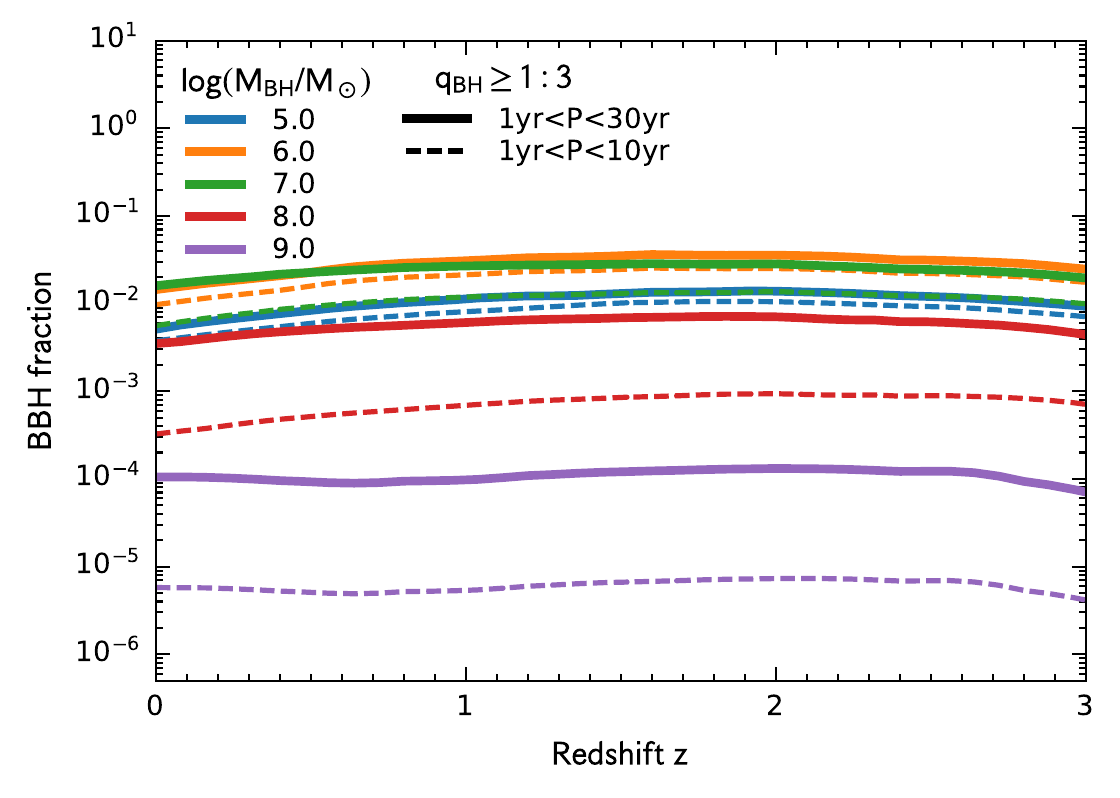}
\includegraphics[width=0.45\textwidth]{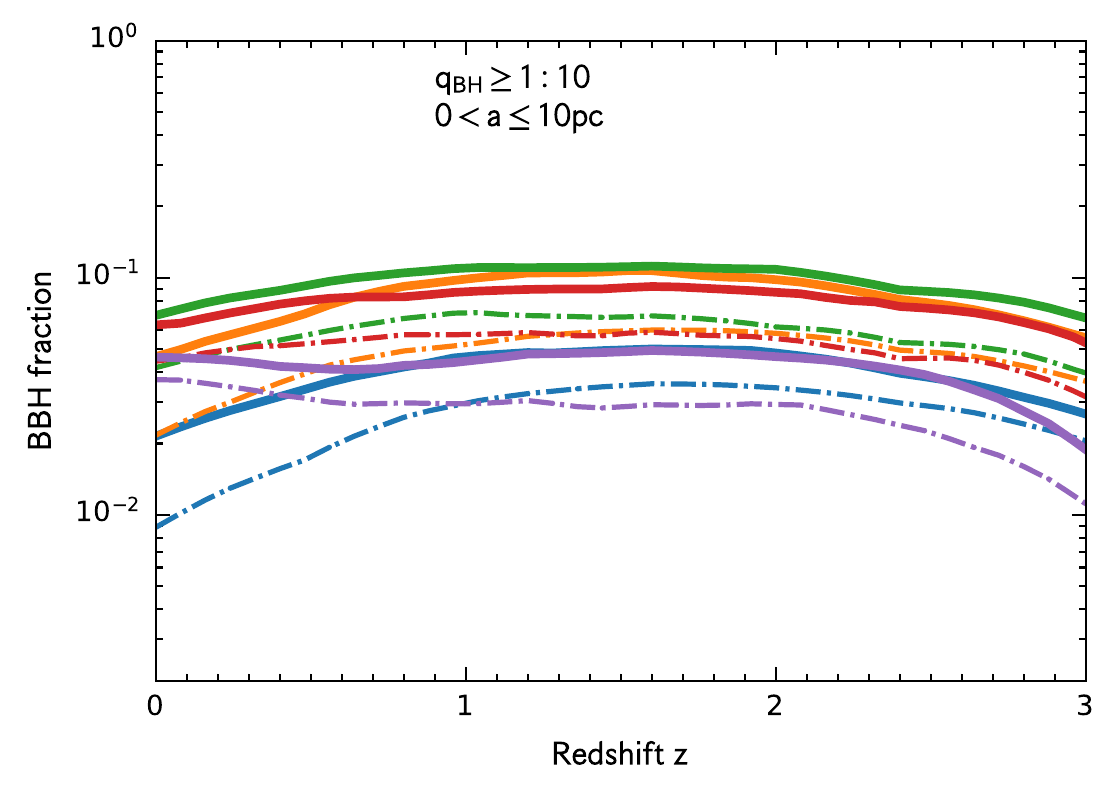}
\includegraphics[width=0.45\textwidth]{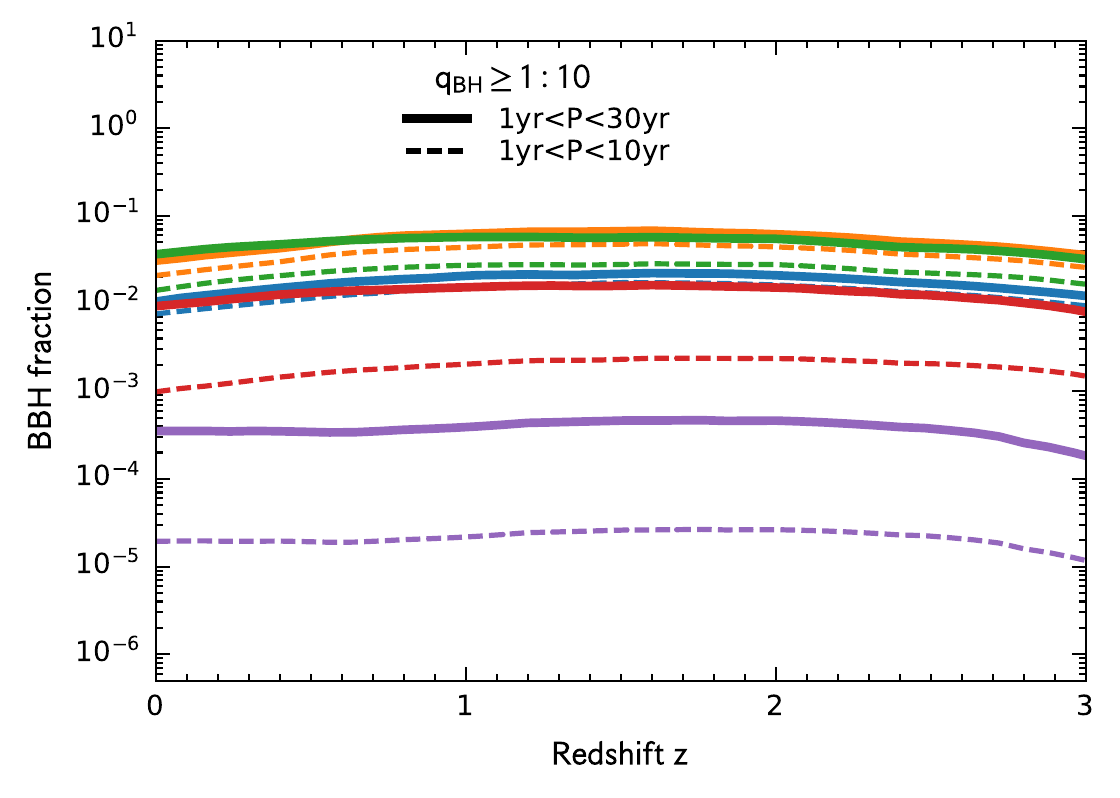}
\includegraphics[width=0.45\textwidth]{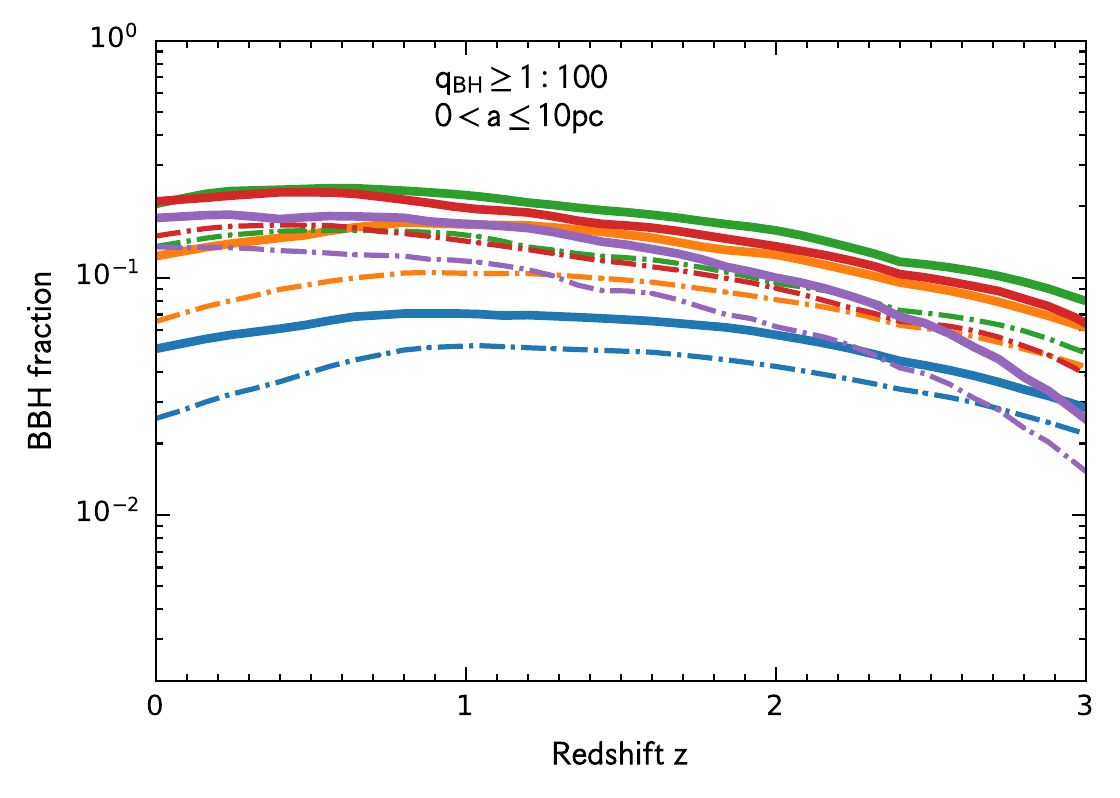}
\includegraphics[width=0.45\textwidth]{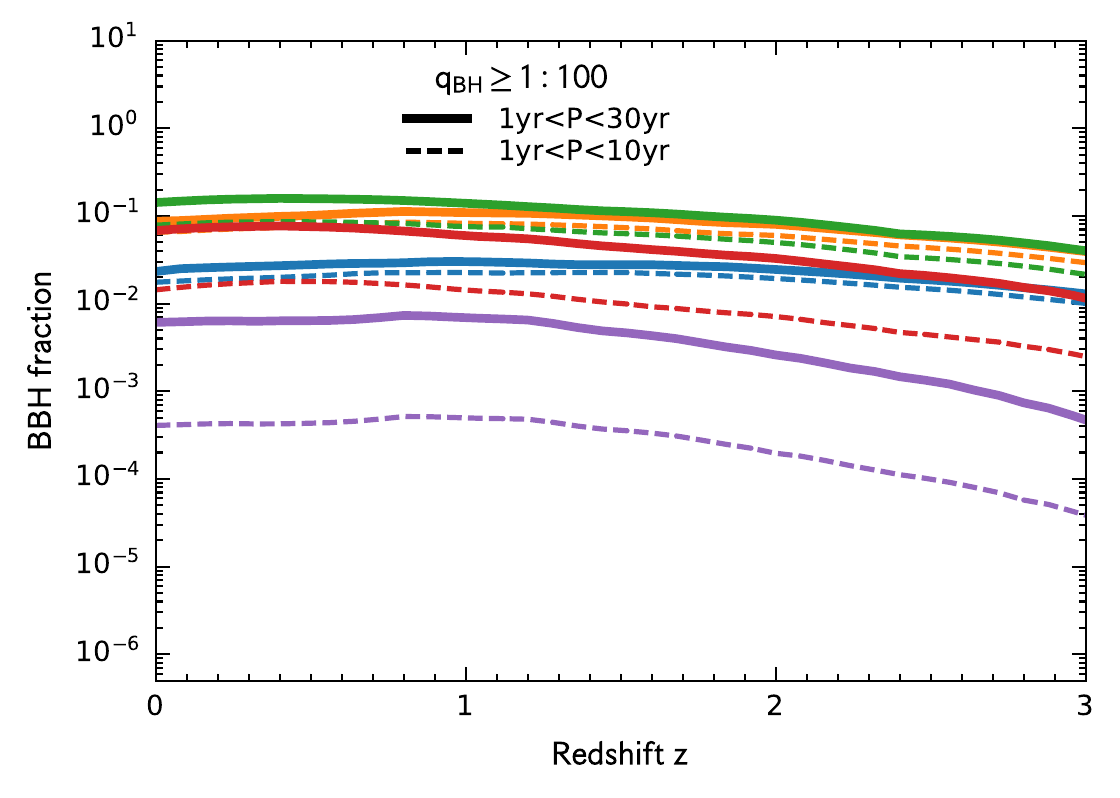}
\caption{The fractions of the MBHs that are surviving BBHs with $q\bh\ge
1/3,1/10,1/100$, as a function of redshift.  In the left panels, only BBHs with
semimajor axis $0<(a/\pc)\leq 10$ are included, and in the right panels, only
BBHs with orbital periods $1\yr<P<10\yr$ or $1\yr<P<30\yr$ are included.  The
different colors represent the different BBH total masses $M\bh$, as denoted by
the texts. The dot-dashed curves represent the contribution only by those BBHs
whose coalescence timescale $\tau_0$ are shorter than the Hubble timescale at
the corresponding redshift, and the solid curves represent all the
contribution.  As seen from the top left panel, the surviving BBH fractions
with $q\bh\ge 1/3$ are around $\sim$1\%--3\% (solid curves) for different BH
masses, and more than about a half of them have total evolution timescales
shorter than the Hubble timescale. The surviving BBH fractions with $q\bh\ge
1/100$ increase to $\sim 10\%$ in the bottom left panel.
As seen from the top right panel, at $z=0$, the fractions of surviving BBHs
with $q\bh\ge 1/3$ and orbital periods $1\yr<P<10\yr$ are expected to be
$\sim$0.5\%--1\% in MBHs with $M\bh\sim 10^{6}$-$10^7\msun$, the fractions of
those with $1\yr<P<30\yr$ are expected to be $\sim$2\% in MBHs with $M\bh\sim
10^{6}$-$10^7\msun$, and the fractions decrease in MBHs with other masses,
down to $10^{-4}$ for $M\bh\sim 10^9\msun$.
The fractions are not sensitive to redshifts at $z\la 3$.
The fractions of surviving BBHs with $q\bh\ge 1/100$ and $1\yr<P<30\yr$
increase to $\sim 10\%$ for $M\bh\sim 10^{6}$-$10^7\msun$ and $\sim 6\times
10^{-3}$ for $M\bh\sim 10^9\msun$, as shown in the bottom right panel.
See Section~\ref{sec:res:surv}.
} \label{fig:surv_frac} \end{figure*}

\begin{figure*}
\centering
\includegraphics[width=0.8\textwidth]{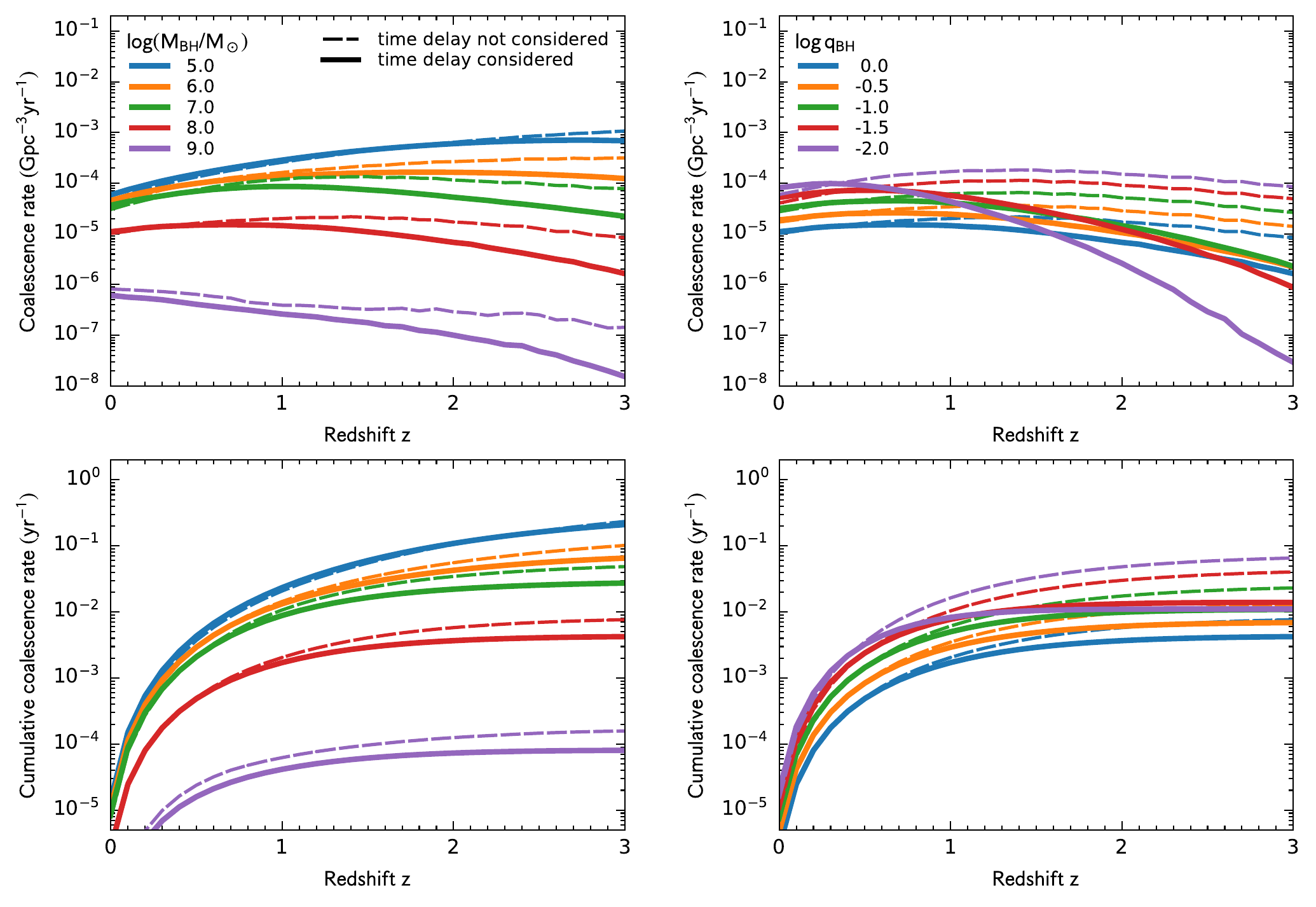}
\caption{The BBH coalescence rates as a function of redshift, and their
dependence on BBH total masses and mass ratios. In the upper panels, the
vertical axis represents the variable $M\bh q\bh (\ln 10)^2 R\bh(M\bh,q\bh,z)$
(see Eq.~\ref{eq:Rbh}).
In the lower panels, the vertical axis represents
the volume-integrated rates for for the corresponding upper panel, i.e.,
$\int_0^z M\bh q\bh (\ln 10)^2 R\bh(M\bh,q\bh,z) |dV_{\rm c}/dz|/(1+z) dz$,
where $dV_{\rm c}$ is the comoving volume of the universe at the
redshift range $z\rightarrow z+dz$.
The line colors have the same meanings as those
shown in the corresponding panels of Figure~\ref{fig:calc_aphi}.
We set $q\bh=1.0$ in the left panels and $M\bh=10^8\msun$ in the right panels,
as shown in Figure~\ref{fig:calc_aphi}.
The solid and the dashed lines represent the results obtained with and without
including the time delays between host galaxy mergers and embedded BBH
coalescences, respectively, and their differences are significant for low
BBH mass ratios and at high redshifts.
In addition, the low $q\bh$ cases illustrated in the right panels suggest that
the ignoration of the time
delays between galaxy mergers and BBH coalescences may lead to an overestimate
of the LISA detection rates.
See Sections~\ref{sec:res:cols} and \ref{sec:res:lisa}.  
} \label{fig:cols_rate} \end{figure*}

\begin{figure*}
\centering
\includegraphics[width=0.8\textwidth]{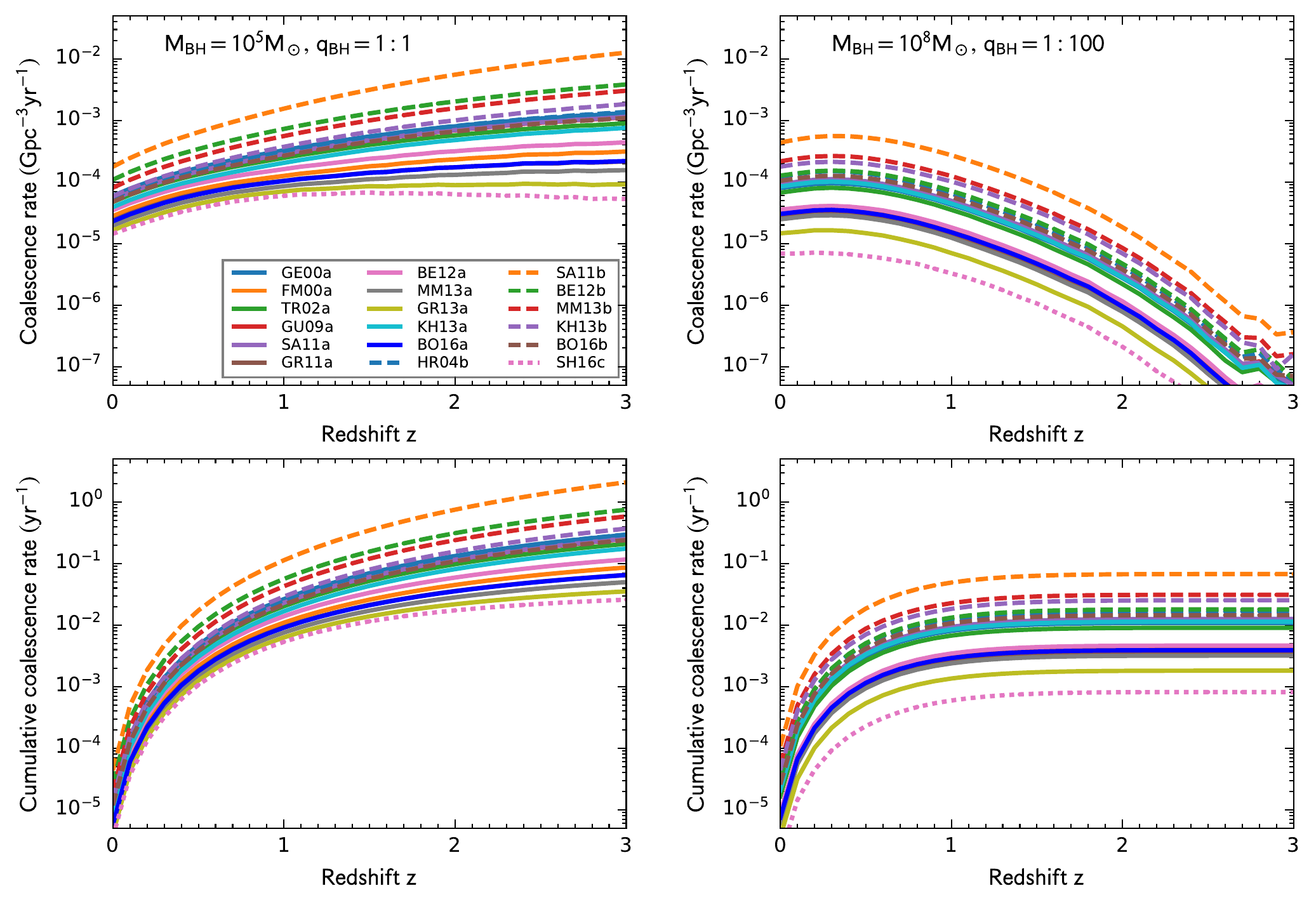}
\caption{The BBH coalescence rates obtained with the different sets of
BH--host galaxy relations listed in Table~\ref{tab:scal}. The vertical axes
represent the same variables as those in the corresponding panels of
Figure~\ref{fig:cols_rate}, and the time delays between host galaxy mergers and
embedded BBH coalescences are included.
The left panels show the results for the case of $(M\bh,q\bh)=(10^5\msun,1)$,
and the right panels show the results for the case of
$(M\bh,q\bh)=(10^8\msun,0.01)$. 
As seen from the figure, the difference in the BBH coalescence rates caused by
the difference in the BH--host galaxy relations can be up to two orders of
magnitude.
See Section~\ref{sec:res:cols}.  } \label{fig:cols_rate_scale} \end{figure*}

\clearpage


\begin{figure*}
\centering
\includegraphics[width=0.75\textwidth]{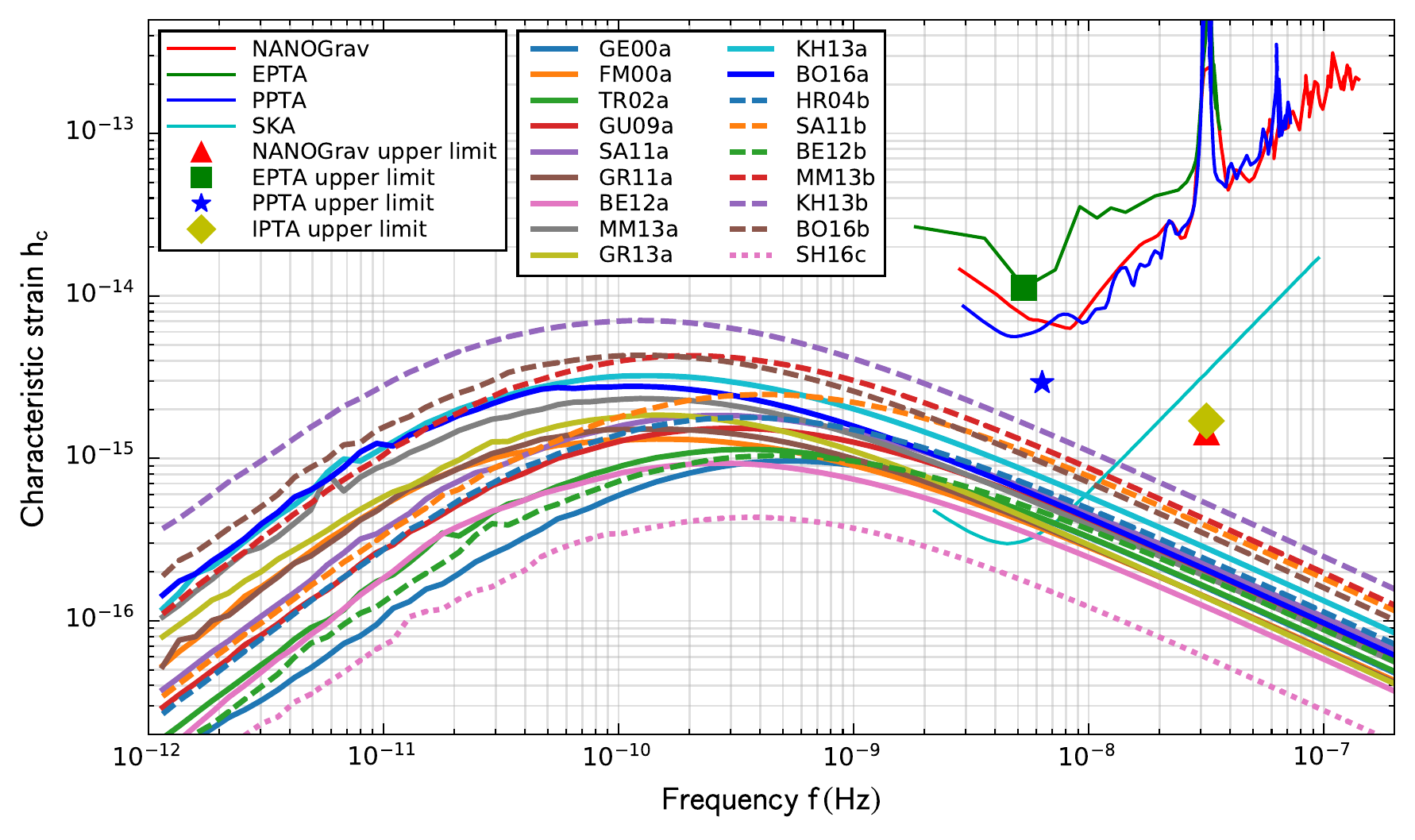}
\includegraphics[width=0.75\textwidth]{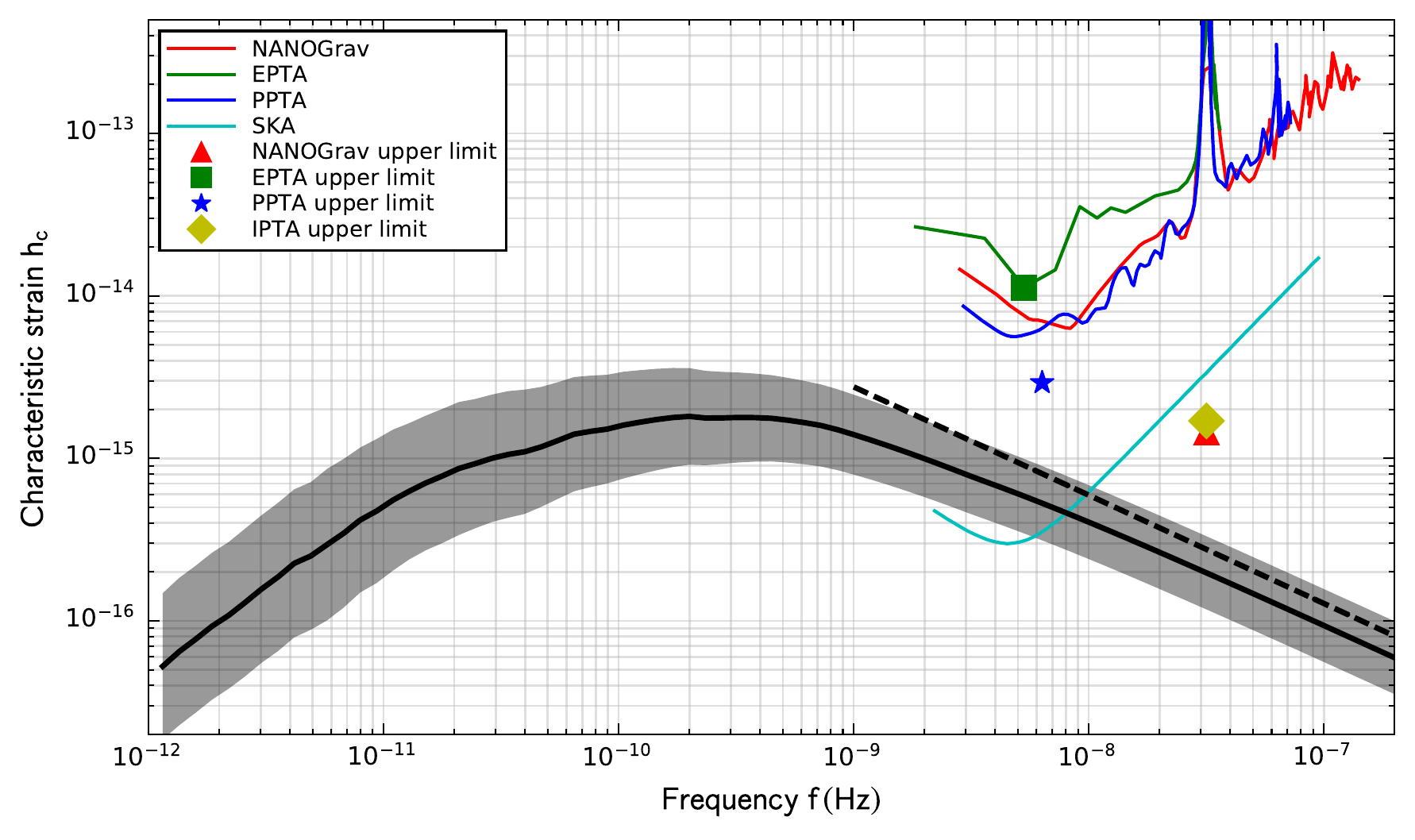}
\caption{
The characteristic strain spectra $h\rmc$ of the stochastic
GWB from the cosmic population of BBHs (cf.\ Eq.~\ref{eq:hc2}),
as a function of the observer-rest frequency $f$ around the PTA band.
The upper panel shows the characteristic strain spectra $h\rmc$ of the
stochastic GWB from the cosmic population of BBHs expected from the model
in this work. The different curves covering the whole frequency range
in the upper panel represent the results
obtained with the different sets of BH--host galaxy relations listed in
Table~\ref{tab:scal}, which differ by one order of magnitude and contribute
significantly to the uncertainty in the estimate of the stochastic GWB.
In the lower panel, the black solid curve and the shaded region represent the
median and the standard deviation of the spectra shown in the upper panel.
The dashed curve represents the median spectrum obtained without including the
time delays between the galaxy mergers and BBH coalescences in the model.
The spectrum at the high-frequency end follows the
unique $-2/3$ power law of the observational frequency.
The bending of the strain spectrum
shown at the low-frequency end ($\la 10^{-9}\hz$) is caused by that a
significant part of the BBH energy loss is driven by three-body interactions
with surrounding stars, and the spectra correspond to the BBH evolution
tracks at $a\ga a\gr$.
In the figure, the red, green, blue, and cyan curves located close to
the upper right corner are the currently most stringent sensitivity curves
obtained by NANOGrav \citep{Arzoumanian18}, EPTA \citep{Lentati15}, PPTA
\citep{Shannon15}, and the expected sensitivity curves of the planned
PTAs based on SKA \citep{Bonetti18}, respectively,
as labeled by the texts.
The red, green, blue, and yellow symbols mark the currently upper limit of the
GWB set by NANOGrav, EPTA, PPTA, and IPTA, respectively.
As seen from the figure, the estimated maximum strain amplitude at $f=1\pyr$ is
below the current upper limit set by the PTA experiments by a factor of $\sim
2$, while the median of the predicted values is smaller than the current
upper limit by a factor of $\sim 5$.
The GWB strain amplitudes at a little lower frequency $\sim 10$~nHz
are expected to be within the detection ability of future experiments (e.g.,
SKA).
See Section~\ref{sec:res:gwb}.  }
\label{fig:strn_fill}
\label{fig:strn_spec}
\label{fig:strn_spec_scale}
\label{fig:strn_line}
\end{figure*}

\begin{figure}
\centering
\includegraphics[width=0.55\textwidth]{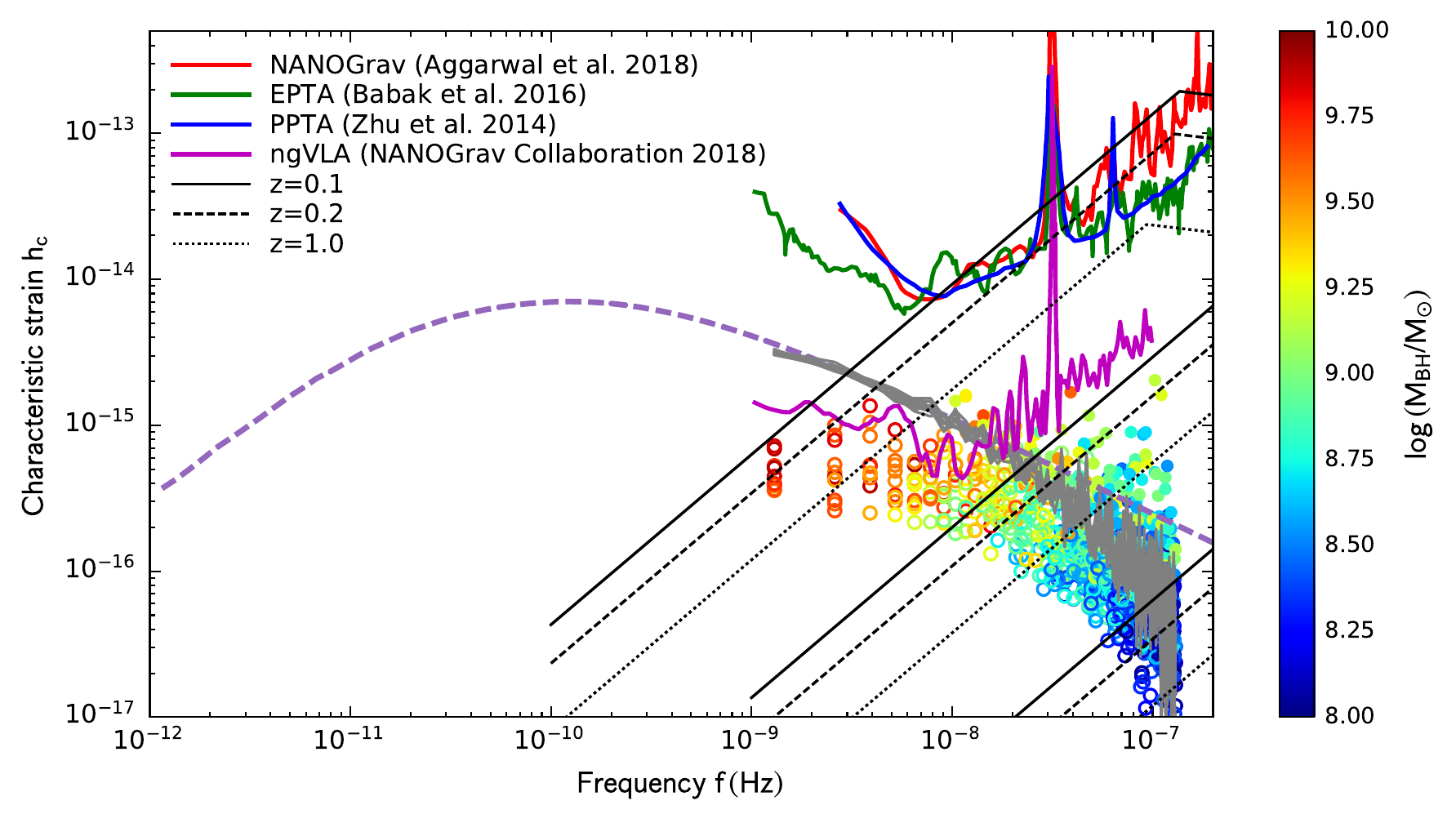}
\includegraphics[width=0.55\textwidth]{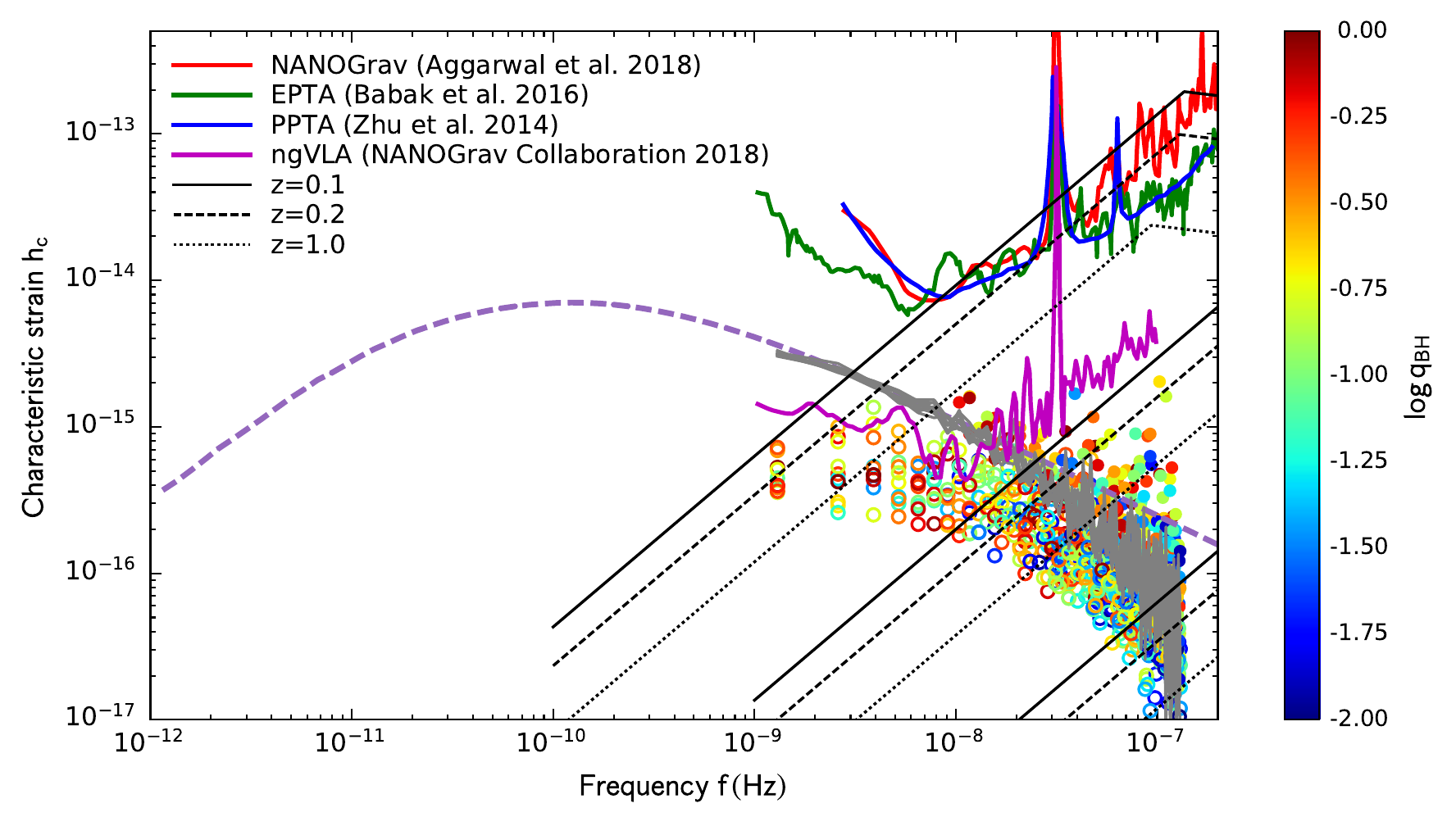}
\includegraphics[width=0.55\textwidth]{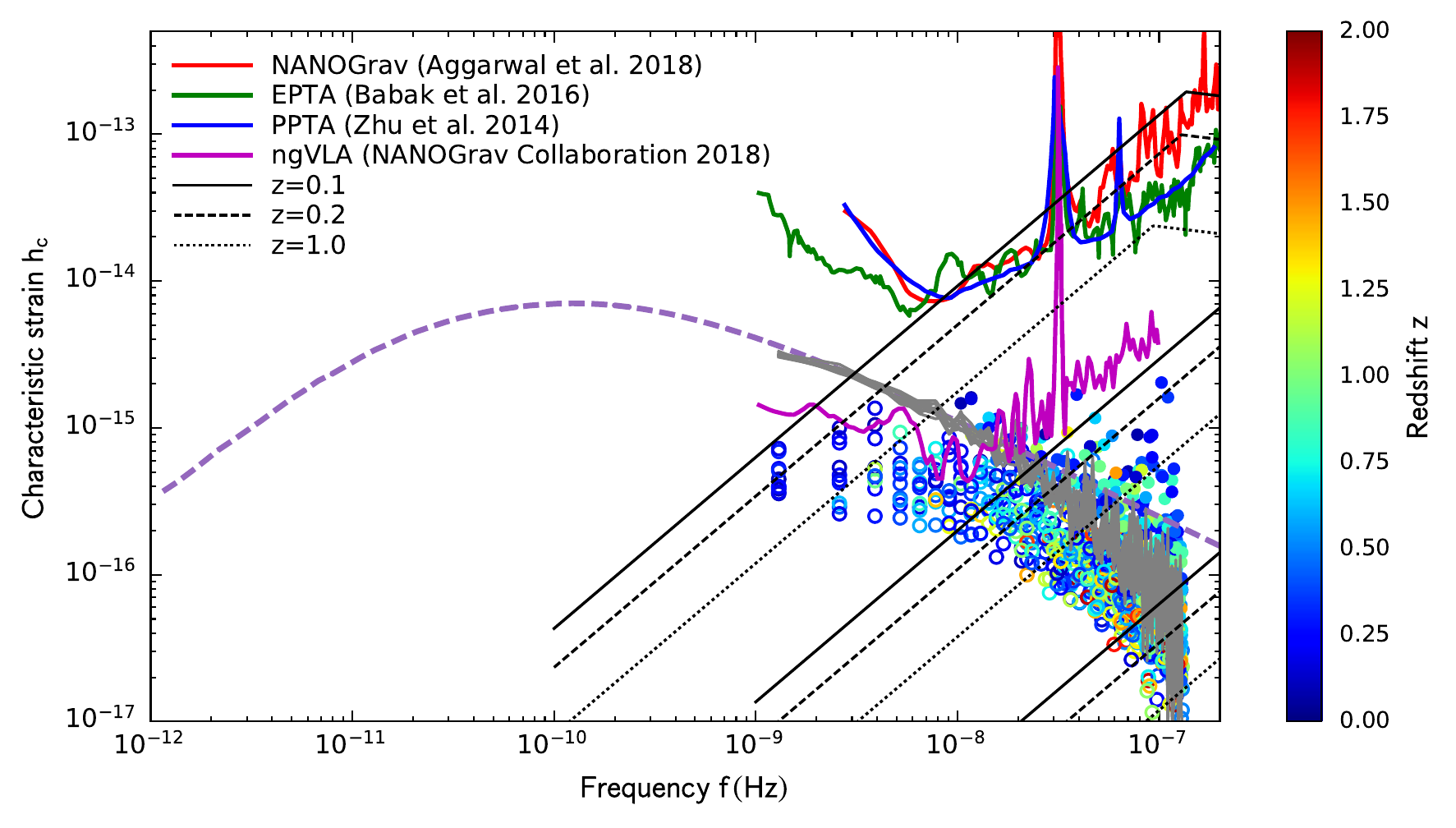}
\caption{\scriptsize Demonstration of the GW strain spectrum fluctuation at the
high-frequency end of the PTA band due to the limited number of BBH systems
within those frequencies, and the prospects for detecting individual BBH
sources with characteristic strain amplitudes standing above the average
background at the PTA band. 
The distributions of the characteristic strain amplitudes of
individual BBH sources are obtained from ten realizations of the
cosmic population of BBHs generated in our model (see
Section~\ref{sec:res:indv}).
In each panel, the purple dashed curve shows the same strain
spectrum as the one in Figure~\ref{fig:strn_spec_scale}, obtained with the
$M\bh{-}M\bulge$ relation of \citet{Kormendy13}.
Each grey curve represents the background strain spectrum of one realization of
the cosmic population of BBHs, obtained by removing the loudest source in each
frequency bin and then adding up the GW signal of the rest sources
(Eq.~\ref{eq:hc}). The grey curves fluctuate at high frequencies, due
to the limited number of the BBH sources within these frequencies. The
loudest source in each frequency bin is shown by a color circle in the figure,
and it is shown by a filled circle if it is loud enough to stand above its
corresponding background (grey curve).
The different colors in circles represent different total BBH masses, different
BBH mass ratios, and different redshifts in the top, the middle, and the
bottom panels, respectively.
The characteristic strain amplitudes of the individual sources within the given
frequency bins are obtained by assuming an observational run lasting $T\obs=25$
years.
At the upper right corner of the panels, the red curve represents the 95\%
upper limit on the characteristic strain amplitude of individual continuous
sources, based on a Bayesian analysis of the NANOGrav 11-year data set
\citep{Aggarwal18}; the green curve shows the same limit, but based on a
$\calF_p$ statistics analysis of the EPTA data set (see Figure~6 in
\citealt{Babak16}); the blue
curve represents the sensitivity curve towards the median sensitive sky
position shown in Figure 10 of \citet{Zhu14}, based on an analysis of the PPTA
DR1; and the magenta curve represents the expected sensitivity curve of the
planned PTAs based on Next-Generation Very Large Array (ngVLA;
\citealt{nanograv18}).
As a reference, the broken black lines (see Eq.~\ref{eq:hcindividual})
illustrate the characteristic strain amplitude of some example BBH systems as a
function of observer-rest frequency.  The example BBHs have the same mass
ratios $q\bh=0.1$, different total BBH masses $M\bh=10^{10}$, $10^9$, and
$10^8\msun$ from the top to bottom (given the same line style), and different
redshifts $z=0.1$ (solid), 0.2 (dashed), and 1.0 (dotted)
shown by the different line styles.
The black lines are also obtained by assuming an observational run lasting
$T\obs=25$ years for those example BBH systems.
} \label{fig:strn_indv} \end{figure}

\begin{figure*}
\centering
\includegraphics[width=0.98\textwidth]{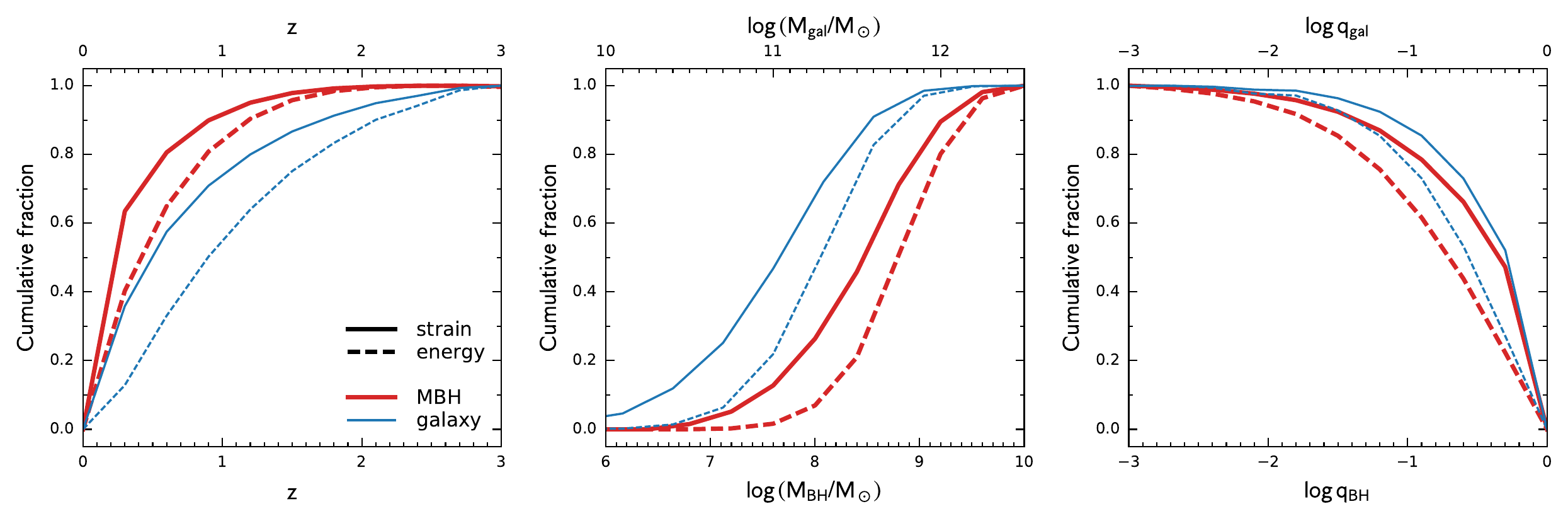}
\caption{
Cumulative fractions of the GWB energy density (dashed) and the characteristic
strain amplitude (solid) at the observer-rest frequency $f=1\pyr$. The left,
the middle, and the right panels show the fractions contributed by
galaxy mergers or BBH mergers at redshift $<z$,
total stellar mass $<M\gal$ or BBH mass $<M\bh$, mass ratios $>q\gal$
or $>q\bh$ (right), respectively. The blue curves represent galaxy mergers,
and the red curves represent BBH mergers.
The $M\bh{-}M\bulge$ relation of \citet{Kormendy13} is adopted in this figure.
As seen from the figure,
most of the GWB energy density is contributed by galaxy or BBH mergers within
redshift lower than 2, with total galaxy mass within $\sim
10^{10}$--$10^{12}\msun$ or BBH mass within $\sim 10^{8}$--$10^{10}\msun$, and
with merging galaxy or BH mass ratio greater than $\sim 0.01$.
See Section~\ref{sec:res:gwb}.
}
\label{fig:gwe_contr}
\end{figure*}

\begin{figure*}
\centering
\includegraphics[width=0.45\textwidth]{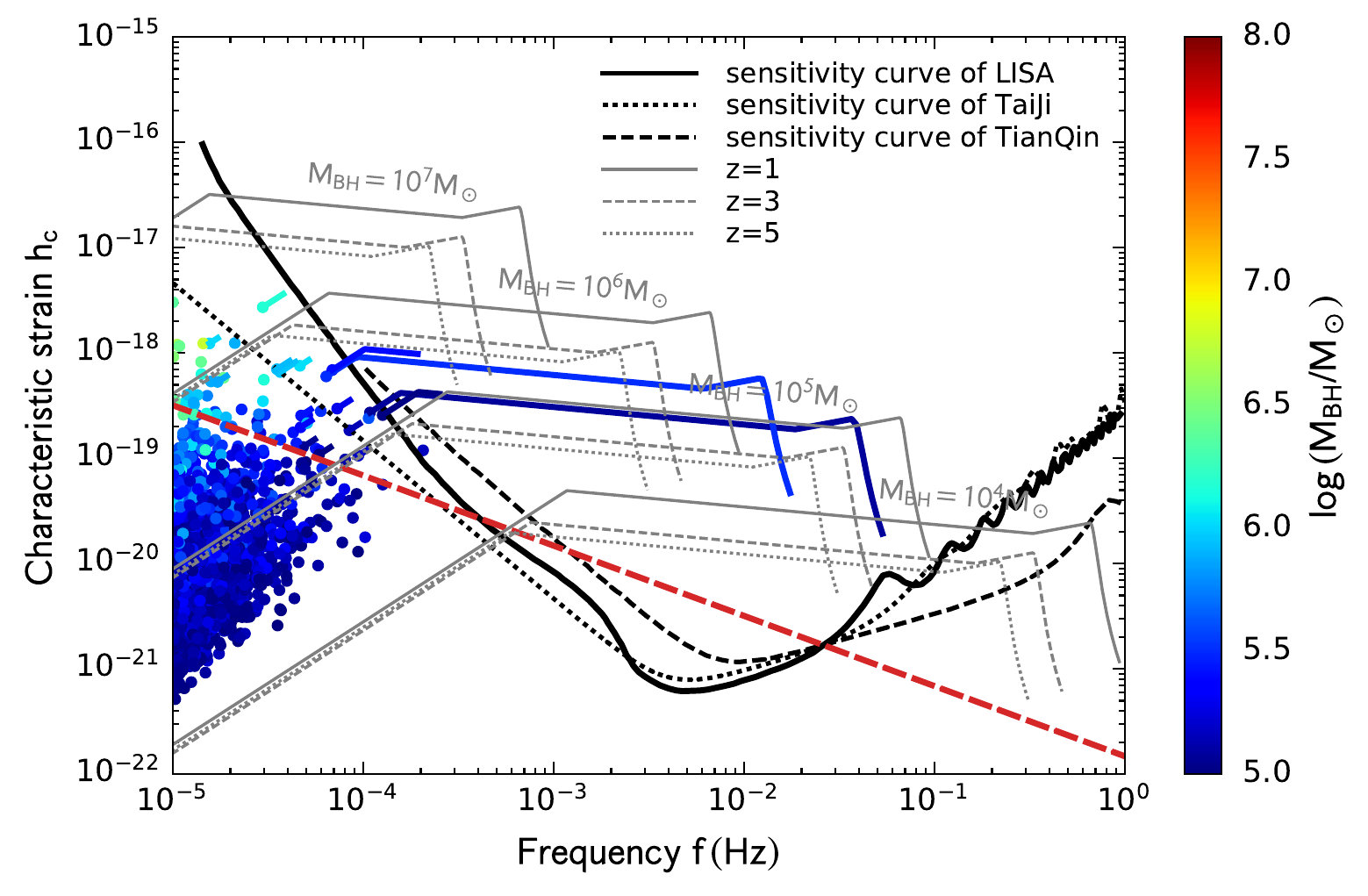}
\includegraphics[width=0.45\textwidth]{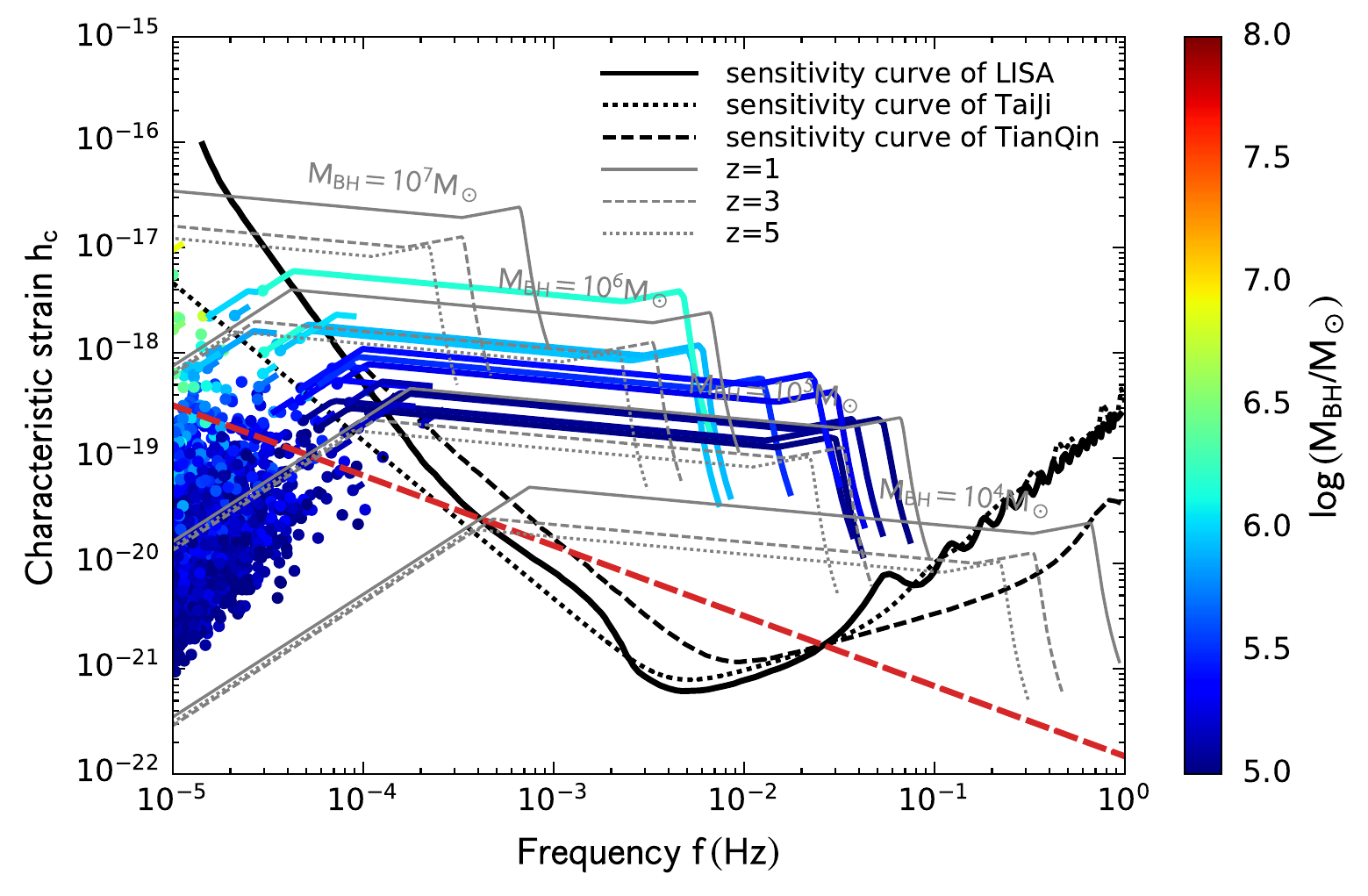}
\includegraphics[width=0.45\textwidth]{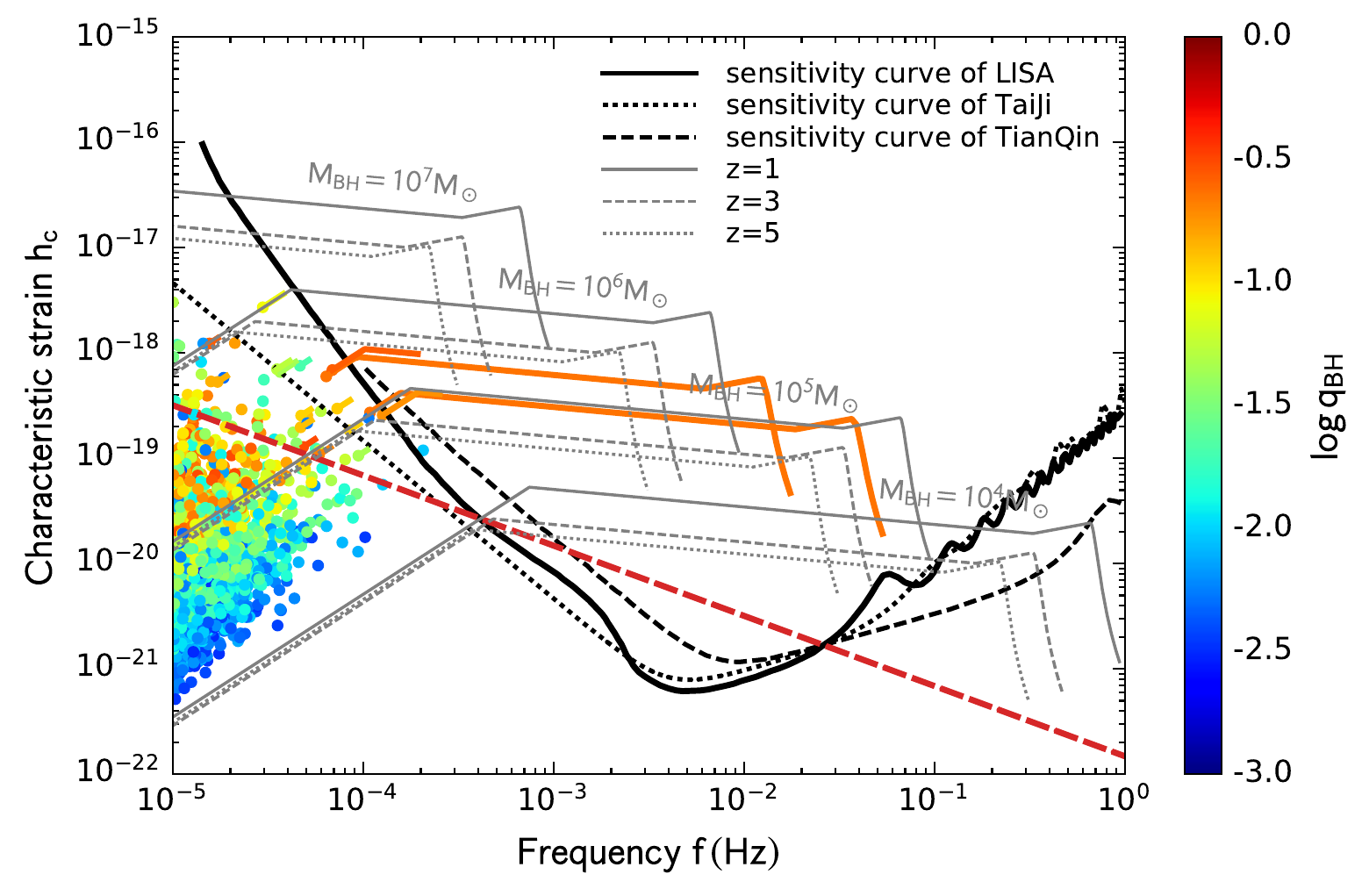}
\includegraphics[width=0.45\textwidth]{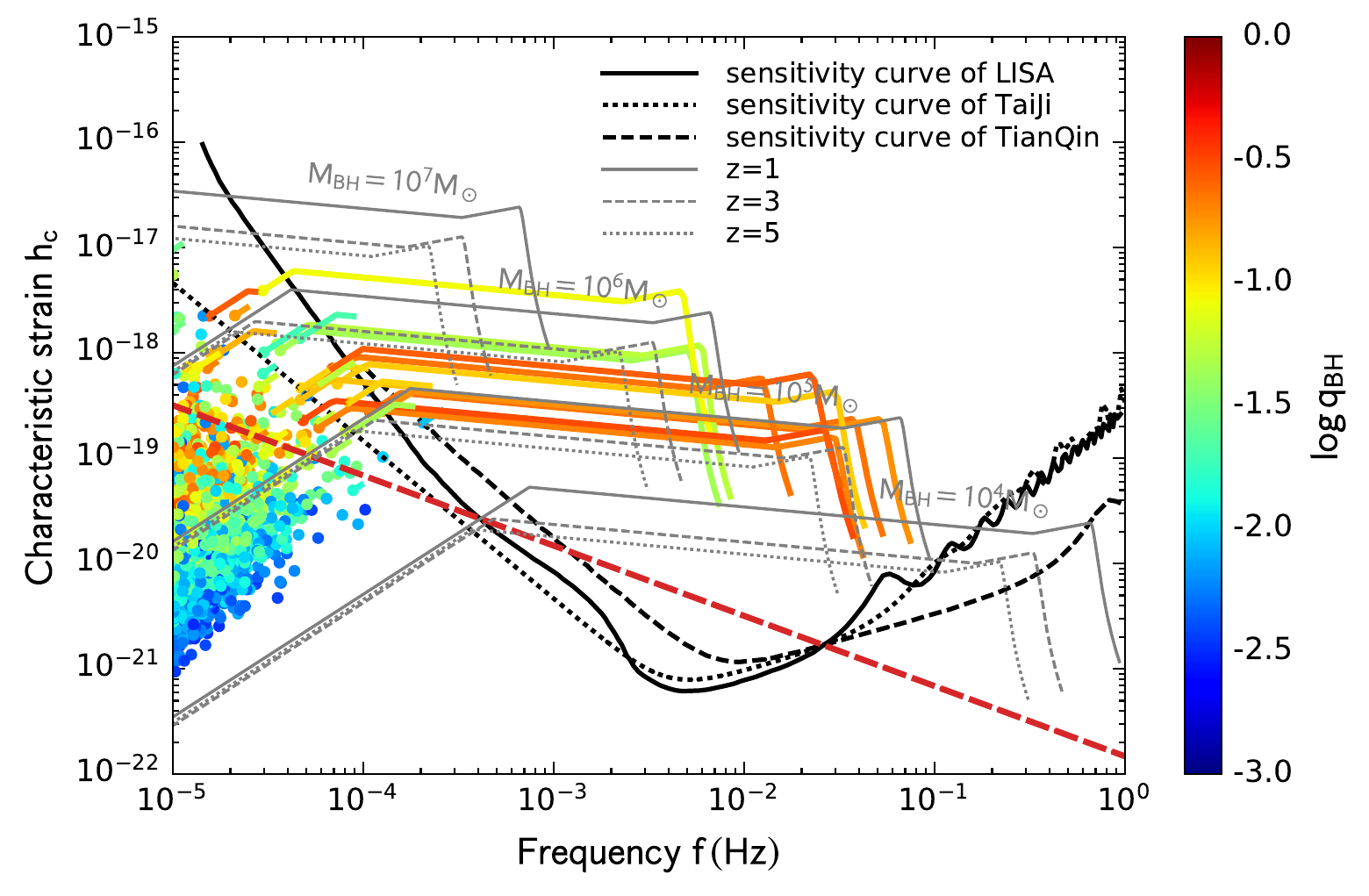}
\includegraphics[width=0.45\textwidth]{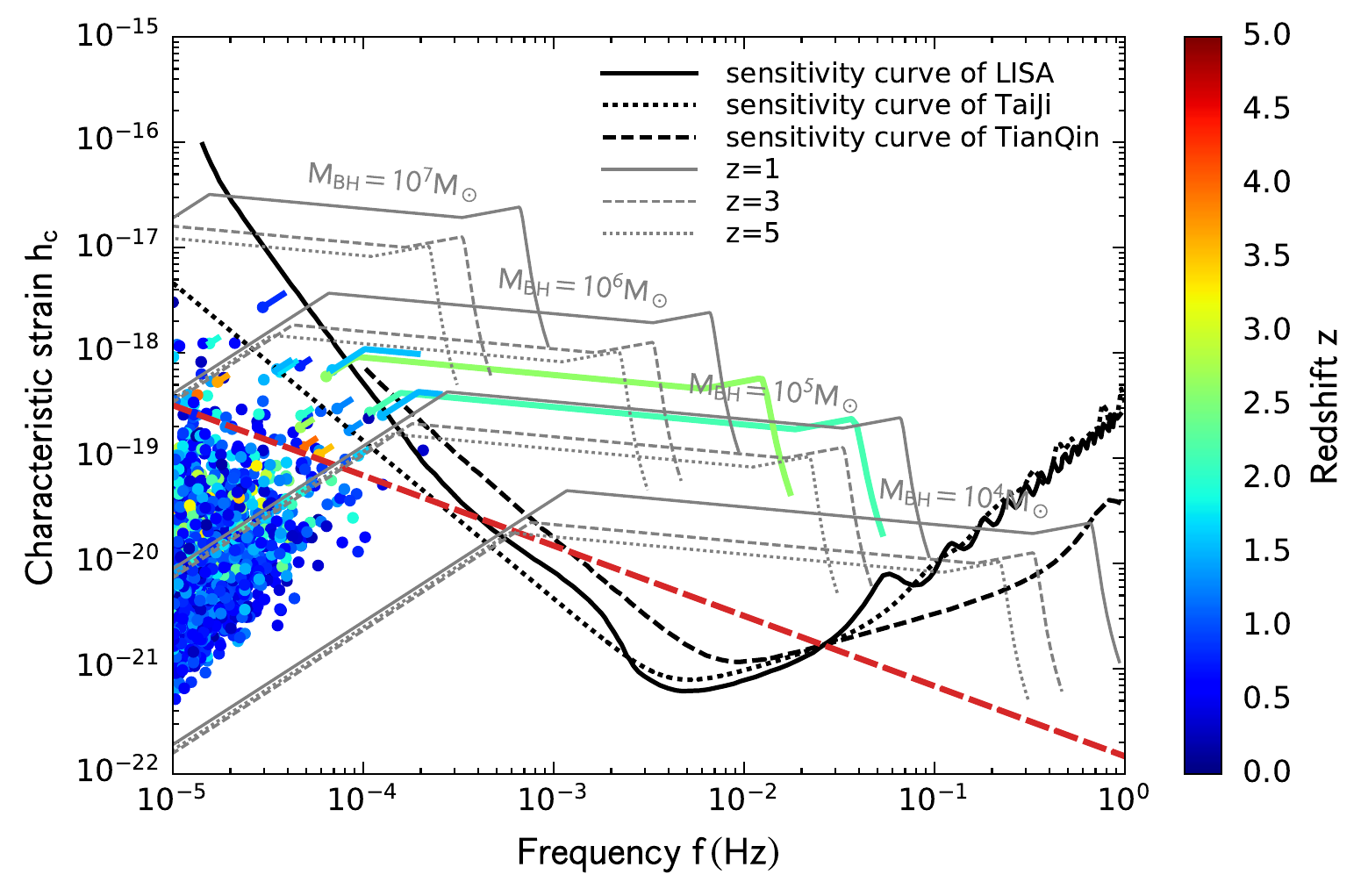}
\includegraphics[width=0.45\textwidth]{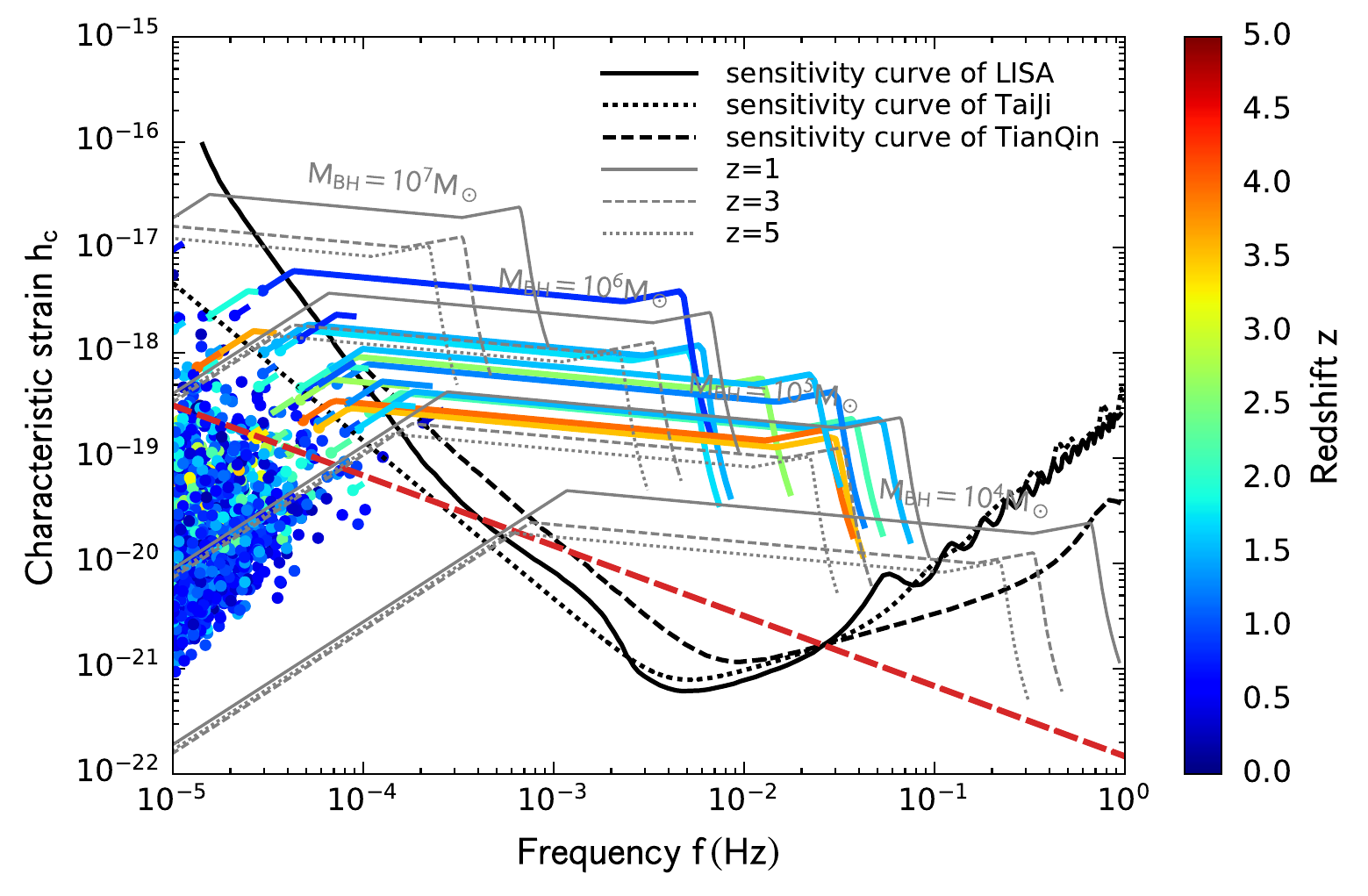}
\caption{Demonstration of prospects for the detection of merging BBHs
around the LISA frequency band. The black solid curve 
is the sensitivity curve of the LISA detector (shown in Figure 1 of
\citealt{Amaro-Seoane17}), the black dotted curve is for Taiji \citep{Ruanetal18}, and the black dashed curve is for Tianqin \citep{Wangetal19}.
The red dashed line represents a reference to the characteristic strain
spectrum of the stochastic background obtained by applying
Equation~(\ref{eq:hc2}) in our model and extending the involved model gradients
to the LISA band.
Also as a reference, the grey curves illustrate the evolution tracks of some
example merging BBH systems in the GW strain-frequency space.  The example BBHs
have the same mass ratio $q\bh=0.1$, different total masses as labeled on the
curves, and different redshifts shown by different line styles and labeled by
the texts.  The GW characteristic strains of individual BBH sources generated
from one realization of the BBH population in our model are shown as color
points at the beginning, with the color scales representing the BBH total
masses, mass ratios, and redshifts in the top, middle, and bottom panels,
respectively. The color solid lines start from the color points and correspond
to the evolution tracks of the same sources during three years' observation in
the left panels and during ten years' observation in the right panels.
The $M\bh{-}M\bulge$ relation of \citet{Kormendy13} is adopted in this figure.
The detection rate demonstrated in this figure is $\sim 0.9\pyr$.
See Section~\ref{sec:res:lisa}. } \label{fig:strn_lisa} \end{figure*}

\begin{deluxetable}{lccccccccccl}
\tablecaption{The BH--host galaxy relations \label{tab:scal}}
\tablewidth{0pt}
\tablehead{\colhead{Abbr} & \colhead{$\tilde{\alpha}$} &
\colhead{$\tilde{\beta}$} & \colhead{$\tilde{\gamma}$} &
\colhead{$\tilde{\epsilon}$} & \colhead{$\log(A_{\rm yr}^{\rm pl})$} &
\colhead{$\log(\ayr)$} & \colhead{Drop} & \colhead{$\log(f_{\rm turn}/{\rm Hz})$} &
\colhead{$\kappa\gw$} & \colhead{$\gamma\gw$} & \colhead{Reference}\\
 & & & & & & & \colhead{dex} & & & & }
\startdata
GE00a &  --  & 3.75 & 8.08 & 0.30 & -15.69 & -15.80 & 0.11 & -9.28 & 3.27 & 0.27 & \citet{Gebhardt00} \\
FM00a &  --  & 4.80 & 8.14 & 0.30 & -15.71 & -15.85 & 0.14 & -9.87 & 4.57 & 0.12 & \citet{Ferrarese00} \\
TR02a &  --  & 4.02 & 8.13 & 0.30 & -15.67 & -15.79 & 0.12 & -9.48 & 3.65 & 0.20 & \citet{Tremaine02} \\
GU09a &  --  & 3.96 & 8.23 & 0.31 & -15.58 & {\bf -15.70} & 0.13 & -9.54 & 3.63& 0.21 & \citet{Gultekin09} \\
SA11a &  --  & 4.00 & 8.29 & 0.33 & -15.52 & -15.65 & 0.13 & -9.60 & 3.78 & 0.19 & \citet{Sani11} \\
GR11a &  --  & 5.13 & 8.13 & 0.34 & -15.71 & -15.85 & 0.14 & -9.84 & 3.94 & 0.17 & \citet{Graham11} \\
BE12a &  --  & 4.42 & 7.99 & 0.36 & -15.79 & -15.91 & 0.12 & -9.61 & 4.18 & 0.15 & \citet{Beifiori12} \\
MM13a &  --  & 5.64 & 8.32 & 0.38 & -15.56 & -15.73 & 0.17 & -9.93 & 3.96 & 0.17 & \citet{McConnell13} \\
GR13a &  --  & 6.08 & 8.15 & 0.41 & -15.68 & -15.86 & 0.18 & -9.84 & 3.38 & 0.25 & \citet{Graham13} \\
KH13a &  --  & 4.42 & 8.50 & 0.29 & -15.40 & -15.55 & 0.15 & -9.90 & 4.25 & 0.15 & \citet{Kormendy13} \\
BO16a &  --  & 5.35 & 8.32 & 0.49 & -15.51 & -15.69 & 0.18 & -9.98 & 3.99 & 0.17 & \citet{van-den-Bosch16} \\
HR04b & 1.12 &  --  & 8.20 & 0.30 & -15.50 & -15.62 & 0.12 & -9.50 & 3.72 & 0.20 & \citet{Haring04} \\
SA11b & 0.79 &  --  & 8.20 & 0.37 & -15.30 & -15.42 & 0.12 & -9.41 & 3.59 & 0.21 & \citet{Sani11} \\
BE12b & 0.91 &  --  & 7.84 & 0.46 & -15.62 & -15.73 & 0.11 & -9.36 & 3.64 & 0.19 & \citet{Beifiori12} \\
MM13b & 1.05 &  --  & 8.46 & 0.34 & -15.24 & -15.38 & 0.14 & -9.72 & {\bf 3.74}& 0.20 & \citet{McConnell13} \\
KH13b & 1.17 &  --  & 8.69 & 0.29 & -15.11 & -15.28 & 0.17 & -9.94 & 3.70 & 0.19 & \citet{Kormendy13} \\
BO16b & 1.21 &  --  & 8.33 & 0.49 & -15.31 & -15.47 & 0.16 & -9.89 & 3.74 & 0.19 & \citet{van-den-Bosch16} \\
SH16c & 0.50 & 4.50 & 7.70 & 0.25 & -16.09 & -16.22 & 0.14 & -9.48 & 3.38 & 0.24 & \citet{Shankar16}
\enddata
\tablecomments{The BH--host galaxy relations in different works and the 
characteristic strain amplitudes of the stochastic GWBs obtained with the relations.  The
BH--host galaxy relations are described by the parameters $\tilde{\alpha}$,
$\tilde{\beta}$, $\tilde{\gamma}$ (see Eq.~\ref{eq:scal}), and the intrinsic
scatter of the relation $\tilde{\epsilon}$ (see Section~\ref{sec:data:msigma}),
and the references to the relations are listed in the last column of the table.
Regarding the relation in \citet{Ferrarese00} (FM00a), an intrinsic scatter is
not provided in the original work; and here a scatter of $0.3{\rm\, dex}$ is
assumed, as listed in the table.
The $\log(\ayr)$ and
$\log(A_{\rm yr}^{\rm pl})$ represent the strain amplitudes at $f=1\pyr$,
obtained with and without including the time delays between galaxy mergers and
BBH coalescences in our model, respectively.  The column of
``Drop'' gives the values of $\log(A_{\rm yr}^{\rm pl})-\log(\ayr)$.
The
parameter $f_{\rm turn}$ is the turnover frequency of the obtained GWB spectrum,
$\kappa\gw$ is the power describing the spectrum at the low-frequency end of
the spectrum, and $\gamma\gw$ is a parameter to describe the shape transition of the
GWB spectrum around the turnover frequency (see Eqs.~\ref{eq:fbendkappa} and \ref{eq:fturn}). 
The numbers in bold are about the medians of $\log(\ayr)$ and $\kappa\gw$, respectively. 
}
\end{deluxetable}


\begin{thebibliography}{0}
\expandafter\ifx\csname natexlab\endcsname\relax\def\natexlab#1{#1}\fi

\bibitem[Aggarwal et al.(2018)]{Aggarwal18}
Aggarwal, K., Arzoumanian, Z., Baker, P.~T., et al.\ 2018,
\href{https://ui.adsabs.harvard.edu/abs/2018arXiv181211585A}
{arXiv e-prints , arXiv:1812.11585.}

\bibitem[Amaro-Seoane et al.(2010)]{Amaro-Seoane10}
Amaro-Seoane, P., Eichhorn, C., Porter, E.~K., \& Spurzem, R.\ 2010,
\href{http://adsabs.harvard.edu/abs/2010MNRAS.401.2268A}
{\mnras, 401, 2268}

\bibitem[Amaro-Seoane et al.(2017)]{Amaro-Seoane17}
Amaro-Seoane, P., Audley, H., Babak, S., et al.\ 2017,
\href{https://ui.adsabs.harvard.edu/#abs/arXiv:1702.00786}
{arXiv e-prints , arXiv:1702.00786.}

\bibitem[Arzoumanian et al.(2014)]{Arzoumanian14}
Arzoumanian, Z., Brazier, A., Burke-Spolaor, S., et al.\ 2014,
\href{http://adsabs.harvard.edu/abs/2014ApJ...794..141A}
{\apj, 794, 141}

\bibitem[Arzoumanian et al.(2016)]{Arzoumanian16}
Arzoumanian, Z., Brazier, A., Burke-Spolaor, S., et al.\ 2016,
\href{http://adsabs.harvard.edu/abs/2016ApJ...821...13A}
{\apj, 821, 13}

\bibitem[Arzoumanian et al.(2018)]{Arzoumanian18}
Arzoumanian, Z., Baker, P.~T., Brazier, A., et al.\ 2018,
\href{http://adsabs.harvard.edu/abs/2018ApJ...859...47A}
{\apj, 859, 47.}

\bibitem[Babak \& Sesana(2012)]{Babak12}
Babak, S., \& Sesana, A.\ 2012,
\href{http://adsabs.harvard.edu/abs/2012PhRvD..85d4034B}
{\prd, 85, 044034}

\bibitem[Babak et al.(2016)]{Babak16}
Babak, S., Petiteau, A., Sesana, A., et al.\ 2016,
\href{http://adsabs.harvard.edu/abs/2016MNRAS.455.1665B}
{\mnras, 455, 1665}

\bibitem[Baker(2006)]{Baker06}
Baker, R.~M.~L., Jr.\ 2006,
\href{http://adsabs.harvard.edu/abs/2006AN....327..710B}
{Astronomische Nachrichten, 327, 710}

\bibitem[Begelman et al.(1980)]{Begelman80}
Begelman, M.~C., Blandford, R.~D., \& Rees, M.~J.\ 1980,
\href{http://adsabs.harvard.edu/abs/1980Natur.287..307B}
{\nat, 287, 307}

\bibitem[Behroozi et al.(2013)]{Behroozi13}
Behroozi, P.~S., Wechsler, R.~H., \& Conroy, C.\ 2013,
\href{http://adsabs.harvard.edu/abs/2013ApJ...770...57B}
{\apj, 770, 57}

\bibitem[Behroozi et al.(2019)]{Behroozi19}
Behroozi, P., Wechsler, R.~H., Hearin, A.~P., et al.\ 2019,
\href{https://ui.adsabs.harvard.edu/abs/2019MNRAS.488.3143B}
{\mnras, 488, 3143}

\bibitem[Beifiori et al.(2012)]{Beifiori12}
Beifiori, A., Courteau, S., Corsini, E.~M., \& Zhu, Y.\ 2012,
\href{http://adsabs.harvard.edu/abs/2012MNRAS.419.2497B}
{\mnras, 419, 2497}

\bibitem[Benson et al.(2010)]{Benson10}
Benson, A., Holley-Bockelmann, K., \& Gultekin, K.\ 2010,
\href{http://adsabs.harvard.edu/abs/2010AAS...21540416B}
{Bulletin of the American Astronomical Society, 42, 404.16 }

\bibitem[Berczik et al.(2006)]{Berczik06}
Berczik, P., Merritt, D., Spurzem, R., \& Bischof, H.-P.\ 2006,
\href{http://adsabs.harvard.edu/abs/2006ApJ...642L..21B}
{\apjl, 642, L21}

\bibitem[Bernardi et al.(2013)]{Bernardi13}
Bernardi, M., Meert, A., Sheth, R.~K., et al.\ 2013,
\href{http://adsabs.harvard.edu/abs/2013MNRAS.436..697B}
{\mnras, 436, 697}

\bibitem[Bevington \& Robinson(2003)]{BR03}
Bevington, P.~R., \& Robinson, D.~K.\ 2003, Data reduction and error analysis
for the physical sciences, 3rd ed., by Philip R.~Bevington, and Keith
D.~Robinson.~Boston, MA: McGraw-Hill, ISBN 0-07-247227-8, 2003

\bibitem[Binney \& Merrifield(1998)]{BM98}
Binney, J., \& Merrifield, M.\ 1998, Galactic astronomy / James Binney and
Michael Merrifield. Princeton

\bibitem[Binney \& Tremaine(2008)]{BT08}
Binney, J., \& Tremaine, S.\ 2008, Galactic Dynamics: Second Edition,
by James Binney and Scott Tremaine.~ISBN 978-0-691-13026-2
(HB).~Published by Princeton University Press, Princeton, NJ USA,
2008.

\bibitem[Bonetti et al.(2018)]{Bonetti18}
Bonetti, M., Sesana, A., Barausse, E., et al.\ 2018,
\href{https://ui.adsabs.harvard.edu/#abs/arXiv:1709.06095}
{\mnras, 477, 2599.}

\bibitem[Boyle \& Pen(2012)]{Boyle12}
Boyle, L., \& Pen, U.-L.\ 2012,
\href{http://adsabs.harvard.edu/abs/2012PhRvD..86l4028B}
{\prd, 86, 124028}

\bibitem[Bundy et al.(2009)]{Bundy09}
Bundy, K., Fukugita, M., Ellis, R.~S., et al.\ 2009,
\href{http://adsabs.harvard.edu/abs/2009ApJ...697.1369B}
{\apj, 697, 1369}

\bibitem[Cappellari et al.(2011)]{Cappellari11}
Cappellari, M., Emsellem, E., Krajnovi{\'c}, D., et al.\ 2011,
\href{http://adsabs.harvard.edu/abs/2011MNRAS.413..813C}
{\mnras, 413, 813}

\bibitem[Cappellari et al.(2013)]{Cappellari13}
Cappellari, M., Scott, N., Alatalo, K., et al.\ 2013,
\href{http://adsabs.harvard.edu/abs/2013MNRAS.432.1709C}
{\mnras, 432, 1709}

\bibitem[Chen et al.(2016)]{Chen16}
Chen, S., Sesana, A., \& Del Pozzo, W.\ 2016,
\href{https://arxiv.org/abs/1612.00455}
{arXiv:1612.00455}

\bibitem[Conselice et al.(2009)]{Conselice09}
Conselice, C.~J., Yang, C., \& Bluck, A.~F.~L.\ 2009,
\href{http://adsabs.harvard.edu/abs/2009MNRAS.394.1956C}
{\mnras, 394, 1956}

\bibitem[Conselice et al.(2014)]{Conselice14}
Conselice, C.~J., Bluck, A.~F.~L., Mortlock, A., Palamara, D.,
\& Benson, A.~J.\ 2014,
\href{http://adsabs.harvard.edu/abs/2014MNRAS.444.1125C}
{\mnras, 444, 1125}

\bibitem[Cui \& Yu(2014)]{Cui14}
Cui, X., \& Yu, Q.\ 2014,
\href{http://adsabs.harvard.edu/abs/2014MNRAS.437..777C}
{\mnras, 437, 777}

\bibitem[de Ravel et al.(2009)]{de-Ravel09}
de Ravel, L., Le F{\`e}vre, O., Tresse, L., et al.\ 2009,
\href{http://adsabs.harvard.edu/abs/2009A%26A...498..379D}
{\aap, 498, 379}

\bibitem[Desvignes et al.(2016)]{Desvignes16}
Desvignes, G., Caballero, R.~N., Lentati, L., et al.\ 2016,
\href{http://adsabs.harvard.edu/abs/2016MNRAS.458.3341D}
{\mnras, 458, 3341}

\bibitem[Ellis et al.(2012)]{Ellis12}
Ellis, J.~A., Siemens, X., \& Creighton, J.~D.~E.\ 2012,
\href{https://ui.adsabs.harvard.edu/abs/2012ApJ...756..175E}
{\apj, 756, 175}

\bibitem[Emsellem et al.(2007)]{Emsellem07}
Emsellem, E., Cappellari, M., Krajnovi{\'c}, D., et al.\ 2007,
\href{http://adsabs.harvard.edu/abs/2007MNRAS.379..401E}
{\mnras, 379, 401}

\bibitem[Ferrarese \& Merritt(2000)]{Ferrarese00}
Ferrarese, L., \& Merritt, D.\ 2000,
\href{https://ui.adsabs.harvard.edu/abs/2000ApJ...539L...9F}
{\apjl, 539, L9}

\bibitem[Gallazzi et al.(2006)]{Gallazzi06}
Gallazzi, A., Charlot, S., Brinchmann, J., \& White, S.~D.~M.\ 2006,
\href{http://adsabs.harvard.edu/abs/2006MNRAS.370.1106G}
{\mnras, 370, 1106}

\bibitem[Gebhardt et al.(2000)]{Gebhardt00}
Gebhardt, K., Bender, R., Bower, G., et al.\ 2000,
\href{https://ui.adsabs.harvard.edu/abs/2000ApJ...539L..13G}
{\apjl, 539, L13}

\bibitem[Graham et al.(2011)]{Graham11}
Graham, A.~W., Onken, C.~A., Athanassoula, E., \& Combes, F.\ 2011,
\href{http://adsabs.harvard.edu/abs/2011MNRAS.412.2211G}
{\mnras, 412, 2211}

\bibitem[Graham \& Scott(2013)]{Graham13}
Graham, A.~W., \& Scott, N.\ 2013,
\href{http://adsabs.harvard.edu/abs/2013ApJ...764..151G}
{\apj, 764, 151}

\bibitem[Graham et al.(2015)]{Graham15}
Graham, M.~J., Djorgovski, S.~G., Stern, D., et al.\ 2015,
\href{http://adsabs.harvard.edu/abs/2015Natur.518...74G}
{\nat, 518, 74}

\bibitem[G{\"u}ltekin et al.(2009)]{Gultekin09}
G{\"u}ltekin, K., Richstone, D.~O., Gebhardt, K., et al.\ 2009,
\href{http://adsabs.harvard.edu/abs/2009ApJ...698..198G}
{\apj, 698, 198}

\bibitem[Guo et al.(2011)]{Guo11}
Guo, Q., White, S., Boylan-Kolchin, M., et al.\ 2011,
\href{https://ui.adsabs.harvard.edu/abs/2011MNRAS.413..101G}
{\mnras, 413, 101}

\bibitem[Haiman et al.(2009)]{Haimanetal09}
Haiman, Z., Kocsis, B., \& Menou, K.\ 2009, \apj, 700, 1952

\bibitem[H{\"a}ring \& Rix(2004)]{Haring04}
H{\"a}ring, N., \& Rix, H.-W.\ 2004,
\href{http://adsabs.harvard.edu/abs/2004ApJ...604L..89H}
{\apjl, 604, L89}

\bibitem[Heggie(1975)]{Heggie75}
Heggie, D.~C.\ 1975,
\href{http://adsabs.harvard.edu/abs/1975MNRAS.173..729H}
{\mnras, 173, 729}

\bibitem[Hinshaw et al.(2013)]{Hinshawetal13}
Hinshaw G.\ et al., 2013, ApJS,
208, 19

\bibitem[Hobbs(2013)]{Hobbs13}
Hobbs, G.\ 2013,
\href{http://adsabs.harvard.edu/abs/2013CQGra..30v4007H}
{Classical and Quantum Gravity, 30, 224007}

\bibitem[Holley-Bockelmann \& Khan(2015)]{Holley-Bockelmann15}
Holley-Bockelmann, K., \& Khan, F.~M.\ 2015,
\href{http://adsabs.harvard.edu/abs/2015ApJ...810..139H}
{\apj, 810, 139}

\bibitem[Hopkins et al.(2010a)]{Hopkins10a}
Hopkins, P.~F., Bundy, K., Croton, D., et al.\ 2010,
\href{http://adsabs.harvard.edu/abs/2010ApJ...715..202H}
{\apj, 715, 202}

\bibitem[Hopkins et al.(2010b)]{Hopkins10b}
Hopkins, P.~F., Croton, D., Bundy, K., et al.\ 2010,
\href{http://adsabs.harvard.edu/abs/2010ApJ...724..915H}
{\apj, 724, 915}

\bibitem[Ilbert et al.(2013)]{Ilbert13}
Ilbert, O., McCracken, H.~J., Le F{\`e}vre, O., et al.\ 2013,
\href{http://ads.bao.ac.cn/abs/2013A%26A...556A..55I}
{\aap, 556, A55}

\bibitem[Katz et al.(2019)]{Katz19}
Katz, M.~L., Kelley, L.~Z., Dosopoulou, F., et al.\ 2019,
\href{https://ui.adsabs.harvard.edu/abs/2019MNRAS.tmp.2700K}
{\mnras, 2700}

\bibitem[Kelley et al.(2017)]{Kelley17}
Kelley, L.~Z., Blecha, L., Hernquist, L., Sesana, A., \& Taylor, S.~R.\ 2017,
\href{http://adsabs.harvard.edu/abs/2017MNRAS.471.4508K}
\mnras, 471, 4508

\bibitem[Kelley et al.(2018)]{Kelley18}
Kelley, L.~Z., Blecha, L., Hernquist, L., Sesana, A.,
\& Taylor, S.~R.\ 2018,
\href{http://adsabs.harvard.edu/abs/2018MNRAS.477..964K}
{\mnras, 477, 964}

\bibitem[Kelvin et al.(2014)]{Kelvin14}
Kelvin, L.~S., Driver, S.~P., Robotham, A.~S.~G., et al.\ 2014,
\href{http://adsabs.harvard.edu/abs/2014MNRAS.444.1647K}
{\mnras, 444, 1647}

\bibitem[Khan et al.(2011)]{Khan11}
Khan, F.~M., Just, A., \& Merritt, D.\ 2011,
\href{http://adsabs.harvard.edu/abs/2011ApJ...732...89K}
{\apj, 732, 89}

\bibitem[Khan et al.(2013)]{Khan13}
Khan, F.~M., Holley-Bockelmann, K., Berczik, P., \& Just, A.\ 2013,
\href{http://adsabs.harvard.edu/abs/2013ApJ...773..100K}
{\apj, 773, 100}

\bibitem[Kitzbichler \& White(2008)]{Kitzbichler08}
Kitzbichler, M.~G., \& White, S.~D.~M.\ 2008,
\href{http://adsabs.harvard.edu/abs/2008MNRAS.391.1489K}
{\mnras, 391, 1489}

\bibitem[Kormendy \& Ho(2013)]{Kormendy13}
Kormendy, J., \& Ho, L.~C.\ 2013,
\href{http://adsabs.harvard.edu/abs/2013ARA%26A..51..511K}
{\araa, 51, 511}

\bibitem[Krajnovi{\'c} et al.(2011)]{Krajnovic11}
Krajnovi{\'c}, D., Emsellem, E., Cappellari, M., et al.\ 2011,
\href{http://adsabs.harvard.edu/abs/2011MNRAS.414.2923K}
{\mnras, 414, 2923}

\bibitem[Krajnovi{\'c} et al.(2013)]{Krajnovic13}
Krajnovi{\'c}, D., Karick, A.~M., Davies, R.~L., et al.\ 2013,
\href{http://adsabs.harvard.edu/abs/2013MNRAS.433.2812K}
{\mnras, 433, 2812}

\bibitem[Kulier et al.(2015)]{Kulier15}
Kulier, A., Ostriker, J.~P., Natarajan, P., et al.\ 2015,
\href{https://ui.adsabs.harvard.edu/abs/2015ApJ...799..178K}
{\apj, 799, 178}

\bibitem[Lackner et al.(2012)]{Lackner12}
Lackner, C.~N., Cen, R., Ostriker, J.~P., et al.\ 2012,
\href{https://ui.adsabs.harvard.edu/abs/2012MNRAS.425..641L}
{\mnras, 425, 641}

\bibitem[Lauer et al.(1995)]{Lauer95}
Lauer, T.~R., Ajhar, E.~A., Byun, Y.-I., et al.\ 1995,
\href{http://adsabs.harvard.edu/abs/1995AJ....110.2622L}
{\aj, 110, 2622}

\bibitem[Lauer et al.(2007a)]{Lauer07a}
Lauer, T.~R., Gebhardt, K., Faber, S.~M., et al.\ 2007,
\href{http://adsabs.harvard.edu/abs/2007ApJ...664..226L}
{\apj, 664, 226}

\bibitem[Lauer et al.(2007b)]{Lauer07b}
Lauer, T.~R., Faber, S.~M., Richstone, D., et al.\ 2007,
\href{http://adsabs.harvard.edu/abs/2007ApJ...662..808L}
{\apj, 662, 808}

\bibitem[Lee et al.(2011)]{Leeetal11}
Lee, K.~J., Wex, N., Kramer, M., et al.\ 2011, \mnras, 414, 3251

\bibitem[Lentati et al.(2015)]{Lentati15}
Lentati, L., Taylor, S.~R., Mingarelli, C.~M.~F., et al.\ 2015,
\href{http://adsabs.harvard.edu/abs/2015MNRAS.453.2576L}
{\mnras, 453, 2576}

\bibitem[Lim et al.(2017)]{Lim17}
Lim, S.~H., Mo, H.~J., Lan, T.-W., \& M{\'e}nard, B.\ 2017,
\href{http://adsabs.harvard.edu/abs/2017MNRAS.464.3256L}
{\mnras, 464, 3256}

\bibitem[L{\'o}pez-Sanjuan et al.(2012)]{Lopez12}
L{\'o}pez-Sanjuan, C., Le F{\`e}vre, O., Ilbert, O., et al.\ 2012,
\href{http://adsabs.harvard.edu/abs/2012A%26A...548A...7L}
{\aap, 548, A7}

\bibitem[Lotz et al.(2008)]{Lotz08}
Lotz, J.~M., Jonsson, P., Cox, T.~J., et al.\ 2008,
\href{https://ui.adsabs.harvard.edu/abs/2008MNRAS.391.1137L}
{\mnras, 391, 1137}

\bibitem[Lotz et al.(2010)]{Lotz10}
Lotz, J.~M., Jonsson, P., Cox, T.~J., \& Primack, J.~R.\ 2010,
\href{http://adsabs.harvard.edu/abs/2010MNRAS.404..575L}
{\mnras, 404, 575}

\bibitem[Lotz et al.(2011)]{Lotz11}
Lotz, J.~M., Jonsson, P., Cox, T.~J., et al.\ 2011,
\href{http://adsabs.harvard.edu/abs/2011ApJ...742..103L}
{\apj, 742, 103}

\bibitem[Magorrian et al.(1998)]{Magorrian98}
Magorrian, J., Tremaine, S., Richstone, D., et al.\ 1998,
\href{https://ui.adsabs.harvard.edu/abs/1998AJ....115.2285M}
{\aj, 115, 2285}

\bibitem[Magorrian \& Tremaine(1999)]{Magorrian99}
Magorrian, J., \& Tremaine, S.\ 1999,
\href{http://adsabs.harvard.edu/abs/1999MNRAS.309..447M}
{\mnras, 309, 447}

\bibitem[Mayer et al.(2007)]{Mayeretal07}
Mayer, L., Kazantzidis, S., Madau, P., et al.\ 2007, Science, 316, 1874

\bibitem[McConnell \& Ma(2013)]{McConnell13}
McConnell, N.~J., \& Ma, C.-P.\ 2013,
\href{http://adsabs.harvard.edu/abs/2013ApJ...764..184M}
{\apj, 764, 184}

\bibitem[McLaughlin(2013)]{McLaughlin13}
McLaughlin, M.~A.\ 2013,
\href{http://adsabs.harvard.edu/abs/2013CQGra..30v4008M}
{Classical and Quantum Gravity, 30, 224008}

\bibitem[McWilliams et al.(2014)]{McWilliams14}
McWilliams, S.~T., Ostriker, J.~P., \& Pretorius, F.\ 2014,
\href{http://adsabs.harvard.edu/abs/2014ApJ...789..156M}
{\apj, 789, 156}

\bibitem[Milosavljevi{\'c} \& Merritt(2001)]{Milosavljevic01}
Milosavljevi{\'c}, M., \& Merritt, D.\ 2001,
\href{http://adsabs.harvard.edu/abs/2001ApJ...563...34M}
{\apj, 563, 34}

\bibitem[Mingarelli et al.(2017)]{Mingarelli17}
Mingarelli, C.~M.~F., Lazio, T.~J.~W., Sesana, A., et al.\ 2017,
\href{https://ui.adsabs.harvard.edu/#abs/arXiv:1708.03491}
{Nature Astronomy, 1, 886.}

\bibitem[Moore et al.(2015)]{Moore15}
Moore, C.~J., Cole, R.~H., \& Berry, C.~P.~L.\ 2015,
\href{http://adsabs.harvard.edu/abs/2015CQGra..32a5014M}
{Classical and Quantum Gravity, 32, 015014}

\bibitem[Muzzin et al.(2013)]{Muzzin13}
Muzzin, A., Marchesini, D., Stefanon, M., et al.\ 2013,
\href{http://adsabs.harvard.edu/abs/2013ApJ...777...18M}
{\apj, 777, 18}

\bibitem[NANOGrav Collaboration(2018)]{nanograv18}
NANOGrav Collaboration\ 2018,
\href{https://arxiv.org/abs/1810.06594}
{ArXiv e-prints , arXiv:1810.06594.}

\bibitem[Padilla \& Strauss(2008)]{Padilla08}
Padilla, N.~D., \& Strauss, M.~A.\ 2008,
\href{http://adsabs.harvard.edu/abs/2008MNRAS.388.1321P}
{\mnras, 388, 1321}

\bibitem[Peters \& Mathews(1963)]{Peters63}
Peters, P.~C., \& Mathews, J.\ 1963,
\href{http://adsabs.harvard.edu/abs/1963PhRv..131..435P}
{Physical Review, 131, 435}

\bibitem[Peters(1964)]{Peters64}
Peters, P.~C.\ 1964,
\href{http://adsabs.harvard.edu/abs/1964PhRv..136.1224P}
{Physical Review, 136, 1224}

\bibitem[Phinney(2001)]{Phinney01}
Phinney, E.~S.\ 2001,
\href{http://adsabs.harvard.edu/abs/2001astro.ph..8028P}
{arXiv:astro-ph/0108028}

\bibitem[Planck Collaboration et al.(2016)]{Planck16}
Planck CollaborationXIII,2016,
\href{http://adsabs.harvard.edu/abs/2016A%26A...594A..13P}
{A\&A, 594, A13}

\bibitem[Polnarev \& Rees(1994)]{PR94}
Polnarev, S., \& Rees, M.\ J.\ 1994,
A\&A, 283, 301

\bibitem[Prada et al.(2012)]{Prada12}
Prada, F., Klypin, A.~A., Cuesta, A.~J., et al.\ 2012,
\href{https://ui.adsabs.harvard.edu/abs/2012MNRAS.423.3018P}
{\mnras, 423, 3018}

\bibitem[Preto et al.(2011)]{Preto11}
Preto, M., Berentzen, I., Berczik, P., \& Spurzem, R.\ 2011,
\href{http://adsabs.harvard.edu/abs/2011ApJ...732L..26P}
{\apjl, 732, L2}

\bibitem[Qu et al.(2017)]{Qu17}
Qu, Y., Helly, J.~C., Bower, R.~G., et al.\ 2017,
\href{https://ui.adsabs.harvard.edu/abs/2017MNRAS.464.1659Q}
{\mnras, 464, 1659}

\bibitem[Quinlan(1996)]{Quinlan96}
Quinlan, G.~D.\ 1996,
\href{http://adsabs.harvard.edu/abs/1996NewA....1...35Q}
{\na, 1, 35}

\bibitem[Quinlan \& Hernquist(1997)]{Quinlan97}
Quinlan, G.~D., \& Hernquist, L.\ 1997,
\href{http://adsabs.harvard.edu/abs/1997NewA....2..533Q}
{\na, 2, 533}

\bibitem[Ravi et al.(2012)]{Ravi12}
Ravi, V., Wyithe, J.~S.~B., Hobbs, G., et al.\ 2012,
\href{http://adsabs.harvard.edu/abs/2012ApJ...761...84R}
{\apj, 761, 84}

\bibitem[Ravi et al.(2014)]{Ravi14}
Ravi, V., Wyithe, J.~S.~B., Shannon, R.~M., et al.\ 2014,
\href{https://ui.adsabs.harvard.edu/abs/2014MNRAS.442...56R}
{\mnras, 442, 56}

\bibitem[Ravi et al.(2015)]{Ravi15}
Ravi, V., Wyithe, J.~S.~B., Shannon, R.~M., \& Hobbs, G.\ 2015,
\href{http://adsabs.harvard.edu/abs/2015MNRAS.447.2772R}
{\mnras, 447, 2772}

\bibitem[Riebe et al.(2011)]{Riebe11}
Riebe, K., Partl, A.~M., Enke, H., et al.\ 2011,
\href{https://ui.adsabs.harvard.edu/abs/2011arXiv1109.0003R}
{arXiv e-prints, arXiv:1109.0003}

\bibitem[Robotham et al.(2014)]{Robotham14}
Robotham, A.~S.~G., Driver, S.~P., Davies, L.~J.~M., et al.\ 2014,
\href{http://adsabs.harvard.edu/abs/2014MNRAS.444.3986R}
{\mnras, 444, 3986}

\bibitem[Rodr{\'{\i}}guez \& Padilla(2013)]{Rodriguez13}
Rodr{\'{\i}}guez, S., \& Padilla, N.~D.\ 2013,
\href{http://adsabs.harvard.edu/abs/2013MNRAS.434.2153R}
{\mnras, 434, 2153}

\bibitem[Rodriguez-Gomez et al.(2015)]{Rodriguez-Gomez15}
Rodriguez-Gomez, V., Genel, S., Vogelsberger, M., et al.\ 2015,
\href{http://adsabs.harvard.edu/abs/2015MNRAS.449...49R}
{\mnras, 449, 49}

\bibitem[Roebber et al.(2016)]{Roebber16}
Roebber, E., Holder, G., Holz, D.~E., \& Warren, M.\ 2016,
\href{http://adsabs.harvard.edu/abs/2016ApJ...819..163R}
{\apj, 819, 163}

\bibitem[Rosado et al.(2015)]{Rosado15}
Rosado, P.~A., Sesana, A., \& Gair, J.\ 2015,
\href{http://adsabs.harvard.edu/abs/2015MNRAS.451.2417R}
{\mnras, 451, 2417}

\bibitem[Ruan et al.(2018)]{Ruanetal18} Ruan, W.-H., Guo, Z.-K., Cai,
R.-G., et al.\ 2018, arXiv e-prints, arXiv:1807.09495

\bibitem[Ruan et al.(2019)]{Ruanetal19}
Ruan, W.-H., Liu, C., Guo, Z.-K., et al.\ 2019,
\href{https://ui.adsabs.harvard.edu/abs/2019arXiv190907104R}
{arXiv e-prints, arXiv:1909.07104}

\bibitem[Salcido et al.(2016)]{Salcido16}
Salcido, J., Bower, R.~G., Theuns, T., et al.\ 2016,
\href{https://ui.adsabs.harvard.edu/#abs/arXiv:1601.06156}
{\mnras, 463, 870.}

\bibitem[Sani et al.(2011)]{Sani11}
Sani, E., Marconi, A., Hunt, L.~K., \& Risaliti, G.\ 2011,
\href{http://adsabs.harvard.edu/abs/2011MNRAS.413.1479S}
{\mnras, 413, 1479}

\bibitem[Schaye et al.(2015)]{Schaye15}
Schaye, J., Crain, R.~A., Bower, R.~G., et al.\ 2015,
\href{http://adsabs.harvard.edu/abs/2015MNRAS.446..521S}
{\mnras, 446, 521}

\bibitem[Schutz(2011)]{Schutz11}
Schutz, B.~F.\ 2011,
\href{http://adsabs.harvard.edu/abs/2011CQGra..28l5023S}
{Classical and Quantum Gravity, 28, 125023}

\bibitem[Scott et al.(2013)]{Scott13}
Scott, N., Graham, A.~W., \& Schombert, J.\ 2013,
\href{http://adsabs.harvard.edu/abs/2013ApJ...768...76S}
{\apj, 768, 76}

\bibitem[Sesana et al.(2007)]{Sesana07}
Sesana, A., Volonteri, M., \& Haardt, F.\ 2007,
\href{http://adsabs.harvard.edu/abs/2007MNRAS.377.1711S}
{\mnras, 377, 1711}

\bibitem[Sesana et al.(2009)]{Sesana09}
Sesana, A., Vecchio, A., \& Volonteri, M.\ 2009,
\href{http://adsabs.harvard.edu/abs/2009MNRAS.394.2255S}
{\mnras, 394, 2255}

\bibitem[Sesana et al.(2011)]{Sesana11}
Sesana, A., Gualandris, A., \& Dotti, M.\ 2011,
\href{http://adsabs.harvard.edu/abs/2011MNRAS.415L..35S}
{\mnras, 415, L35}

\bibitem[Sesana(2013)]{Sesana13}
Sesana, A.\ 2013,
\href{http://adsabs.harvard.edu/abs/2013MNRAS.433L...1S}
{\mnras, 433, L1}

\bibitem[Sesana et al.(2016)]{Sesana16}
Sesana, A., Shankar, F., Bernardi, M., \& Sheth, R.~K.\ 2016,
\href{https://arxiv.org/abs/1603.09348}
{arXiv:1603.09348}

\bibitem[Shankar et al.(2016)]{Shankar16}
Shankar, F., Bernardi, M., Sheth, R.~K., et al.\ 2016,
\href{http://adsabs.harvard.edu/abs/2016MNRAS.tmp..465S}
{\mnras, 460, 3119}

\bibitem[Shannon et al.(2015)]{Shannon15}
Shannon, R.~M., Ravi, V., Lentati, L.~T., et al.\ 2015,
\href{http://adsabs.harvard.edu/abs/2015Sci...349.1522S}
{Science, 349, 1522}

\bibitem[Sijacki et al.(2015)]{Sijacki15}
Sijacki, D., Vogelsberger, M., Genel, S., et al.\ 2015,
\href{http://adsabs.harvard.edu/abs/2015MNRAS.452..575S}
{\mnras, 452, 575}

\bibitem[Snyder et al.(2017)]{Snyder17}
Snyder, G.~F., Lotz, J.~M., Rodriguez-Gomez, V., et al.\ 2017,
\href{https://ui.adsabs.harvard.edu/abs/2017MNRAS.468..207S}
{\mnras, 468, 207}

\bibitem[Somerville et al.(2008)]{Somerville08}
Somerville, R.~S., Hopkins, P.~F., Cox, T.~J., Robertson, B.~E.,
\& Hernquist, L.\ 2008,
\href{http://adsabs.harvard.edu/abs/2008MNRAS.391..481S}
{\mnras, 391, 481}

\bibitem[Springel(2005)]{Springel05}
Springel, V.\ 2005,
\href{https://ui.adsabs.harvard.edu/abs/2005MNRAS.364.1105S}
{\mnras, 364, 1105}

\bibitem[Stone \& Metzger(2016)]{Stone16}
Stone, N.~C., \& Metzger, B.~D.\ 2016,
\href{http://adsabs.harvard.edu/abs/2016MNRAS.455..859S}
{\mnras, 455, 859}

\bibitem[Tomczak et al.(2014)]{Tomczak14}
Tomczak, A.~R., Quadri, R.~F., Tran, K.-V.~H., et al.\ 2014,
\href{http://adsabs.harvard.edu/abs/2014ApJ...783...85T}
{\apj, 783, 85}

\bibitem[Torrey et al.(2015)]{Torrey15}
Torrey, P., Wellons, S., Machado, F., et al.\ 2015,
\href{http://adsabs.harvard.edu/abs/2015MNRAS.454.2770T}
{\mnras, 454, 2770}

\bibitem[Tremaine et al.(2002)]{Tremaine02}
Tremaine, S., Gebhardt, K., Bender, R., et al.\ 2002,
\href{https://ui.adsabs.harvard.edu/abs/2002ApJ...574..740T}
{\apj, 574, 740}

\bibitem[Valtonen et al.(2008)]{Valtonen08}
Valtonen, M.~J., Lehto, H.~J., Nilsson, K., et al.\ 2008,
\href{http://adsabs.harvard.edu/abs/2008Natur.452..851V}
{\nat, 452, 851}

\bibitem[van den Bosch(2016)]{van-den-Bosch16}
van den Bosch, R.~C.~E.\ 2016,
\href{https://ui.adsabs.harvard.edu/abs/2016ApJ...831..134V}
{\apj, 831, 134}

\bibitem[Vasiliev et al.(2015)]{Vasiliev15}
Vasiliev, E., Antonini, F., \& Merritt, D.\ 2015,
\href{http://adsabs.harvard.edu/abs/2015ApJ...810...49V}
{\apj, 810, 49}

\bibitem[Verbiest et al.(2016)]{Verbiest16}
Verbiest, J.~P.~W., Lentati, L., Hobbs, G., et al.\ 2016,
\href{http://adsabs.harvard.edu/abs/2016MNRAS.458.1267V}
{\mnras, 458, 1267}

\bibitem[Vincent \& Ryden(2005)]{Vincent05}
Vincent, R.~A., \& Ryden, B.~S.\ 2005,
\href{http://adsabs.harvard.edu/abs/2005ApJ...623..137V}
{\apj, 623, 137}

\bibitem[Vogelsberger et al.(2014)]{Vogelsberger14}
Vogelsberger, M., Genel, S., Springel, V., et al.\ 2014,
\href{http://adsabs.harvard.edu/abs/2014MNRAS.444.1518V}
{\mnras, 444, 1518}

\bibitem[Wang et al.(2019)]{Wangetal19}
Wang, H.-T., Jiang, Z., Sesana, A., et al.\ 2019,
\href{https://ui.adsabs.harvard.edu/abs/2019PhRvD.100d3003W}
{\prd, 100, 043003}

\bibitem[Warren(2013)]{Warren13}
Warren, M.~S.\ 2013,
\href{https://ui.adsabs.harvard.edu/abs/2013arXiv1310.4502W}
{arXiv e-prints,arXiv:1310.4502}

\bibitem[Weijmans et al.(2014)]{Weijmans14}
Weijmans, A.-M., de Zeeuw, P.~T., Emsellem, E., et al.\ 2014,
\href{http://adsabs.harvard.edu/abs/2014MNRAS.444.3340W}
{\mnras, 444, 3340}

\bibitem[Wyithe \& Loeb(2003)]{Wyithe03}
Wyithe, J.~S.~B., \& Loeb, A.\ 2003,
\href{http://adsabs.harvard.edu/abs/2003ApJ...590..691W}
{\apj, 590, 691}

\bibitem[Xu et al.(2012)]{Xu12}
Xu, C.~K., Zhao, Y., Scoville, N., et al.\ 2012,
\href{http://adsabs.harvard.edu/abs/2012ApJ...747...85X}
{\apj, 747, 85}

\bibitem[Yan et al.(2015)]{Yan15}
Yan, C.-S., Lu, Y., Dai, X., \& Yu, Q.\ 2015,
\href{http://adsabs.harvard.edu/abs/2015ApJ...809..117Y}
{\apj, 809, 117}

\bibitem[Yu(2002)]{Yu02}
Yu, Q.\ 2002,
\href{http://adsabs.harvard.edu/abs/2002MNRAS.331..935Y}
{\mnras, 331, 935}

\bibitem[Zhu et al.(2014)]{Zhu14}
Zhu, X.-J., Hobbs, G., Wen, L., et al.\ 2014,
\href{http://adsabs.harvard.edu/abs/2014MNRAS.444.3709Z}
{\mnras, 444, 3709}

\end{thebibliography}
\end{document}